\documentclass[a4paper,draft=false,toc=bibliography]{scrartcl}
\usepackage{lmodern}
\usepackage[T1]{fontenc}
\usepackage[latin1]{inputenc}
\usepackage[english]{babel}
\usepackage{xint,xinttools,xintexpr,etoolbox}
\usepackage{amsmath,amsthm,braket,amssymb,tikz,caption,subcaption,color,booktabs,units,bbold,microtype}
\usetikzlibrary{decorations.pathmorphing}
\usetikzlibrary{decorations.markings}
\usetikzlibrary{knots}
\usetikzlibrary{patterns}
\usetikzlibrary{calc}
\usepackage{csquotes}
\usepackage{hyperref}
\usepackage{todonotes}
\usepackage{dsfont}
\usepackage{txfonts} 
\theoremstyle{plain}
\newtheorem{thm}{Theorem}

\newtheorem{lem}[thm]{Lemma}

\theoremstyle{remark}

\newcommand{\supp}{\operatorname{supp}}

\newcommand{\de}{\, \mathrm{d}}

\newcommand{\asterisknum}{\addtocounter{equation}{1} \tag{\theequation}}
\DeclareMathOperator{\Tr}{Tr}
\DeclareMathOperator{\tr}{tr}

\DeclareMathOperator{\e}{e}
\DeclareMathOperator{\Li}{Li}
\numberwithin{equation}{subsection}
\newcommand{\Rr}{\mathbb{R}}
\newcommand{\Nn}{\mathbb{N}}
\newcommand{\Zz}{\mathbb{Z}}
\newcommand{\Cc}{\mathbb{C}}

\title{The free energy of the two-dimensional dilute Bose gas. I. Lower bound}

\author{Andreas Deuchert\thanks{\texttt{andreas.deuchert@math.uzh.ch}} \ \quad Simon Mayer\thanks{\texttt{simon.mayer@ist.ac.at}} \ \quad Robert Seiringer\thanks{\texttt{robert.seiringer@ist.ac.at}}\\[0.3cm]
	$^{*,\dagger,\ddagger}$Institute of Science and Technology Austria (IST Austria)\\
	\vspace{0.34cm} Am Campus 1, 3400 Klosterneuburg, Austria \\ 
	$^*$Institute of Mathematics, University of Zurich \\
	Winterthurerstrasse 190, 8057 Zurich, Switzerland} 

\begin{document}

\maketitle

\begin{abstract}
	We prove a lower bound for the free energy (per unit volume) of the two-dimensional Bose gas in the thermodynamic limit. We show that the free energy at density $\rho$ and inverse temperature $\beta$ differs from the one of the non-interacting system by the correction term $4 \pi \rho^2 |\ln a^2 \rho|^{-1} (2 - [1 - \beta_{\mathrm{c}}/\beta]_+^2)$. Here $a$ is the scattering length of the interaction potential, $[\cdot]_+ = \max\{ 0, \cdot \}$ and $\beta_{\mathrm{c}}$ is the inverse Berezinskii--Kosterlitz--Thouless critical temperature for superfluidity. The result is valid in the dilute limit $a^2\rho \ll 1$ and if $\beta \rho \gtrsim 1$.
\end{abstract}

\section{Introduction and main result}

\subsection{Introduction}
Dilute quantum gases have proven a fruitful field of research for several decades in both experiment and theory. One of the milestones in the field was the experimental observation of Bose--Einstein condensation in alkali gases \cite{AEMWC95,DMAvDDKK95}, which was followed by an impressive activity in the field and also by a re-examination of fundamental properties of interacting Bose and Fermi systems. Since the dilute setting is characterized by a small parameter it allows for an investigation of the many-body problem with rigorous mathematical techniques.

One of the fundamental quantities of a quantum gas is its ground state energy per unit volume in the thermodynamic limit. In case of a three-dimensional dilute Bose gas the leading order asymptotics is given by
\begin{equation}
e^\text{3D}(\rho) = 4 \pi a \rho^2 (1 + o(1)).
\end{equation}
Here $a$ denotes the scattering length of the interaction potential and $\rho$ is the density of the gas. The above formula becomes exact in the dilute limit $a^3 \rho \to 0$. An upper bound for the case of the hard sphere gas was obtained in 1957 by Dyson \cite{Dyson1957}. The corresponding lower bound was established only much later by Lieb and Yngvason in 1998 \cite{LY1998} and can be considered as a major mathematical breakthrough. An upper bound for general interaction potentials can be found in \cite{LSY00}. Rigorously proving the form of the next order correction term for the ground state energy (the Lee--Huang--Yang formula), predicted to equal 
\begin{equation}\label{eq:9}
	4 \pi a \rho^2 \frac{128}{15 \sqrt{\pi}} \sqrt{a^3 \rho}
\end{equation}
in \cite{LHY57,LY57}, has been an open problem in mathematical physics for a long time and was recently achieved in \cite{YY09} (upper bound) and \cite{FS19} (lower bound); see also \cite{ErdSchlYau08,GiuSei09,BrieSol} for partial results in this direction, and \cite{benjaminetal} for related work on the Gross--Pitaevskii limit. For predictions of higher order corrections to these formulas we refer to \cite{Wu59,Liu63,Lieb64}. 

In two dimensions, the leading order term for the ground state energy per unit volume is given by
\begin{equation} \label{eqn-2d-gse-asymptotics}
e^\text{2D}(\rho) = \frac{4 \pi \rho^2}{|\ln a^2 \rho|} (1 + o(1))
\end{equation}
as proved in \cite{LY2001}. In this case, the $o(1)$ correction term is small when $a^2 \rho$ is small, which is the dimensionless small parameter characterizing the diluteness of the system in two dimensions. In contrast to the three-dimensional case, the two-dimensional ground state energy is \emph{not} the sum of the ground state energy of $N(N-1)/2$ pairs of particles; it is much larger. In particular, the coupling parameter $|\ln a^2 \rho |^{-1}$ depends on the density. The first prediction of \eqref{eqn-2d-gse-asymptotics} can be found in \cite{Schick71}. The next order correction to \eqref{eqn-2d-gse-asymptotics} is expected to be of the form
\begin{equation}\label{eqn-2d-gse-asymptoticsnextorder}
	\frac{- 4 \pi \rho^2 \ln |\ln a^2 \rho|}{|\ln a^2 \rho|^2},
\end{equation}
see e.g. \cite{Andersen02,CM09}.

At positive temperature, the natural analogue of the ground state energy is the free energy. In three dimensions, the free energy per unit volume of a dilute Bose gas in the thermodynamic limit satisfies the following asymptotic formula
\begin{equation} \label{eqn-3d-free-energy-asymptotics}
f^{\text{3D}}(\beta,\rho)  = f_0^\text{3D}(\beta,\rho) + 4 \pi a \rho^2 \left( 2 - \left[ 1 - \left( \frac{\beta_\mathrm{c}^\text{3D}(\rho)}{\beta}\right)^{3/2} \right]_+^2 \right) (1 + o(1)).
\end{equation}
Here, $f_0^\text{3D}(\beta,\rho)$ is the free energy of non-interacting bosons, $[ \, \cdot \, ]_+ = \max \{0, \, \cdot \, \}$ denotes the positive part, $\beta$ is the inverse temperature and $\beta_\mathrm{c}^\text{3D}(\rho) = \zeta(3/2)^{2/3}/(4 \pi \rho^{2/3})$ is the inverse critical temperature for Bose--Einstein condensation of the ideal Bose gas in three dimensions. The form of the interaction term results from the bosonic nature of the particles. Two bosons in different one-particle wavefunctions feel an exchange effect that increases their interaction energy by a factor of two compared to the case when they are in the same one-particle wavefunction. The $[\cdot]_+$-bracket in \eqref{eqn-3d-free-energy-asymptotics} equals the condensate fraction of the ideal gas, which is, to leading order also the fraction of those particles that do not feel an exchange effect. The free energy asymptotics \eqref{eqn-3d-free-energy-asymptotics} was proved in \cite{Seiringer2008} (lower bound) and \cite{Yin2010} (upper bound). It is valid in case $\beta \rho^{2/3} \gtrsim 1$, that is, if $\beta$ is of the order of the critical temperature of the ideal gas or larger (as $a^3 \varrho \to 0$).

Corresponding formulas for the ground state energy and the free energy of the two- and the three-dimensional dilute Fermi gas have been proven in \cite{LSS05} and \cite{Seiringer06}. We also mention the series of works \cite{NRS18,NRS17,NRS18A,FNRS19}, where the ground state energy and the free energy of the dilute Bose gas in two and three spatial dimensions were investigated by restricting attention to quasi-free states. These articles contain  formulas for the energy and  critical temperature that are conjecturally valid in a combined dilute and weak-coupling limit. 

In this work we consider the free energy per unit volume of the two-dimensional dilute Bose gas. More precisely, we are going to prove a lower bound of the form
\begin{equation} \label{eqn-2d-free-energy-asymptotics}
f^{\text{2D}}(\beta, \rho) \geq f_0^{\text{2D}}(\beta,\rho) + \frac{4 \pi \rho^2}{|\ln a^2 \rho|} \left( 2 - \left[ 1 - \frac{\beta_\mathrm{c}^\text{2D}(\rho,a)}{\beta} \right]_+^2 \right) (1 - o(1)).
\end{equation}
Here, $\beta_\mathrm{c}^\text{2D}(\rho,a)$ is the inverse Berezinskii--Kosterlitz--Thouless critical temperature for superfluidity given by
\begin{equation} \label{eqn-def-crit-temp}
\beta_\text{c}^\text{2D}(\rho,a) = \frac{\ln |\ln a^2 \rho|}{4 \pi \rho},
\end{equation}
see \cite{B71,B72,KT73,K74}. The term $\rho [1 - \beta_{\mathrm{c}}^\text{2D}(\rho,a)/\beta]_+$ in \eqref{eqn-2d-free-energy-asymptotics} has the physical interpretation of the superfluid density \cite{FH88}. For a thorough discussion of the physics of the superfluid phase transition in the two-dimensional Bose gas we refer to \cite{Popov83}. We emphasize that the inverse critical temperature $\beta_\text{c}^\text{2D}(\rho,a)$ depends on the interaction potential via its scattering length. This has to be contrasted with the situation in three dimensions,  where the critical temperature for Bose--Einstein condensation of the ideal gas appears in the formula \eqref{eqn-3d-free-energy-asymptotics} for the free energy. A comparable behavior cannot be expected in two space dimensions because the Mermin--Wagner--Hohenberg theorem \cite{MerminWagner66,Hohenberg67} excludes Bose-Einstein condensation at positive temperatures in this case. To the best of our knowledge the formula \eqref{eqn-2d-free-energy-asymptotics} does not seem to  have appeared explicitly in the literature before. It ought to be possible, however, to obtain it from the analysis  in \cite{FH88}. The corresponding upper bound for $f^{\text{2D}}(\beta, \rho)$ is of the same form as \eqref{eqn-2d-free-energy-asymptotics} and is given in  \cite{MS19}. In combination, \eqref{eqn-2d-free-energy-asymptotics} and this upper bound establish the first two terms in the free energy asymptotics of the two-dimensional dilute Bose gas. 

In the following we will exclusively deal with the two-dimensional system and therefore drop the superscript ``2D'' on the free energies $f^\text{2D}$ and $f_0^\text{2D}$, as well as on the inverse critical temperature $\beta_{\mathrm{c}}^\text{2D}(\rho,a)$.

\subsection{The model} \label{sec:model}

We consider the Hamiltonian for $N$ bosons in a two-dimensional torus $\Lambda$, given by
\begin{equation} \label{eqn-def-H_N}
H_N = -\sum_{i=1}^N \Delta_i + \sum_{i<j}^N v(d(x_i,x_j)),
\end{equation}
where $\Delta_i$ is the Laplacian on $\Lambda$ acting on the $i$-th particle, $d(x,y)$ is the distance function on the torus and $v \geq 0$ is a measurable two-body interaction potential with finite scattering length $a$ (to be defined properly below). The interaction potential is allowed to take the value $+\infty$ on a set of nonzero measure, which in particular permits to model the interaction between hard disks. This Hamiltonian acts on the symmetric tensor product of square integrable functions on the torus
\begin{equation}
\mathcal{H}_N = \bigotimes_{\text{sym}}^N L^2(\Lambda).
\end{equation}
We will describe the torus $\Lambda$ as a square of side length $L$ embedded in the plane with opposing sides identified, i.e., we have $\Lambda = [0,L]^2 \subset \Rr^2$. Then $\Delta$ is the usual Laplacian on $[0,L]^2$ with periodic boundary conditions and the distance function $d(x,y)$ is explicitly given as
\begin{equation}
d(x,y) = \min_{k \in \Zz^2} |x - y - k L|.
\end{equation}

The quantity of interest is the free energy per unit volume of the system as a function of the inverse temperature $\beta = 1/T$ and density $\rho$ defined by
\begin{equation}
f(\beta,\rho) = - \frac 1 \beta \lim_{\substack{N,L \to \infty\\ N/L^2 = \rho}} \frac{1}{L^2} \ln \Tr_{\mathcal{H}_N} \e^{-\beta H_N}.
\end{equation}
The limit is the usual thermodynamic limit\footnote{Existence of this limit (and independence of the boundary conditions used) can be shown by standard techniques, see, e.g., \cite{Robinson71,Ruelle69}.} of large particle number $N$ and large volume $L^2$ (area, really) while keeping the density $\rho = N/L^2$ fixed. The free energy asymptotics we will give applies to the setting of a dilute gas, where the parameter $a^2 \rho$ is small while $\beta \rho$ is of order one or larger. In other words, the scattering length is supposed to be small compared to the average particle distance while the thermal wave length of the particles is of the same order as the average particle distance or larger.

\subsection{The ideal Bose gas}
\label{sec:freegas}
For non-interacting bosons, the free energy density can be calculated explicitly. One has to solve the maximization problem
\begin{equation}
f_0(\beta,\rho) = \sup_{\mu \leq 0} \left\{ \mu \rho + \frac{1}{4 \pi^2 \beta} \int_{\Rr^2} \ln \left( 1 - \e^{- \beta(p^2 - \mu)} \right) \de p  \right\}.
\end{equation}
The chemical potential $\mu_0$ that maximizes the free energy satisfies the equation
\begin{equation}
\frac{1}{4 \pi^2} \int_{\Rr^2} \frac{\de p}{\e^{\beta(p^2 - \mu_0)} - 1} = \rho
\end{equation}
and therefore reads
\begin{equation} \label{eqn-chemical-potential-ideal-gas}
\mu_0(\beta, \rho) = \frac 1 \beta \ln \left( 1 - \e^{- 4 \pi \beta \rho} \right).
\end{equation}
This corresponds to the following explicit form of the free energy
\begin{equation}
f_0(\beta, \rho) = \rho^2 \left[ \frac{1}{\beta \rho} \ln \left( 1 - \e^{- 4 \pi \beta \rho} \right) - \frac{1}{4 \pi (\beta \rho)^2} \Li_2 \left( 1 - \e^{- 4 \pi \beta \rho} \right) \right],
\end{equation}
where
\begin{equation}
\Li_2(z) = -\int_0^z \frac{\ln(1 - t)}{t} \de t
\end{equation}
is the polylogarithm of order 2 (also called the dilogarithm). From this expression for the free energy of free bosons we directly obtain the scaling relation
\begin{equation}
f_0(\beta,\rho) = \rho^2 f_0(\beta \rho, 1).
\end{equation}
In particular, we see that for the free system the dimensionless parameter $\beta \rho$ completely determines (up to a factor of $\rho^2$) the free energy. We have the asymptotic behavior
\begin{align*} \label{eqn-freegas-asymptotics}
f_0(x,1) &= - \frac{\pi}{24 x^2} \left( 1 + O(\e^{-4 \pi x}) \right) &&\text{as } x \to \infty,\\
f_0(x,1) &= -\frac{1}{x} \left( 1 - \ln(4 \pi x) \right) - \pi + O(x) &&\text{as } x \to 0. \asterisknum
\end{align*}

\subsection{Scattering length}

The scattering length $a$ is defined by a variational principle, see \cite[Appendix A]{LY2001}. Let us first  assume  that the potential $v : \mathbb{R}_+ \to \mathbb{R}_+$ has a finite range $R_0$, i.e., we have $v(r) = 0$ for $r > R_0$. Then for $R>R_0$, we define the scattering length of $v$ by
\begin{equation} \label{eqn-scattering-length-1}
\frac{2\pi}{\ln(R/a)} = \inf_g \left\{ \int_{B_R} |\nabla g|^2 + \frac v 2 |g|^2 \right\},
\end{equation}
where the infimum is taken over functions $g \in H^1(B_R)$ with value one on the boundary, i.e., they satisfy $g|_{|x|=R} =1$. Here, $B_R \subset \Rr^2$ denotes the disk of radius $R$ centered at the origin. The unique function $g_0$, for which the infimum on the right-hand side of \eqref{eqn-scattering-length-1}  is attained,  is nonnegative, radially symmetric and satisfies the equation
\begin{equation}
-2 \Delta g_0 + v g_0 = 0
\label{eq:2}
\end{equation}
in the sense of quadratic forms, i.e.,  when integrated against any test function $\varphi \in H^{1}_0(B_R)$ with $\int_{B_R} | \varphi(x) |^2 v(x) \de x < + \infty$. Outside the range of the potential, i.e. for $R_0 < r < R$, the minimizer $g_0$ is explicitly given by
\begin{equation}
g_0(r) = \frac{\ln(r/a)}{\ln(R/a)}.
\end{equation}

As noted in the remark after the proof of \cite[Lemma~A.1]{LY2001}, the definition of the scattering length can be extended to potentials of infinite range by cutting off the potential at a finite range and then letting the cutoff grow to infinity. From \cite[Lemma~1]{LandonSeiringer12}, we know that finiteness of the scattering length is equivalent to a certain integrability condition of the potential. More precisely, if $a<\infty$, then
\begin{equation} \label{eqn-scattering-length-finite}
\int_{|x| > a} v(|x|) \ln^2(|x|/a) \de x < \infty
\end{equation}
holds. Conversely, if \eqref{eqn-scattering-length-finite} holds with $a$ replaced by some $b>0$, then the scattering length of the potential is finite. 

We remark  that defining the scattering length via this variational principle also makes sense for potentials that are not necessarily nonnegative. One has to assume  that $-\Delta + v/2$ as an operator on $L^2(\Rr^2)$ has no negative spectrum, however.

\subsection{Main theorem}
\label{sec:maintheorem}

The main result of this work is an asymptotic lower bound on the free energy in terms of the free energy of non-interacting bosons and a correction term coming from the interaction. It is the two-dimensional analogue of \cite[Theorem~1]{Seiringer2008}. The bound becomes useful for small $a^2 \rho$ and if $\beta \rho \gtrsim 1$. We use the standard  notation $x \lesssim y$ to indicate that there exists a constant $C > 0$, independently of $x$ and $y$, such that $x \leq C y$ (and analogously for ``$\gtrsim$''). If $x \lesssim y$ and $y \lesssim x$ we write $x \sim y$.

\begin{thm}[Free energy asymptotics of two-dimensional dilute Bose gas] \label{thm-lb}
	Assume that the interaction potential satisfies $v \geq 0 $ and has a finite scattering length. As  $a^2 \rho \to 0$ with $\beta \rho \gtrsim 1$, we have
	\begin{equation} \label{eqn-thm-free-energy-asymptotics}
	f(\beta,\rho) \geq f_0(\beta,\rho) + \frac{4 \pi \rho^2}{|\ln a^2 \rho|} \left(2 - \left[ 1 - \frac{\beta_{\mathrm{c}}(\rho,a)}{\beta} \right]_+^2 \right) (1 - o(1)),
	\end{equation}
	with
	\begin{equation} \label{eqn-thm-main-error-rate}
	o(1) \lesssim \frac{\ln \ln |\ln a^2 \rho|}{\ln |\ln a^2 \rho |}.
	\end{equation}
	Here, $[ \, \cdot \, ]_+ = \max\{ \, \cdot \, , 0\}$ denotes the positive part and the inverse critical temperature $\beta_{\mathrm{c}}(\rho,a)$ is defined in \eqref{eqn-def-crit-temp}.
\end{thm}

\subsubsection*{Remarks}

\begin{enumerate}

	\item The proof of a corresponding upper bound of the same form as \eqref{eqn-thm-free-energy-asymptotics} is given in \cite{MS19}. In combination with our result here this establishes  \eqref{eqn-thm-free-energy-asymptotics} as an equality, i.e.,  the first two terms in the asymptotic expansion of the free energy of the two-dimensional Bose gas in the dilute limit.
	
	\item The lower bound on the $o(1)$ error term given here is uniform in $\beta \rho$ as long as $\beta \rho \gtrsim 1$. The proof will show that the actual error rate is much better for $\beta \rho$ some distance away from $\beta_{\mathrm{c}} \rho$ (either above or below), see \eqref{eqn-total-error}.  For very low temperatures, we utilize the proof method of \cite{LY2001}; this way, we recover the ground state energy error rate $|\ln a^2 \rho|^{-1/5}$ for very low temperatures, which was proved for $T=0$ in \cite{LY2001}.
	
	\item The statement is uniform in the interaction potential in the following sense. In case of finite range potentials the error term depends on the interaction potential only through its scattering length $a$ and its range $R_0$. This dependence could be displayed explicitly. To prove the theorem for infinite range potentials with a finite scattering length one has to cut the potential at some radius $R_0$, which results in an  error term (contained in the $o(1)$ in \eqref{eqn-thm-free-energy-asymptotics}) of the form
	\begin{equation}
		\frac{1}{|\ln a^2 \rho |} \int_{|x| > R_0} v(|x|) \ln^2(|x|/a_{R_0}) \de x,
		\label{eq:errorinfiniterange}
	\end{equation}
	where $a_{R_0}$ is the scattering length of the potential with cutoff. When $R_0$ is chosen such that $a_{R_0} \neq 0$, this term is much smaller than the main error term \eqref{eqn-thm-main-error-rate}, but is non-uniform in the potential since $a_{R_0}$ depends on $v$. Note that in contrast to the three-dimensional case one does not need to choose $R_0/a \gg 1$. How one obtains \eqref{eq:errorinfiniterange} is explained in  detail in Lemma~\ref{lem-finite-range} below.

	\item Even though the temperature dependence of the correction term in \eqref{eqn-thm-free-energy-asymptotics} looks very similar to the three-dimensional case \eqref{eqn-3d-free-energy-asymptotics}, the two-dimensional case is actually rather different. While in three dimensions  it is possible to obtain a term of the correct form
	 by naive perturbation theory (with $(8\pi)^{-1} \int v$ in place of the scattering length), this  fails to be the case in two dimensions, for two reasons. First, one would similarly obtain the integral of the potential as a factor in the correction term, which does {\em not} yield the correct behavior in the density (namely the inverse logarithmic factor $|\ln a^2 \rho |^{-1}$). Secondly, the temperature dependence in the correction term would come out wrong, as the critical temperature for Bose--Einstein condensation in two dimensions is equal to zero, hence a factor 2 (compared to zero temperature) would appear at any $T>0$. In other words, in two dimensions a naive perturbation theory would yield
	\begin{equation}
		f_0(\beta,\rho) +  \rho^2 \int v(|x|) \de x,
	\end{equation}
	which differs from the true result in the two instances just described.

	\item The origin of the temperature dependence in the interaction term in \eqref{eqn-thm-free-energy-asymptotics} can be understood from the variational principle 
	\begin{align}
		&\inf _{0 \leq \rho_0 \leq \rho} \left\{ f_0(\beta, \rho - \rho_0) + \frac{4 \pi}{|\ln a^2 \rho|} \left(2 \rho^2 - \rho_0^2 \right) \right\} \nonumber \\
		&\hspace{5cm} = f_0(\beta, \rho) + \frac{4 \pi}{|\ln a^2 \rho|} \left(2 \rho^2 - \rho_{\mathrm{s}}^2 \right)(1-o(1)) \label{eq:variationalprinciple}
	\end{align}
	as $a^2\rho\to 0$. 
	To leading order, the optimal choice of $\rho_0$ turns out to be  $\rho_{\mathrm{s}} = \rho [1 - \beta_{\mathrm{c}}(\rho,a)/\beta]_+$, which coincides with the superfluid density of the system \cite{FH88}. One key ingredient of the proof of the lower bound for the free energy is a $c$-number substitution for low momentum modes. These modes are described by coherent states that do not experience an exchange effect, which decreases their energy relative to the energy of the high momentum modes that have not been substituted. The $c$-number substituted momentum modes take the role of $\rho_0$ and one obtains a formula for the  energy that is approximately given by the left-hand side of \eqref{eq:variationalprinciple}. 
		
\end{enumerate}

The proof of Theorem~\ref{thm-lb} is given in Section~\ref{sec:proof} below. It suitably adapts  the technique used to prove the related formula in the three-dimensional case \cite{Seiringer2008} and, for ease of comparison, we shall use the same section numbers and names as in that reference. For the convenience of the reader we give a short sketch of the proof highlighting the main ideas before we start with the detailed analysis.

\subsubsection*{The proof strategy}

A key ingredient in the proof of the lower bound for the free energy of the interacting gas is the observation that the second term on the right-hand side of \eqref{eqn-thm-free-energy-asymptotics} (the interaction energy) is, in the dilute limit, much smaller than the first term $f_0(\beta,\rho)$. As remarked above, a naive version of first order perturbation theory fails, however, for two reasons: First, the interaction potential is so strong that the interaction energy of the Gibbs state of the ideal gas is too large (it is even infinite in case of hards discs). Secondly, the temperature dependence of the interaction term comes out wrong, as $\rho [1 - \beta_{\mathrm{c}}(\rho,a) / \beta ]_+$ depends on the scattering length, which clearly cannot be captured by an ideal gas state. 

The first problem is overcome with the aid of a version of the Dyson Lemma \cite{Dyson1957}. This Lemma allows  to replace the strong interaction potential $v$ by a softer potential with a longer range that can later be treated using a rigorous version of first order perturbation theory. The price one has to pay is a certain amount of the kinetic energy. It is important that only modes with momenta much larger than $\beta^{-1/2}$ are used in this procedure because the other modes are needed to build up the free energy $f_0(\beta,\rho)$ of the ideal gas. A version of the Dyson Lemma fulfilling such requirements was for the first time proven in \cite{LSS05} to treat the ground state energy of the dilute Fermi gas.

After this replacement we utilize a rigorous version of first order perturbation theory at positive temperature, which was developed in \cite{Seiringer2008}. 
The method is based on a correlation inequality \cite{Seiringer06B} that applies to fermionic systems at all temperatures and to bosonic systems at sufficiently large temperatures. The main ingredient needed for this method to work is that the reference state in the perturbative analysis (usually the Gibbs state of the corresponding ideal gas) shows an approximate tensor product structure with respect to localization in different regions in space. In case of a quasi-free state this is true if its one-particle density matrix shows sufficiently fast decay (in position space). In order to overcome this restriction, highly occupied low-momentum modes leading to long-range correlations have to be treated with a $c$-number substitution. I.e.,  coherent states on the bosonic Fock space are used to replace creation and annihilation operators of the  low momentum modes by complex numbers. Since coherent states show an exact tensor product structure with respect to localization in different regions in space they fit seamlessly into the  framework. Although there is no Bose--Einstein condensation in the two-dimensional  Bose gas, we are also faced with highly occupied low momentum modes at very low temperatures. 
As explained in Remark~5 above, the use of coherent states for the low momentum modes naturally leads to the correct temperature dependence of the interaction energy in \eqref{eqn-thm-free-energy-asymptotics}, whose origin is non-perturbative.

In order to be able to use a Fock space formalism, which is essential for the formalism of the $c$-number substitution, it will be necessary to replace the interaction potential $v$ by an integrable potential $\tilde{v}$ with uniformly bounded Fourier transform. In contrast to the three-dimensional case, we will need that the integral of $\tilde{v}$ is suitably small in order to control various error terms. This replacement will be done in the first step of the proof.

\section{Proof of Theorem~\ref{thm-lb}} \label{sec:proof}

We will frequently use the Heaviside step function in the proof and use the convention
\begin{equation}
\theta(x) = \begin{cases}
1 & \text{if } x \geq 0,\\
0 & \text{if } x < 0.
\end{cases}
\end{equation}
In particular, $\theta(0) = 1$. 

\subsection{Reduction to integrable potentials with finite range}

The statement of Theorem~\ref{thm-lb} is general in the sense that it allows interaction potentials that are infinitely ranged and possibly have infinite integral (e.g., in the case of a hard disc potential), while still having finite scattering length. In the following it will be convenient to work with integrable potentials with finite range. The first condition is of importance because for the Fock space formalism we need to assume that the interaction potential has a bounded Fourier transform. Since we want to prove a lower bound we can replace the original potential by a smaller one. The scattering length of the new potential is smaller, however. The following two Lemmas quantify the change of the scattering length if we do such a replacement. We start with a lemma that quantifies the change of the scattering length when the potential is replaced by one that is cut off at some finite radius $R_0$.

\begin{lem} \label{lem-finite-range}
	Let $v$ be a nonnegative radial potential with finite scattering length $a$. We denote by $v_{R_0}$ the potential with cutoff at $R_0 > 0$ (i.e., $v_{R_0}(r) = \theta(R_0 - r) v(r)$) and its scattering length by $a_{R_0}$. Then 
	\begin{equation} \label{eqn-lem-finite-range-statement}
	\frac{1}{\ln(R/a_{R_0})} \geq \left( \ln(R/a) + \frac{1}{4\pi} \int_{|x|>R_0} v(|x|) \ln^2(|x|/a_{R_0}) \de x \right)^{-1}
	\end{equation}
	for all $R>R_0$.
	\begin{proof}
		The claim is equivalent to the inequality
		\begin{equation} \label{eqn-lem-finite-range-claim}
		\ln(a_{R_0}/a) \geq - \frac{1}{4\pi}\int_{|x| \geq R_0} v(|x|) \ln^2(|x|/a_{R_0}) \de x.
		\end{equation}
		To show \eqref{eqn-lem-finite-range-claim}, we use the variational principle for the scattering length of the potential with cutoff at $R_1$, where $R_1$ is such that $R_0 < R_1 < R$. Let $\phi_{v_{R_0}}$ denote the minimizer of the energy functional \eqref{eqn-scattering-length-1} with potential $v_{R_0}$. Then we have
		\begin{align*}
		\frac{2\pi}{\ln(R/a_{R_1})} &\leq \int_{B_R} \left( |\nabla \phi_{v_{R_0}}|^2 + \frac{v_{R_1}}{2} |\phi_{v_{R_0}}|^2 \right) = \frac{2\pi}{\ln(R/a_{R_0})} + \pi \int_{R_0}^{R_1} v(r) |\phi_{v_{R_0}}(r)|^2 r \de r\\
		&= \frac{2\pi}{\ln(R/a_{R_0})} \left( 1 + \frac{1}{2\ln(R/a_{R_0})} \int_{R_0}^{R_1} v(r) \ln^2(r/a_{R_0}) r \de r \right). \asterisknum
		\end{align*}
		This implies
		\begin{equation}
		- \ln a_{R_1} \geq \frac{\ln(R/a_{R_0})}{1 + \frac{1}{2 \ln(R/a_{R_0})} \int_{R_0}^{R_1} v(r) \ln^2(r/a_{R_0}) r \de r} - \ln R
		\end{equation}
		and by taking the limit $R \to \infty$, we obtain
		\begin{equation}
		\ln(a_{R_0}/a_{R_1}) \geq - \frac 1 2 \int_{R_0}^{R_1} v(r) \ln^2(r/a_{R_0}) r \de r.
		\end{equation}
		We can now take the limit $R_1 \to \infty$ and obtain \eqref{eqn-lem-finite-range-claim}.
	\end{proof}
\end{lem}

When we apply  Lemma~\ref{lem-finite-range}, the cutoff parameter $R_0$ has to be chosen such that $a_{R_0} > 0$, which is the case if $v_{R_0}\not\equiv 0$. We shall  choose $R$ such that $\ln (R/a) \sim |\ln a^2\rho| \gg 1$, hence the second term on the right side of \eqref{eqn-lem-finite-range-statement} is indeed a small correction to the first term. The relative error term we obtain this way is proportional to
\begin{equation}
\frac 1 { |\ln a^2 \rho|} \int_{|x| > R_0} v(|x|) \ln^2(|x|/a_{R_0}) \de x \,,
\end{equation}
which is much smaller than other error terms we shall obtain below, see \eqref{eqn-largest-error-rate}. 

From now on we can thus assume that the interaction potential $v$ has a fixed finite range $R_0$. For simplicity of notation, we shall drop the subscript $R_0$ from $v$ and $a$.

The next lemma quantifies the change of the scattering length if we replace a potential $v$ with finite range $R_0$ by a smaller potential $\tilde v$ whose integral is bounded by some number $4 \pi \varphi > 0$. The error term we obtain is small as long as $\varphi$ is much greater than $1/\ln(R/a)$. In particular, $\varphi$ can be chosen as a small parameter, which is different from the corresponding three-dimensional case.

\begin{lem} \label{lem-scattering-length}
	Let $v$ be a nonnegative radial potential with finite range $R_0$ and scattering length $a$. For any $0<\delta <1$ and any $\varphi > 0$, there exists a potential $\tilde v$ with $0 \leq \tilde v \leq v$ such that $\int_{\Rr^2} \tilde v(|x|) \de x \leq 4 \pi \varphi$ and the scattering length $\tilde a$ of $\tilde v$ satisfies 
	\begin{equation} \label{eqn-statement-lem-scattering-length}
	\frac{1}{\ln(R/\tilde a)} \geq \frac{1}{\ln(R/a)} \left( 1 - \frac{1}{\sqrt{\varphi \ln(R/a)}} + \frac{\ln(1-\delta)}{\ln(R/a)} \right)
	\end{equation}
	for all $R > R_0$.
	\begin{proof}
		Let
		\begin{equation}
		t = \inf \left\{ s : \int_s^\infty r v(r) \de r < \infty \right\}
		\end{equation}
		and note that $t \leq a$ holds. To see this let $s>a$ and bound
		\begin{align*}
		\int_s^\infty r v(r) \de r &\leq \frac{1}{\ln^2(s/a)} \int_s^\infty r v(r) \ln^2(r/a) \de r\\
		&\leq \frac{1}{\ln^2(s/a)} \int_a^\infty r v(r) \ln^2(r/a) \de r \leq \frac{2 \ln(R_0/a)}{\ln^2(s/a)}, \asterisknum
		\end{align*}
		where the last inequality follows from an easy calculation, compare with \cite[Eqs.~(34)--(36)]{LandonSeiringer12}. From this calculation we see that $\int_s^\infty r v(r) \de r$ is finite for all $s>a$. 
		
		Now we distinguish two cases. Assume first that $\int_t^\infty r v(r) \de r \geq 2 \varphi$ (which includes the possibility that $v \to \infty$ in a non-integrable sense as $r \to t$). Then we choose $s \geq t$ such that $\int_s^\infty r v(r) \de r = 2 \varphi$ and define $\tilde v(r) = v(r) \theta(r-s)$. Let $\phi_v$ denote the minimizer of the energy functional \eqref{eqn-scattering-length-1} and define the function
		\begin{equation}
		\phi(r) = \left( \phi_{\tilde v}(r) - \phi_{\tilde v}(s) \frac{\ln(R/r)}{\ln(R/s)} \right) \theta(r-s),
		\end{equation}
		which is nonnegative and continuous. We use $\phi$ as test function in the variational principle for the scattering length and obtain the upper bound
		\begin{align*}
		\frac{2\pi}{\ln(R/a)} &\leq \int_{B_R} \left(|\nabla \phi|^2 + \frac{v}{2} |\phi|^2\right) = \int_{B_R} \overline{\phi} \left( - \Delta + \frac v 2 \right) \phi + \int_{\partial B_R} \overline{\phi} \nabla \phi \cdot n\\
		&= - \frac{\phi_{\tilde v}(s)}{2\ln(R/s)} \int_{B_R} \overline{\phi}(|x|) v(|x|) \ln(R/|x|) \theta(|x|-s) \de x + \int_{\partial B_R} \overline{\phi} \nabla \phi \cdot n, \asterisknum
		\end{align*}
		where we integrated by parts and used the zero-energy scattering equation \eqref{eq:2} for $\tilde v$ as well as the fact that the function $r \mapsto \ln(R/r)$ is harmonic away from zero. In the boundary integral, we denoted by $n$ the outward facing unit normal vector of the disk (which is in this case just the unit vector pointing in the radial direction). We note that the first term on the right-hand side is negative and can be dropped for an upper bound. Since $R > R_0$, the boundary term can be explicitly computed as
		\begin{equation}
		\int_{\partial B_R} \overline{\phi} \nabla \phi \cdot n = \frac{2\pi}{\ln(R/\tilde a)} + \frac{2\pi \phi_{\tilde v}(s)}{\ln(R/s)}.
		\end{equation}
		Hence,
		\begin{equation} \label{eqn-lem-int-ub1}
		\frac{1}{\ln(R/a)} \leq \frac{1}{\ln(R/\tilde a)} + \frac{\phi_{\tilde v}(s)}{\ln(R/s)}.
		\end{equation}
		Using the fact that $\phi_{\tilde v}(s)$ is always greater or equal than the asymptotic solution given by $\ln(s/\tilde a) / \ln(R/\tilde a)$, we obtain
		\begin{equation}
		\frac{\phi_{\tilde v}(s)}{\ln(R/s)} \leq \frac{1}{\ln(R/\tilde a)} \cdot \frac{1}{1/\phi_{\tilde v}(s) - 1}.
		\end{equation}
		We get an upper bound on $\phi_{\tilde v}(s)$ via the monotonicity of $\phi_{\tilde v}(r)$:
		\begin{equation} \label{eqn-lem-int-ub2}
		\frac{1}{\ln(R/a)} \geq \frac{1}{\ln(R/\tilde a)} \geq \frac 1 2 \int_s^\infty r v(r) \phi_{\tilde v}(r)^2 \de r \geq \phi_{\tilde v}(s)^2 \varphi.
		\end{equation}
		Therefore,
		\begin{equation}
		\phi_{\tilde v}(s) \leq \frac{1}{\sqrt{\varphi \ln(R/a)}}.
		\end{equation}
		In conclusion, we have shown that
		\begin{equation}
		\frac{1}{\ln(R/\tilde a)} \geq \frac{1}{\ln(R/a)} \left(1 - \frac{1}{\sqrt{\varphi \ln(R/a)}} \right),
		\end{equation}
		which proves the statement (for $\delta = 0$) in the first case.
		
		It remains to consider the second case: Assume $\int_t^\infty r v(r) \de r = 2 \varphi - T$ for some $T > 0$. We may assume further that $t>0$, since if $t=0$ we can take $\tilde v = v$ and there is nothing to prove. By the definition of $t$, we have that for any $0<\delta <1$
		\begin{equation}
		\int_{(1-\delta)t}^t r v(r) \de r = \infty.
		\end{equation}
		Therefore there exists a $\tau = \tau(T,\delta)$ such that
		\begin{equation}
		\int_{(1-\delta)t}^t r \min \{ v(r), \tau \} \de r = T.
		\end{equation}
		We define
		\begin{equation}
		\tilde v(r) = \begin{cases}
		v(r) & \text{if } r \geq t,\\
		\min\{ v(r), \tau \} & \text{if } (1-\delta) t \leq r < t,\\
		0 & \text{otherwise.}
		\end{cases}
		\end{equation}
		Note that
		\begin{equation}
		\int_0^\infty r \tilde v(r) \de r = \int_{(1-\delta)t}^\infty r \tilde v(r) \de r = 2 \varphi.
		\end{equation}
		By the same argument as before (cf.~equation \eqref{eqn-lem-int-ub1} with $s=t$) and with this definition of $\tilde v$, we obtain
		\begin{equation} \label{eqn-lem-int-case2}
		\frac{1}{\ln(R/a)} \leq \frac{1}{\ln(R/\tilde a)} + \frac{\phi_{\tilde v}(t)}{\ln(R/t)}.
		\end{equation}
		Similarly to \eqref{eqn-lem-int-ub2}, we have
		\begin{equation}
		\frac{1}{\ln(R/a)} \geq \frac{1}{\ln(R/\tilde a)} \geq \frac 1 2 \int_{(1-\delta)t}^\infty r \tilde v(r) \phi_{\tilde v}(r)^2 \de r \geq \phi_{\tilde v}((1-\delta)t)^2 \varphi.
		\end{equation}
		Therefore,
		\begin{equation} \label{eqn-lem-int-ub3}
		\phi_{\tilde v} ((1-\delta)t) \leq \frac{1}{\sqrt{\varphi \ln(R/a)}}.
		\end{equation}
		From \eqref{eq:2} we deduce that $\Delta\phi_{\tilde v}$ defines a positive measure, and using Gauss' theorem, we have
		\begin{equation}
		\int_{|x| \leq r} \Delta \phi_{\tilde v}  = \int_{|x| = r} \nabla \phi_{\tilde v} \cdot n  =  2 \pi r \phi_{\tilde v}'(r).
		\end{equation}
		Since the left-hand side is increasing in $r$,  we conclude  that $r \mapsto r \phi_{\tilde v}'(r)$ is monotone increasing. This implies for any $s \leq r$ and for $r \geq R_0$
		\begin{equation}
		s \phi_{\tilde v}'(s) \leq r \phi_{\tilde v}'(r) = \frac{1}{\ln(R/ \tilde a)}.
		\end{equation}
		Thus, using the fundamental theorem of calculus,
		\begin{align*} \label{eqn-lem-int-ub4}
		\phi_{\tilde v} (t) - \phi_{\tilde v}((1-\delta)t) &= \delta t \int_0^1 \phi_{\tilde v}'((1-\delta w)t) \de w\\
		&\leq \frac{\delta}{\ln(R/\tilde a)} \int_0^1 \frac{\de w}{1 - \delta w} = - \frac{\ln(1-\delta)}{\ln(R/\tilde a)}. \asterisknum
		\end{align*}
		Putting \eqref{eqn-lem-int-case2}, \eqref{eqn-lem-int-ub3} and \eqref{eqn-lem-int-ub4} together as well as using $t \leq a$ and $\tilde a \leq a$, we obtain
		\begin{align*}
		\frac{1}{\ln(R/a)} &\leq \frac{1}{\ln(R/\tilde a)} + \frac{\phi_{\tilde v}(t)}{\ln(R/t)}\\
		&\leq \frac{1}{\ln(R/\tilde a)} + \frac{1}{\ln(R/t)} \left( \phi_{\tilde v}(t) - \phi_{\tilde v}((1-\delta)t) \right) + \frac{1}{\ln(R/t)} \frac{1}{\sqrt{\varphi \ln(R/a)}}\\
		&\leq \frac{1}{\ln(R/\tilde a)} - \frac{\ln(1-\delta)}{\ln(R/a)^2} + \frac{1}{\ln(R/a)} \frac{1}{\sqrt{\varphi \ln(R/a)}}. \asterisknum
		\end{align*}
		Rearranging the terms, we obtain \eqref{eqn-statement-lem-scattering-length}.
	\end{proof}
\end{lem}
In the following we denote by $\tilde{v}$ the interaction potential that is obtained from $v$ (which is assumed to have finite range $R_0$ as discussed after Lemma~\ref{lem-finite-range}) by cutting it as indicated by Lemma~\ref{lem-scattering-length}, such that its integral is bounded by $4 \pi \varphi > 0$. As mentioned already before we have $H_N \geq \tilde{H}_N$, where $\tilde{H}_N$ denotes the Hamiltonian with $v$ replaced by $\tilde{v}$.
\subsection{Fock space}
\label{sec:fockspace}

In our proof we relax the restriction on the number of particles, which is possible for a lower bound and is  motivated by the fact that this allows us to use the formalism of the $c$-number substitution, as detailed in the next subsection. We denote by $\mathcal{F}$ the bosonic Fock space and define the Fock space Hamiltonian
\begin{equation}
\mathbb{H} = \mathbb{T} + \mathbb{V} + \mathbb{K} + \mu_0 N
\label{eq:fockspace1}
\end{equation}
with
\begin{equation}
\mathbb{T} = \sum_p \left( p^2 - \mu_0 \right) a_p^{\dagger} a_p, \quad \mathbb{V} = \frac{1}{2 | \Lambda |} \sum_{p,k,\ell} \hat{v}(p) a^{\dagger}_{k+p} a^{\dagger}_{\ell-p} a_k a_{\ell}
\label{eq:fockspace2}
\end{equation}
and 
\begin{equation}
\mathbb{K} = \frac{4 \pi C}{| \Lambda | | \ln a^2 \rho |} \left( \mathbb{N} - N \right)^2.
\label{eq:fockspace2a}
\end{equation}
Here, the chemical potential $\mu_0$ is given by \eqref{eqn-chemical-potential-ideal-gas} and $a_p^\dagger$ and $a_p$ are the usual creation and annihilation operators that create or annihilate a plane wave with momentum $p$, respectively. The sums over $p$, $k$ and $\ell$ are taken over $\frac{2\pi}{L} \Zz^2$.  By $\hat{v}$ we denote the Fourier transform of $\tilde{v}$ (we drop the \textasciitilde{} in the Fourier transform for notational clarity), which is given by $\hat{v}(p) = \int_{\Lambda} \tilde{v}(d(x,0)) \e^{-ipx} \de x = \int_{\Rr^2} \tilde{v}(|x|) \e^{-ipx} \de x$. Here and in the following we assume that $L > 2 R_0$, which is no restriction since we are interested in the thermodynamic limit $L\to\infty$. Note that $\hat v$ is uniformly bounded, which is one reason we introduced $\tilde v$: We have
\begin{equation}
|\hat v(p)| \leq \hat v(0) \leq 4 \pi \varphi.
\end{equation} 
The number operator is defined by
\begin{equation}
\Nn = \sum_p a_p^\dagger a_p
\end{equation}
and the operator $\mathbb{K}$ was  introduced to have a  control on the number of particles in the system after the extension to Fock space. Note that  $\mathbb{K}$ vanishes on all states with exactly $N$ particles. The parameter $C>0$ in the definition of $\mathbb{K}$ will be suitably chosen later.

Recall that we defined the total Hamiltonian for $N$ particles by $H_N$ (in Eq.~\eqref{eqn-def-H_N}) and that we denote by $\tilde{H}_N$ the operator $H_N$ where $v$ is replaced by $\tilde{v}$. We then have $H_N \geq \tilde{H}_N = \mathbb{H} P_N$, where $P_N$ is the projection on the Fock space sector with $N$ particles. This implies in particular that
\begin{equation}
\Tr_{\mathcal{H}_N} \exp ( - \beta H_N ) \leq \Tr_{\mathcal{H}_N} \exp ( - \beta \tilde{H}_N ) \leq \Tr_{\mathcal{F}} \exp( -\beta \mathbb{H} ).
\label{eq:fockspace4}
\end{equation}
We will proceed deriving an upper bound for the expression on the right-hand side. 

\subsection{Coherent states}
\label{sec:coherentstates}

We use the method of coherent states (see, e.g., \cite{LSY05}) in order to obtain an upper bound for the partition function $\Tr_{\mathcal{F}} \exp( -\beta \mathbb{H} )$. This method is based on the fact that coherent states are eigenfunctions of the annihilation operators, which can be used to replace the operators $a_p$ and $a_p^\dagger$ by complex numbers. This procedure is also called  $c$-number substitution. Although we have no condensate in our system, this separate treatment of a certain number of low momentum modes is necessary for low temperatures, as  pointed out in the proof strategy in Section~\ref{sec:maintheorem}. We start by introducing the necessary notation related to the $c$-number substitution.

Pick some $p_{\text{c}} \geq 0$ and write $\mathcal{F} = \mathcal{F}_< \otimes \mathcal{F}_>$. Here $\mathcal{F}_<$ and $\mathcal{F}_>$ denote the Fock spaces corresponding to the modes $| p | < p_{\text{c}}$ and $| p | \geq p_{\text{c}}$, respectively. We define $M = \sum_{|p | < p_{\text{c}}} 1 = \# \{ p \in \frac {2\pi}L \mathbb{Z}^2 : |p| < p_\mathrm{c}\}$ and introduce for $z \in \mathbb{C}^M$ the coherent state $\ket{z} \in \mathcal{F}_<$ by
\begin{equation} \label{eq:coherentstates8}
\ket{z} = \exp \left( \sum_{|p| < p_{\text{c}}} z_p a_p^\dagger - \bar z_p a_p \right) | 0 \rangle =: U(z) | 0 \rangle.
\end{equation}
Here $| 0 \rangle$ is the vacuum vector in $\mathcal{F}_<$ and  the last equality defines the Weyl operator $U(z)$. The lower symbol $\mathbb{H}_\text{s}(z)$ of $\mathbb{H}$ is the operator on $\mathcal{F}_>$ given by the partial inner product
\begin{equation}
\mathbb{H}_\text{s}(z) = \braket{z | \mathbb{H} | z} = \mathbb{T}_{\mathrm{s}}(z) + \mathbb{V}_{\mathrm{s}}(z) + \mathbb{K}_{\mathrm{s}}(z).
\label{eq:10}
\end{equation}
We can use the fact that $a_p \ket{z} = z_p \ket{z}$ and obtain the lower symbol by simply replacing all $a_p$ by $z_p$ and $a_p^\dagger$ by $\bar z_p$ for $|p| < p_{\text{c}}$ in the normal-ordered form of the Hamiltonian. To display it explicitly,  let us introduce the notation
\begin{equation}
A_p = z_{p} \mathds{1}\left( |p| < p_{\mathrm{c}} \right) + a_{p} \mathds{1}\left( |p| \geq p_{\mathrm{c}} \right)
\end{equation}
with adjoint $A_p^{\dagger}$. The lower symbols of the operators on the right-hand side of \eqref{eq:10} are given by 
\begin{align}
	\mathbb{T}_{\mathrm{s}}(z) &= \sum_{p} \left( p^2 - \mu_0 \right) A_p^{\dagger} A_p, \quad \mathbb{V}_{\mathrm{s}}(z) = \frac{1}{2 | \Lambda |} \sum_{p,k,\ell} \hat{v}(p) A^{\dagger}_{k+p} A^{\dagger}_{\ell-p} A_{k} A_{\ell} \quad \text{ and } \\
	\mathbb{K}_{\mathrm{s}}(z) &= \frac{4 \pi C}{| \Lambda | | \ln a^2 \rho |} \left( \sum_{p,q} A_p^{\dagger} A_q^{\dagger} A_q A_p - \sum_{p} A_p^{\dagger} A_p (2N-1) + N^2 \right).
\end{align}
The upper symbol of an operator is the operator-valued function that is obtained by starting from the anti-normal ordered form of the operator and then replacing $a_p$ by $z_p$ and $a_p^\dagger$ by $\bar z_p$ for $|p| < p_\text{c}$. This implies that the upper symbol can be calculated from the lower symbol by replacing for example $|z_p|^2$ by $|z_p|^2 - 1$ and similarly for other polynomials in $z_p$ (see \cite{LSY05} for more details). The upper symbol $\mathbb{H}^\text{s}(z)$ of $\mathbb{H}$ satisfies
\begin{equation}
\label{eq:1}
\mathbb{H} = \int_{\Cc^M} \mathbb{H}^\text{s}(z) | z \rangle \langle z | \de z,
\end{equation}
where $\de z = \prod_{i=1}^M \frac{\de z_i}{\pi}$, $\de z_i = \de x_i \de y_i$ is the product measure related to the real and imaginary part of $z_i \in \Cc$. The Berezin--Lieb inequality \cite{bere,bere2,lieb,LSY05} implies
\begin{equation}
\Tr_{\mathcal{F}} \exp (-\beta \mathbb{H}) \leq \int_{\mathbb{C}^M} \Tr_{\mathcal{F}_>} \exp (-\beta \mathbb{H}^\text{s}(z)) \de z.
\label{eq:coherentstates1}
\end{equation}
We prefer to work with the lower symbol instead, and therefore will replace the upper by the lower symbol on the right-hand side of \eqref{eq:coherentstates1}. Let $\Delta \mathbb{H}(z) = \mathbb{H}_\text{s}(z) - \mathbb{H}^\text{s}(z)$ be the difference between the two symbols, which reads
\begin{align}
\Delta \mathbb{H}(z) =& \sum_{| p | < p_{\text{c}}} \left( p^2 - \mu_0 \right) + \frac{1}{2 | \Lambda |} \bigg[ \hat{v}(0) \left( 2 M \mathbb{N}_s(z) - M^2 \right) \nonumber \\
&+ 2 \sum_{| \ell | <p_{\text{c}}, |k| \geq p_{\text{c}}} \hat{v}(\ell - k) a_k^{\dagger} a_k + \sum_{| \ell |, |k| <p_{\text{c}}} \hat{v}(\ell - k) \left( 2 |z_k|^2 - 1 \right) \bigg] \nonumber \\
&+ \frac{4 \pi C}{| \Lambda | | \ln a^2 \rho |} \left[ 2 | z |^2 + M \left( 2 \mathbb{N}_s(z) - 2 N - M \right) \right], \label{eq:coherentstates2}
\end{align}
where $|z|^2 = \sum_{|p| < p_{\mathrm{c}}} |z_p|^2$ and $\mathbb{N}_{\mathrm{s}}(s) = |z|^2 + \sum_{|p| \geq p_{\mathrm{c}}} a_p^{\dagger} a_p$. Using the bound $| \hat{v}(p) | \leq \hat v (0) \leq  4 \pi \varphi$ we  have 
\begin{equation}
\Delta \mathbb{H}(z) \leq M\left( p_{\text{c}}^2 - \mu_0 \right) + \frac{8 \pi \varphi}{| \Lambda |} M \mathbb{N}_s(z) + \frac{8 \pi C}{|\Lambda| |\ln a^2\rho |} \left[ |z|^2 + M \left( \mathbb{N}_s(z) - N  \right) \right].
\label{eq:coherentstates3}
\end{equation}
The lower symbol of $\mathbb{K}$ reads
\begin{equation}
\mathbb{K}_\text{s}(z) =  \frac{4 \pi C}{|\Lambda| |\ln a^2\rho |} \left( \left( \mathbb{N}_\text{s}(z) - N \right)^2 + |z|^2 \right) \geq \frac{4 \pi C}{|\Lambda| |\ln a^2\rho |} \left( \mathbb{N}_\text{s}(z) - N \right)^2
\label{eq:coherentstates4}
\end{equation}
and allows us to estimate
\begin{align} \label{eq:coherentstates5}
\tfrac{1}{2} \mathbb{K}_\text{s}(z) - \Delta \mathbb{H}(z) &\geq - M (p_{\text{c}}^2 - \mu_0) - \frac{8 \pi N}{|\Lambda|} \left( \varphi M + \frac{C}{| \ln a^2 \rho | } \right) - \frac{32 \pi C (M+1)^2}{| \Lambda | | \ln a^2 \rho |} \left( 1 + \frac{\varphi | \ln a^2 \rho | }{C} \right)^2 \nonumber \\
&=: - Z^{(1)}.
\end{align}
Note that $M \sim p_{\text{c}}^2 | \Lambda |$ in the thermodynamic limit. We will choose the parameters $p_{\text{c}}$, $\varphi$ and $C$ such that $Z^{(1)} \ll | \Lambda | \rho^2/| \ln a^2 \rho |$ for small $a^2\rho$. 
We also define
\begin{equation}
F_z(\beta) = -\frac{1}{\beta} \ln \Tr_{\mathcal{F}_>} \exp \left( -\beta \left( \mathbb{T}_\text{s}(z) + \mathbb{V}_\text{s}(z) + \tfrac{1}{2} \mathbb{K}_\text{s}(z) \right) \right).
\label{eq:coherentstates6}
\end{equation}
Eq.~\eqref{eq:coherentstates1} and the above estimates imply the bound
\begin{equation}
-\frac{1}{\beta} \ln \Tr_{\mathcal{F}} \exp\left( -\beta \mathbb{H} \right) \geq \mu_0 N - \frac{1}{\beta} \ln \int_{\mathbb{C}^M} \exp \left( -\beta F_z(\beta) \right) \de z - Z^{(1)}.
\label{eq:coherentstates7}
\end{equation}
In the following subsections we will derive a lower bound for $F_z(\beta)$. 

The free energy $F_z(\beta)$ can also be written in terms of the free energy of a Gibbs state. In fact, let  $\Gamma^z$ be the Gibbs state of $\mathbb{T}_\text{s}(z) + \mathbb{V}_\text{s}(z) + \tfrac{1}{2} \mathbb{K}_\text{s}(z)$ on $\mathcal{F}_>$, i.e.,
\begin{equation} \label{eqn-def-Gamma-z}
\Gamma^z = \frac{\exp\left( - \beta \left[ \mathbb{T}_\text{s}(z) + \mathbb{V}_\text{s}(z) + \frac 1 2 \mathbb{K}_\text{s}(z) \right]\right)}{\Tr_{\mathcal{F}_>} \exp\left( - \beta\left[ \mathbb{T}_\text{s}(z) + \mathbb{V}_\text{s}(z) + \frac 1 2 \mathbb{K}_\text{s}(z) \right]\right)}\,,
\end{equation}
and define the state
\begin{equation}\label{def:Upsilon}
\Upsilon^z = U(z) \Pi_0 U(z)^{\dagger} \otimes \Gamma^z
\end{equation}
on $\mathcal{F}$, where $\Pi_0 = | 0 \rangle \langle 0 |$ denotes the vacuum state on $\mathcal{F}_<$. With these definitions we obtain the identity
\begin{equation}
F_z(\beta) = \Tr_{\mathcal{F}} \left[ \left( \mathbb{T}+ \mathbb{V}+ \tfrac{1}{2} \mathbb{K} \right) \Upsilon^z \right] - \frac{1}{\beta} S(\Upsilon^z),
\label{eq:coherentstates9}
\end{equation}
where $S(\Upsilon^z) = - \Tr_{\mathcal{F}} [\Upsilon^z \ln \Upsilon^z]$ is the von Neumann entropy of the state $\Upsilon^z$ (which equals the one of $\Gamma^z$).

\subsection{Relative entropy and a priori bounds}
\label{sec:relativeentropyandaprioribounds}
To prove a lower bound for $F_z(\beta)$ we will need some  information on the state $\Upsilon^z$ defined in \eqref{def:Upsilon} above. The  a-priori information that is being used is a bound on the relative entropy (to be defined below) of $\Upsilon^z$ with respect to a suitable reference state describing non-interacting bosons and a bound on the expected number of particles in the system. To obtain this a-priori information we will assume that a certain upper bound for $F_z(\beta)$ holds. This does not lead to a loss of generality because there will be nothing to prove if the assumption is not fulfilled. That is, the statement will hold independently of the assumption.

Let $\Gamma_0$ be the Gibbs state on $\mathcal{F}_>$ for the kinetic energy operator $\mathbb{T}_\text{s}(z)$ (which is independent of $z$) and define the state $\Omega_0^z$ on $\mathcal{F}$ by $\Omega_0^z = U(z) \Pi_0 U(z)^{\dagger} \otimes \Gamma_0$. Since $\mathbb{V} \geq 0$ we have
\begin{equation}
F_z(\beta) \geq - \frac 1 \beta \ln\left( \Tr_{\mathcal{F}_>}\left[ \e^{-\beta \mathbb{T}_\text{s}(z)} \right] \right) + \frac{1}{2} \Tr_{\mathcal{F}} \left[ \mathbb{K} \Upsilon^z \right] + \frac 1 \beta S(\Upsilon^z,\Omega^z_0), 
\label{eq:relativeentropyandaprioribounds1}
\end{equation}
where 
\begin{equation}
	S(\Upsilon^z,\Omega^z_0) = \Tr_{\mathcal{F}}\left[ \Upsilon^z \left( \ln\Upsilon^z - \ln\Omega^z_0 \right) \right]
	\label{eq:11}
\end{equation}
denotes the relative entropy of $\Upsilon^z$ with respect to $\Omega^z_0$. Since $\Upsilon^z$ and $\Omega^z_0$ are equal on $\mathcal{F}_<$ we have $S(\Upsilon^z,\Omega^z_0) = S(\Gamma^z,\Gamma_0)$. 
We distinguish two cases: Either
\begin{equation}
F_z(\beta) \geq - \frac 1 \beta \ln\left( \Tr_{\mathcal{F}_>}\left[ \e^{-\beta \mathbb{T}_\text{s}(z)} \right] \right) + \frac{8 \pi \vert \Lambda \vert \rho^2}{|\ln a^2\rho |} \label{eq:relativeentropyandaprioribounds2}
\end{equation}
holds or it does not hold. In the latter case we have
\begin{equation}
S(\Upsilon^z,\Omega_0^z) = S(\Gamma^z,\Gamma_0) \leq \frac{8 \pi | \Lambda | \beta \rho^2}{|\ln a^2\rho |}
\label{eq:relativeentropyandaprioribounds4}
\end{equation}
as well as
\begin{equation}
\Tr_{\mathcal{F}}\left[ \mathbb{K} \Upsilon^z \right] \leq \frac{16 \pi | \Lambda | \rho^2}{|\ln a^2\rho |}.
\label{eq:relativeentropyandaprioribounds5}
\end{equation}
From now on we will assume to be in the second case. The lower bound we are going to derive on $F_z(\beta)$ will actually be worse than \eqref{eq:relativeentropyandaprioribounds2} above, that is, the bound is true in any case, irrespective of whether the assumptions \eqref{eq:relativeentropyandaprioribounds4} and \eqref{eq:relativeentropyandaprioribounds5} hold.

Eq.~\eqref{eq:relativeentropyandaprioribounds5} implies the following upper bound on $|z|^2$:
\begin{align}
|z|^2 - N \leq \Tr_{\mathcal{F}} \left[ \left( \mathbb{N} - N \right) \Upsilon^z \right] &\leq \left( \Tr_{\mathcal{F}} \left[ \left( \mathbb{N} - N \right)^2 \Upsilon^z \right] \right)^{1/2} \nonumber \\
&= \left( \frac{| \Lambda | | \ln a^2 \rho | }{4 \pi C} \right)^{1/2} \left( \Tr_{\mathcal{F}}\left[ \mathbb{K} \Upsilon^z \right] \right)^{1/2} \leq \frac{2 }{\sqrt{C}} | \Lambda | \rho. \label{eq:relativeentropyandaprioribounds6}
\end{align}
In other words,
\begin{equation}
\rho_z := \frac{|z|^2}{|\Lambda|} \leq \rho \left( 1 + \frac{2}{\sqrt{C}} \right). \label{eq:relativeentropyandaprioribounds7}
\end{equation}
We will choose $C \gg 1$ below.
\subsection{Replacing vacuum}
\label{sec:replcaingvacuum}
In this section we replace the vacuum state $\Pi_0$ in the definition of $\Upsilon^z$  in \eqref{def:Upsilon} by a more general quasi-free state $\Pi$ on $\mathcal{F}_<$ and estimate the effect of this replacement on \eqref{eq:coherentstates9}. The replacement will become relevant in Subsection~\ref{sec:effectofcutoff} when we estimate the relative entropy of the above state with respect to a certain quasi-free state describing non-interacting bosons. For that purpose, we require the momentum distribution to be sufficiently smooth, and do not want it to jump to zero for momenta less than $p_c$. 

Let $\Pi$ be the unique quasi-free state on $\mathcal{F}_<$ whose one-particle density matrix is given by
\begin{equation}
\pi = \sum_{|p|<p_{\text{c}}} \pi_p | p \rangle\langle p |.
\label{eq:replacingvacuum1}
\end{equation} 
The coefficients $\pi_p$ will be chosen later. We  denote the trace of $\pi$ by $P$. Define the state $\Upsilon_{\pi}^z$ on $\mathcal{F}$ by 
\begin{equation}\label{def:Upsilonpi}
\Upsilon_{\pi}^z = U(z) \Pi U(z)^{\dagger} \otimes \Gamma^z \,.
\end{equation}
 Using $| \hat{v}(p) | \leq 4 \pi \varphi$, we see that\footnote{We note that in \cite[first line of (2.5.4)]{Seiringer2008} there is an erroneous term $- 2 \sum_{|k|<p_\text{c}} \pi_k |z_k|^2$. Since it is negative it was dropped for the following estimate, which resulted in an analogous upper bound on $\Tr_{\mathcal{F}}[\mathbb{V}(\Upsilon^z_\pi - \Upsilon^z)]$.}
\begin{align} \label{eq:replacingvacuum2}
\Tr_{\mathcal{F}}\left[ \mathbb{V} \left( \Upsilon_{\pi}^z - \Upsilon^z \right) \right] &= \frac{1}{2 | \Lambda |} \hat{v}(0) \left( P^2 + 2 P \Tr_{\mathcal{F}_>}\left[ \mathbb{N}_\text{s}(z) \Gamma^z \right] \right) + \frac{1}{2 | \Lambda |} \sum_{|k|,|\ell| < p_{\text{c}}} \hat{v}(k-\ell) \left[ \pi_k \pi_{\ell} + 2 |z_k|^2 \pi_{\ell} \right] \nonumber \\
&\qquad + \frac{1}{| \Lambda |} \sum_{|k| < p_{\text{c}} , |\ell| \geq p_{\text{c}}}  \hat{v}(k-\ell) \pi_k \Tr_{\mathcal{F}_>}\left[ a_{\ell}^{\dagger} a_{\ell} \Gamma^z \right] \nonumber \\
&\leq \frac{4 \pi \varphi}{| \Lambda |} \left( P^2 + 2 P \Tr_{\mathcal{F}_>} \left[ \mathbb{N}_{\mathrm{s}}(z) \Gamma^z \right] \right) = \frac{4 \pi \varphi}{| \Lambda |} \left( P^2 + 2 P \Tr_{\mathcal{F}} \left[ \mathbb{N} \Upsilon^z \right] \right).
\end{align}
To obtain the bound, we used that the last term in the first line plus the term in the second line are bounded from above by the first term on the right-hand side in the first line. In \eqref{eq:relativeentropyandaprioribounds6} we have shown that $\Tr_{\mathcal{F}}\left[ \mathbb{N} \Upsilon^z \right] \leq N ( 1 + 2/\sqrt{C} )$ and we therefore obtain from \eqref{eq:replacingvacuum2}
\begin{equation}
\Tr_{\mathcal{F}} \left[ \mathbb{V} \Upsilon^z \right] \geq \Tr_{\mathcal{F}} \left[ \mathbb{V} \Upsilon_{\pi}^z  \right] - Z^{(2)}
\label{eq:replacingvacuum3}
\end{equation}
with
\begin{equation}
Z^{(2)} := \frac{4 \pi \varphi P^2}{| \Lambda |} + \frac{8 \pi P \varphi}{| \Lambda |} N \left( 1 + \frac{2}{\sqrt{C}} \right).
\label{eq:replacingvacuum4}
\end{equation}
We will later choose $\varphi \gg |\ln a^2 \rho|^{-1}$ and $C \gg 1$. Hence, $Z^{(2)} \ll |\Lambda| \rho^2/| \ln a^2 \rho| $ as long as $\varphi P \ll N/| \ln a^2 \rho|$.

The replacement of $\Upsilon^z$ by $\Upsilon_{\pi}^z$ causes also a change in the kinetic energy given by
\begin{equation}
\Tr_{\mathcal{F}}\left[ \mathbb{T} \Upsilon^z \right] = \Tr_{\mathcal{F}}\left[ \mathbb{T} \Upsilon_{\pi}^z \right] - \sum_{|p| < p_{\text{c}}} \left( p^2 - \mu_0 \right) \pi_p.
\label{eq:replacingvacuum5}
\end{equation}
By combining \eqref{eq:coherentstates9}, \eqref{eq:replacingvacuum3} and \eqref{eq:replacingvacuum5} we therefore obtain the lower bound
\begin{equation}
F_z(\beta) \geq \Tr_{\mathcal{F}}\left[ \left( \mathbb{T} + \mathbb{V} \right) \Upsilon_{\pi}^z \right] + \tfrac{1}{2} \Tr_{\mathcal{F}} \left[ \mathbb{K} \Upsilon^z \right]- \frac{1}{\beta} S\left( \Upsilon^z \right) - \sum_{|p| < p_{\text{c}}} \left( p^2 - \mu_0 \right) \pi_p  -Z^{(2)}.
\label{eq:replacingvacuum6}
\end{equation}
\subsection{Dyson Lemma}
\label{sec:dysonlemma}
As already mentioned in the proof strategy in Section~\ref{sec:maintheorem}, in order to be in a perturbative regime we have to replace the short ranged and possibly very strong interaction potential $\tilde{v}$ by a softer interaction potential with longer range. To achieve this goal we have to pay with a certain amount of kinetic energy. More precisely, we will only use modes with momenta much larger than $\beta^{-1/2}$ for this procedure because the other momentum modes are needed to obtain the free energy $f_0(\beta,\rho)$ of the ideal gas. 

To separate the high momentum part of the kinetic energy (which is the relevant part contributing to the interaction energy) from the low momentum part, we choose a radial cutoff function $\chi : \Rr^2 \to [0,1]$ and define
\begin{equation}
h(x) = \frac{1}{|\Lambda|} \sum_{p} (1 - \chi(p)) \e^{-i p x}.
\end{equation}
We assume that $\chi(p) \to 1$ sufficiently fast as $|p| \to \infty$ so that $h \in L^1(\Lambda) \cap L^\infty(\Lambda)$. Define further for $R_0 < R < L/2$ 
\begin{equation} \label{eqn-Dyson-lemma-def-f-w}
f_R(x) = \sup_{|y| \leq R} | h(x-y) - h(x)| \quad \text{and} \quad w_R(x) = \frac 2 \pi f_R(x) \int_\Lambda f_R(y) \de y.
\end{equation}
Finally, we introduce the soft potential $U_R$ which is a nonnegative function supported on the interval $[R_0,R]$. Its integral should satisfy
\begin{equation} \label{eqn-U_R-cond}
\int_{R_0}^R U_R(t) \ln(t/ \tilde a) t \de t \leq 1.
\end{equation}
We then have the following statement:
\begin{lem} \label{lem-Dyson-lemma}
	Let $y_1, \ldots{}, y_n$ be $n$ points in $\Lambda$ and denote by $y_{\mathrm{NN}}(x)$ the nearest neighbor of $x \in \Lambda$ among the points $y_i$. Then for any $\epsilon > 0$, we have
	\begin{equation} \label{eqn-Dyson-lemma-statement}
	- \nabla \chi(p)^2 \nabla + \frac 1 2 \sum_{i=1}^n \tilde v(d(x,y_i)) \geq (1 - \epsilon) U_R(d(x,y_{\mathrm{NN}}(x))) - \frac{1}{\epsilon} \int_{\Rr_+} U_R(t) t \de t \sum_{i=1}^n w_R(x - y_i).
	\end{equation}
\end{lem}
We remark that $y_{\mathrm{NN}}(x)$ is well defined except on a set of zero measure. The Lemma above is a two-dimensional version of \cite[Lemma~2]{Seiringer2008}. It is referred to as Dyson Lemma because Dyson was the first to prove a  statement of this kind in his treatment of the dilute Bose gas at $T=0$ in \cite{Dyson1957}. A version of the Dyson Lemma for two and three space dimensions, where only high momentum modes are used to replace the interaction potential by a softer one, appeared for the first time in \cite{LSS05}. The proof of Lemma~\ref{lem-Dyson-lemma} can be obtained by combining the ideas of the proofs of \cite[Lemma~2]{Seiringer2008} and \cite[Lemma~7]{LSS05}. The main differences between Lemma~\ref{lem-Dyson-lemma} and \cite[Lemma~7]{LSS05} are the boundary conditions for the Laplacian and the fact that we do not assume a minimal distance between the particles here. Since the proof of \cite[Lemma~7]{LSS05} was not  spelled out in detail, we include a proof of Lemma~\ref{lem-Dyson-lemma} in Appendix~\ref{sec:proof-dysonlemma}.

We will use Lemma~\ref{lem-Dyson-lemma} for a lower bound on the operator $\mathbb{T} + \mathbb{V}$. In the Fock space sector with $n$ particles this operator reads
\begin{equation}
\tilde{H}_n = \sum_{j=1}^n \Biggl[ -\Delta_j + \frac{1}{2} \sum_{i,\,  i \neq j} \tilde{v}(d(x_i,x_j)) \Biggr].
\label{eq:afterDysonlemma1}
\end{equation}
We want to keep a small part of the total kinetic energy for later use and therefore write for $0 < \kappa < 1$
\begin{equation}
p^2 = p^2 \left( 1 - (1-\kappa) \chi(p)^2 \right) + (1-\kappa)p^2 \chi(p)^2.
\label{eq:afterDysonlemma2}
\end{equation}
The kinetic term in $\tilde H_n$ will be split accordingly and we apply Lemma~\ref{lem-Dyson-lemma} to the last part of the kinetic term plus the potential term. Using also the positivity of $\tilde v$, we obtain for any set $J_j \subseteq \{ 1, \ldots{} j-1, j+1, \ldots{}, n\}$
\begin{align}
-\Delta_j +  \frac{1}{2} \sum_{i,\,i \neq j}& \tilde{v}(d(x_i,x_j)) \geq -\nabla_j (1-(1-\kappa)\chi(p_j)^2) \nabla_j \nonumber \\
&+(1-\epsilon)(1-\kappa) U_R\left ( d\left(x_j, x_{\text{NN}}^{J_j}(x_j) \right) \right) - \frac{1}{\epsilon} \int_{\Rr_+} U_R(t) t \de t \sum_{i \in J_j} w_R(x_j - x_i). \label{eq:afterDysonlemma3}
\end{align}
Here $x_{\text{NN}}^{J_j}(x_j)$ denotes the nearest neighbor of $x_j$ among the points $x_i$ whose index $i$ is contained in $J_j$, and interaction terms for particles $k\not\in J_j$ are simply dropped for a lower bound. The subset $J_j$ is defined via the following construction (which is not unique). Fix $x_j$ and consider those $x_i$ whose distance to the nearest neighbor (among all other $x_k$, $k \neq i,j$) is at least $R/5$, and add the corresponding index $i$ to the set. Next, we go in some order through the set $\{ x_1, \ldots{}, x_{j-1}, x_{j+1}, \ldots{}, x_n\}$ and add $i$ to the set if $d(x_i, x_k) \geq R/5$ for all $k$ that are already in the set $J_j$. Note that this last step depends on the ordering of the $x_i$ and therefore $J_j$ will depend on the ordering as well. Hence, the right side of \eqref{eq:afterDysonlemma3} is not permutation symmetric and strictly speaking it should be replaced by its symmetrization. We do not need to do this, however, as we are only interested in expectation values of this potential in bosonic (permutation symmetric) states anyway.

The motivation to introduce the set $J_j$ is the following. By definition, all particles whose index is contained in $J_j$ have a minimum distance $R/5$ to their nearest neighbor, which is needed in order to control the error terms coming from $w_R$. On the other hand, the set $J_j$ is constructed to be maximal in the sense that if $l \notin J_j$, then there exists a particle $x_k$ with $k \in J_j$ such that $d(x_l,x_k) < R/5$. In other words, we need the disks of radius $R$ centered at the particle coordinates to be able to have sufficient overlap in order to obtain the desired lower bound. For certain values of $z$  the system could be far from being homogeneous\footnote{recall that $z = (z_1, \ldots{}, z_M) \in \Cc^M$ is the complex vector  introduced in Subsection~\ref{sec:coherentstates}.} and many particles could cluster in a relatively small volume; we want to be able to detect this as an increase in the interaction energy.

\subsection{Filling the holes}
\label{sec:fillingholes}
After having applied Lemma~\ref{lem-Dyson-lemma}, we want to replace the resulting interaction potential $U_R$ by a potential without a hole of radius $R_0$ at the origin because it will be advantageous to work with a potential of positive type. To obtain such a potential we use Lemma~\ref{lem:fillingholes} below. Its proof requires a different technique than the corresponding Lemma in the three-dimensional case \cite[Lemma~3]{Seiringer2008}, due to the fact that a sufficiently weak attractive potential in three dimensions has no bound state, while it always does in two dimensions.  

For some unit vector $e \in \mathbb{R}^2$ we define the function $j : \mathbb{R}_+ \to \mathbb{R}_+$ by
\begin{equation}
j(t) = \frac{32}{\pi} \int_{\mathbb{R}^2} \theta\left( \frac{1}{2} - \vert y \vert \right) \theta\left( \frac{1}{2} - \vert y - t e \vert \right) \de y.
\label{eq:fillingholes1}
\end{equation}
Note that the support of the function $j$ is given by the interval $[0,1]$ and that we have $\int_0^1 j(t) t \de t = 1$. An explicit computation yields
\begin{equation}
j(t) = \frac{16}{\pi} \left[ \arccos(t) - t \sqrt{ 1-t^2 } \right] \mathds{1}_{[0,1]}(t),
\label{eq:fillingholes2}
\end{equation}
where $\mathds{1}_{[0,1]}$ denotes the characteristic function of the interval $[0,1]$. The potential we intend to work with is given by $\tilde{U}_R(t) = R^{-2} \ln(R/\tilde{a})^{-1} j(t/R)$. To obtain this potential we choose $U_R(t) = \tilde{U}_R(t) \theta(t - R_0)$ when we apply the Dyson Lemma. This choice indeed satisfies the integral condition \eqref{eqn-U_R-cond}, since
\begin{align*}
\int_{R_0}^R U_R(t) \ln(t/\tilde a) t \de t &= \frac{1}{R^2 \ln(R/\tilde a)} \int_{R_0}^R j(t/R) \ln(t/\tilde a) t \de t\\
&\leq \frac{1}{R^2} \int_{R_0}^R j(t/R) t \de t = \int_{R_0/R}^1 j(t) t \de t \leq \int_0^1 j(t) t \de t = 1. \asterisknum
\end{align*} 
The following lemma will allow us to quantify the error we make when we replace $U_R$ by $\tilde{U}_R$.
\begin{lem} 
	\label{lem:fillingholes}
	Let $y_1,\ldots{},y_n$ denote $n$ points in $\Lambda$, with $d(y_i,y_j) \geq R/5$ for $i \neq j$ and let $R_0 < R/10$. Then
	\begin{equation}
	-\Delta - \frac{1}{R_0^2 \ln(R/R_0)} \sum_{i=1}^n \theta\left( R_0 - d(x,y_i) \right) \geq -  \frac{\tilde{C}}{R^2} \sum_{i=1}^n \theta\left( R/10 - d(x,y_i) \right)
	\label{eq:lemfillingholes1}
	\end{equation}
	holds for a universal constant $\tilde{C}>0$.
	\begin{proof}
		It is sufficient to prove that
		\begin{equation}
		\int_{\vert x \vert \leq R/10} \left( \vert \nabla \phi(x) \vert^2 - \frac{1}{R_0^2 \ln(R/R_0)} \theta \left( R_0 - |x| \right) \vert \phi(x) \vert^2 \right) \de x \geq - \frac{\tilde{C}}{R^2} \int_{\vert x \vert \leq R/10} \vert \phi(x) \vert^2 \de x \label{eq:lemfillingholes2}
		\end{equation}
		holds for any function $\phi \in H^1(\mathbb{R}^2)$ with $\tilde{C}>0$ being independent of that function. In other words, we need to show that the lowest eigenvalue of the quadratic form on the left-hand side of Eq.~\eqref{eq:lemfillingholes2} is bounded from below by a constant times $-R^{-2}$. 
		
		Denote by $E^{\mathrm{N}}_R$ this lowest eigenvalue and by $\phi^{\mathrm{N}}_R$ the corresponding normalized eigenfunction. We will bound $E^{\mathrm{N}}_R$ from below in terms of $E_0$, the lowest eigenvalue of the Schr\"odinger operator 
		\begin{equation}
		h = -\Delta - \frac{1}{R_0^2 \ln(R/R_0)} \theta \left( R_0 - |x| \right) 
		\label{eq:lemfillingholes3}
		\end{equation}
		acting on $L^2(\mathbb{R}^2)$. By  rearrangement $\phi^{\mathrm{N}}_R$ is a radial decreasing function, satisfying Neumann boundary conditions. Choose $\lambda \in C^{\infty}\left( [0,\infty) \right)$ such that $\lambda(0)=1$, $\lambda'(0) = 0$, $\lambda(t) = 0$ for $t \geq 1$ and $\vert \lambda'(t) \vert^2 \leq 2$, $|\lambda(t)| \leq 1$ for all $t \geq 0$. We define
		\begin{equation}
		\tilde{\phi}_R(x) = \begin{cases} \phi^{\mathrm{N}}_R(x) & \text{ if } \vert x \vert \leq R/10, \\ \eta \lambda\left( \frac{ \vert x \vert - R/10}{R} \right) & \text{ if } \vert x \vert > R/10, \end{cases}
		\label{eq:lemfillingholes4}
		\end{equation}
		where $\eta$ is chosen such that $\tilde{\phi}_R(x)$ is continuously differentiable, that is, $\eta = \phi^{\mathrm{N}}_R(e R/10)$ with $e \in \mathbb{R}^2$ a unit vector. We have
		\begin{equation} \label{eq:lemfillingholes5}
		E_0  \leq \frac{ \left\langle \tilde{\phi}_R, h \tilde{\phi}_R \right\rangle }{\left\langle \tilde{\phi}_R, \tilde{\phi}_R \right\rangle} = \frac{1}{ \left\langle \tilde{\phi}_R, \tilde{\phi}_R \right\rangle } \left( E_R^\mathrm{N} 
		+ \frac{\eta^2}{R^2} \int_{| x | >R/10} \left|  \lambda'\left( \frac{ \vert x \vert - R/10}{R} \right) \right|^2 \de x \right). 
		\end{equation}
		With $\vert \lambda'(t) \vert^2 \leq 2$ and $\lambda'(t) = 0$ for $t \geq 1$ we see that the integral on the right-hand side of Eq.~\eqref{eq:lemfillingholes5} is bounded from above by $12 \pi R^2/5$. We therefore have
		\begin{equation}
		E^{\mathrm{N}}_R \geq E_0 \left\Vert \tilde{\phi}_R \right\Vert^{2}  - \frac{12 \pi}{5} \eta^2.
		\label{eq:lemfillingholes6}
		\end{equation}
		With the definition of $\lambda$ we conclude 
		\begin{equation}
		\left\Vert \tilde{\phi}_R \right\Vert^{2} \leq 1 + 2 \pi \eta^2 \int_{R/10}^{R/10 + R} r \de r = 1 + \frac{6 \pi}{5} \eta^2 R^2
		\label{eq:lemfillingholes7}
		\end{equation}
		and since $E_0 < 0$, we have
		\begin{equation}
		E^{\mathrm{N}}_R \geq E_0 \left( 1 + \frac{6 \pi}{5} \eta^2 R^2 \right) - \frac{12 \pi}{5} \eta^2.
		\label{eq:lemfillingholes8}
		\end{equation}
		It remains to derive upper bounds for $\eta$ and $| E_0 |$.
		
		Since $\phi^{\mathrm{N}}_R$ is symmetrically decreasing and has $L^2$-norm equal to one its value at the boundary $\{ x : \vert x \vert = R/10 \}$ is at most $(\pi (R/10)^2)^{-1/2}$, that is, $\eta \leq 10 /(\sqrt{\pi}R)$. On the other hand, we know from \cite[Theorem~3.4]{Simon75} that
		\begin{equation}
		E_0 \sim - \frac{1}{R_0^2} \exp\left( \frac{-4 \pi}{ \frac{1}{R_0^2 \ln(R/R_0)} \int_{\mathbb{R}^2} \theta \left( R_0 - |x| \right) \de x } \right) = -\frac {R_0^2}{R^4} .
		\label{eq:lemfillingholes9}
		\end{equation}
		Here $E_0 \sim -\exp(-b/\delta)$ means that for all $\epsilon > 0$ there exists a $\delta_0>0$ such that $\exp(-(b+\epsilon)/\delta) \leq -E_0 \leq \exp(-(b-\epsilon)/\delta)$ for all $0 < \delta < \delta_0$. Together with Eq.~\eqref{eq:lemfillingholes8} and the upper bound on $\eta$, this shows that for all $\epsilon > 0$ there exists a $\delta_0 > 0$ such that
		\begin{equation}
		E^{\mathrm{N}}_R \geq - \frac{121}{R^2} \left( \frac{R_0}{R} \right)^{2-\epsilon} - \frac{240}{R^2}
		\label{eq:lemfillingholes10}
		\end{equation}
		holds as long as $R_0/R < \delta_0$.
		
		If this is not the case we use the simple bound
		\begin{equation}
		E^\text{N}_R \geq - \frac{1}{R_0^2 \ln(R/R_0)}.
		\label{eq:lemfillingholes11}
		\end{equation}
		Since $R_0 < R/10$ by assumption we know that $\ln(R/R_0) > \ln(10)$. On the other hand, $R_0^2 \geq R^2 \delta_0^2$ implies that
		\begin{equation}
		E^\text{N}_R \geq - \frac{1}{R^2 \delta_0^2 \ln(10)}
		\label{eq:lemfillingholes12}
		\end{equation}
		for $R_0/R \geq \delta_0$. This proves the claim \eqref{eq:lemfillingholes1}.
	\end{proof}
\end{lem}

For the simple step function potential in Lemma~\ref{lem:fillingholes} one can also compute the lowest eigenvalue explicitly in terms of Bessel functions. The method of proof given here is more general, however.

Recall that $d(x_i,x_k) \geq R/5$ for $i,k \in J_j$. With $\tilde{U}_R(t) \leq j(0)/(R^2 \ln(R/\tilde{a})) = 8/(R^2 \ln(R/\tilde{a}))$, as well as using $\tilde a < R_0$, we see that Lemma~\ref{lem:fillingholes} implies
\begin{align}
&\left( \tilde{U}_R - U_R \right) \left( d\left( x_j, x_{\mathrm{NN}}^{J_j}\left(x_j \right) \right) \right) \leq \theta\left( R_0 - d\left( x_j, x_{\mathrm{NN}}^{J_j}\left(x_j \right) \right) \right) \frac{8}{R^2 \ln(\tilde{a}/R)} \label{eq:fillingholes3} \nonumber\\
&\quad = 8 \left( \frac{R_0}{R} \right)^2 \sum_{i \in J_j} \theta\left( R_0 - d\left( x_i, x_j \right) \right) \frac{1}{R_0^2 \ln(\tilde{a}/R)} \leq 8 \left( \frac{R_0}{R} \right)^2 \left[ -\Delta_j + \frac{\tilde{C}}{R^2} \sum_{i \in J_j} \theta\left( R/10 - d(x_i,x_j) \right) \right] \nonumber \\
&\quad= 8 \left( \frac{R_0}{R} \right)^2 \left[ -\Delta_j + \frac{\tilde{C}}{R^2} \theta\left( R/10 - d\left( x_j, x_{\mathrm{NN}}^{J_j}\left(x_j \right) \right) \right) \right].
\end{align}
The constant $\tilde{C}>0$ is determined by Lemma~\ref{lem:fillingholes}. On the other hand, we know that $\tilde U_R(t)$ can be bounded from below as $\tilde{U}_R(t) \geq j(1/10)/(R^2 \ln(R/\tilde{a}))$ for $t \leq R/10$ and this implies
\begin{equation}
\theta\left( R/10 - d\left( x_j, x_{\mathrm{NN}}^{J_j}\left(x_j \right) \right) \right)  \label{eq:fillingholes4}
\leq \frac{\tilde{U}_R\left( d\left( x_j, x_{\mathrm{NN}}^{J_j}\left(x_j \right) \right) \right) R^2 \ln(R/\tilde{a})}{j(1/10)}.
\end{equation}
Eqs.~\eqref{eq:fillingholes3} and \eqref{eq:fillingholes4} together show that
\begin{align} \label{eq:fillingholes5}
&\left( \tilde{U}_R - U_R \right) \left( d\left( x_j, x_{\mathrm{NN}}^{J_j}\left(x_j \right) \right) \right)\nonumber\\ 
&\quad \leq - 8 \left( \frac{R_0}{R} \right)^2  \Delta_j + \frac{8 \tilde{C}}{j(1/10)} \left( \frac{R_0}{R} \right)^2 \ln(R/\tilde{a}) \tilde{U}_R  \left( d\left( x_j, x_{\mathrm{NN}}^{J_j}\left(x_j \right) \right) \right).
\end{align}
Define $a'$ by the equation (assuming that the last factor on the right side is positive)
\begin{equation}
\frac{1}{\ln(R/a')} = \frac{1}{\ln(R/\tilde a)} (1 - \epsilon)( 1- \kappa) \left( 1 - \frac{8 \tilde{C}}{j(1/10)} \left( \frac{R_0}{R} \right)^2 \ln(R/\tilde a) \right)
\label{eq:fillingholes6}
\end{equation}
and let
\begin{equation}
\tilde{U}_R'(t) = \frac{j(t/R)}{R^2 \ln(R/a')}.
\label{eq:fillingholes7}
\end{equation}
We also define 
\begin{equation}
\kappa' = \kappa - 8 \left( \frac{R_0}{R} \right)^2
\label{eq:fillingholes8}
\end{equation}
and write the remaining kinetic energy as (compare with \eqref{eq:afterDysonlemma3})
\begin{align}
&-\nabla_j (1-(1-\kappa)\chi(p)^2) \nabla_j +(1-\epsilon)(1-\kappa)\left(  8 \left( \frac{R_0}{R} \right)^2  \Delta_j  \right) \nonumber \\
&\quad \geq -\nabla_j (1-(1-\kappa)\chi(p)^2) \nabla_j + 8 \left( \frac{R_0}{R} \right)^2  \Delta_j \nonumber \\
&\quad = -\Delta_j \kappa' - (1-\kappa) \nabla_j \left( 1- \chi(p)^2 \right) \nabla_j. \label{eq:fillingholes9}
\end{align}
In the following, we will choose $\kappa \gg R_0^2/R^2$, which, in particular, implies $\kappa' > 0$. Concerning the attractive part of the interaction potential that we obtain after applying Lemma~\ref{lem-Dyson-lemma}, we use the definition of $U_R$ to see that 
\begin{equation}
\int_{\Rr_+} U_R(t) t \de t \leq \frac{1}{\ln(R/\tilde{a})}.
\label{eq:fillingholes10}
\end{equation}
Eqs.~\eqref{eq:afterDysonlemma3}, \eqref{eq:fillingholes5}, \eqref{eq:fillingholes9} and \eqref{eq:fillingholes10} then imply
\begin{equation}
\mathbb{T} + \mathbb{V} \geq \mathbb{T}^{\mathrm{c}} + \mathbb{W},
\label{eq:fillingholes11}
\end{equation}
where
\begin{equation}
\mathbb{T}^{\mathrm{c}} = \sum_p \epsilon(p) a_p^{\dagger} a_p \quad \text{ and } \quad \epsilon(p) = \kappa' p^2 + (1-\kappa) p^2 \left(1 - \chi(p)^2 \right) - \mu_0. 
\label{eq:fillingholes12}
\end{equation}
In the Fock space sector with particle number $n$, the operator $\mathbb{W}$ is given by the (symmetrization of the) multiplication operator
\begin{equation}
\sum_{j=1}^n \left[ \tilde{U}_R' \left( d\left( x_j, x_{\mathrm{NN}}^{J_j}\left(x_j \right) \right) \right) - \frac{1}{\epsilon\ln(R/\tilde{a})} \sum_{i \in J_j} w_R(x_j - x_i) \right].
\label{eq:fillingholes13}
\end{equation}
We recall that the set $J_j$ depends on all particle coordinates $x_i$, $i \neq j$.

We conclude this section with the choice of the cutoff function $\chi$. Let $\nu : \mathbb{R}^2 \rightarrow \mathbb{R}_+$ be a smooth radial function with $\nu(p) = 0$ for $| p | \leq 1$, $\nu(p) = 1$ for $p \geq 2$, and $0 \leq \nu(p) \leq 1$ in-between. For some $s \geq R$ we choose
\begin{equation}
\chi(p) = \nu(sp).
\label{eq:fillingholes14}
\end{equation}
We will choose $p_{\text{c}} \leq 1/s$ below. This implies in particular that $\epsilon(p) = (1-\kappa+\kappa') p^2 - \mu_0$ for $|p| < p_{\text{c}}$.
With $\Upsilon^z$ and $\Upsilon^z_\pi$ defined in \eqref{def:Upsilon} and \eqref{def:Upsilonpi}, respectively, we therefore have
\begin{equation}
\Tr_{\mathcal{F}} \left[ \mathbb{T}^{\text{c}} \Upsilon^z_{\pi} \right] = \Tr_{\mathcal{F}} \left[ \mathbb{T}^{\text{c}} \Upsilon^z \right] + \sum_{|p| < p_{\text{c}}} \left( (1-\kappa+\kappa' ) p^2 - \mu_0 \right) \pi_p.
\label{eq:fillingholes15}
\end{equation}
Using Eqs.~\eqref{eq:replacingvacuum6}, \eqref{eq:fillingholes11}, \eqref{eq:fillingholes15} and further
\begin{equation}
\Tr_{\mathcal{F}}\left[ \mathbb{T}^{\text{c}} \Upsilon^z \right] - \frac{1}{\beta} S\left( \Upsilon^z \right) \geq - \frac{1}{\beta} \ln \Tr_{\mathcal{F_>}} \exp\left( - \beta \mathbb{T}_\text{s}^{\text{c}}(z) \right),
\label{eq:fillingholes16}
\end{equation}
we conclude that
\begin{align}
F_z(\beta) \geq& - \frac{1}{\beta} \ln \Tr_{\mathcal{F}_>} \exp\left( - \beta \mathbb{T}_\text{s}^{\text{c}}(z) \right) + \Tr_{\mathcal{F}}\left[ \mathbb{W} \Upsilon_{\pi}^z \right] + \tfrac{1}{2} \Tr_{\mathcal{F}} \left[ \mathbb{K} \Upsilon^z \right]  \label{eq:fillingholes17} \\
&- (\kappa - \kappa') \sum_{| p | < p_{\text{c}}} p^2 \pi_p - Z^{(2)}. \nonumber
\end{align}
The first term on the right-hand side of \eqref{eq:fillingholes17} can be computed explicitly and reads
\begin{align*}
&-\frac{1}{\beta}  \ln \Tr_{\mathcal{F}_>} \exp\left( - \beta \mathbb{T}_\text{s}^{\text{c}}(z) \right)\\
&\quad = \sum_{| p | < p_{\text{c}}} \left( \left( 1 - \kappa + \kappa' \right) p^2 - \mu_0 \right) |z_p|^2 \label{eq:fillingholes18} + \frac{1}{\beta} \sum_{|p| \geq p_{\text{c}}} \ln\left( 1 - \exp\left( - \beta \epsilon(p) \right)\right). \asterisknum
\end{align*}
In the following, we will derive a lower bound on $\Tr_{\mathcal{F}}\left[ \mathbb{W} \Upsilon_{\pi}^z \right]$.
\subsection{Localization of relative entropy}
\label{sec:localizationofrelative entropy}
In order to compute $\Tr_{\mathcal{F}}\left[ \mathbb{W} \Upsilon_{\pi}^z \right]$ we will replace the  unknown state $\Gamma_z$ in the definition of $\Upsilon_\pi^z = U(z) \Pi U(z)^\dagger \otimes \Gamma^z$ by the quasi-free state $\Gamma_0$, the Gibbs state for the kinetic energy operator $\mathbb{T}_\text{s}(z)$. The error resulting from this replacement will be controlled via the a priori bound on the relative entropy \eqref{eq:relativeentropyandaprioribounds4}. For that purpose we need a local version of the relative entropy bound, which will be derived in this section.

Let us denote by $\Omega_{\pi}$ the unique quasi-free state whose one-particle density matrix is given by
\begin{equation}
\omega_{\pi} = \sum_p \omega_{\pi}(p) | p \rangle \langle p | = \sum_p \frac{1}{\e^{\ell(p)}-1} | p \rangle \langle p |,
\label{eq:locrelentropy1}
\end{equation}
where 
\begin{equation} \label{eqn-def-l(p)}
\ell(p) = \begin{cases}
\ln(1 + 1/\pi_p) & \text{if } |p| < p_\text{c},\\	
\beta ( p^2 - \mu_0) & \text{if } |p| \geq p_\text{c}.
\end{cases}
\end{equation}
In other words, 
\begin{equation}\label{def:Omegapi}
\Omega_{\pi} = \Pi \otimes \Gamma_0.
\end{equation}
We will choose $\pi_p$ such that $\ell(p) \geq \beta(p^2 - \mu_0)$ holds for all $p$. Let $\eta : \mathbb{R}_+ \to [0,1]$ be a function with the following properties:
\begin{itemize}
	\item $\eta \in C^{\infty}(\mathbb{R}_+)$
	\item $\eta(0)=1$, and $\eta(x) = 0$ for $x \geq 1$
	\item $\hat{\eta}(p) = \int_{\mathbb{R}^2} \eta(|x|) \e^{-ipx} \de x \geq 0$ for all $p \in \mathbb{R}^2$.
\end{itemize}
Such a function can be obtained by choosing a smooth radial and nonnegative function on $\mathbb{R}^2$ with compact support and then convolving it with itself. Given a function with these properties, we define $\eta_b(x) = \eta(x/b)$ for some $b \leq L/2$. We also define the one-particle density matrix $\omega_b$ by its integral kernel 
\begin{equation}
\omega_b(x,y) = \omega_{\pi}(x,y) \eta_b(d(x,y)).
\label{eq:locrelentropy2}
\end{equation}
The unique quasi-free state related to $\omega_b$ will be denoted by $\Omega_b$ and
\begin{equation} \label{eqn-def-Omega-b-z}
\Omega_b^z = U(z) \Omega_b U(z)^{\dagger}.
\end{equation}
We also introduce the notation $\rho_{\omega} = \omega_b(x,x) = \omega_{\pi}(x,x)$. 

To state the inequality we are looking for, we need to define  spatial restriction of states. To that end, we denote for $r < L/2$ by $\chi_{r,\xi}(x)= \theta(r - d(x,\xi))$ the characteristic function of a disk of radius $r$ centered at $\xi \in \Lambda$. Since $\chi_{r,\xi}$ defines a projection on the one-particle Hilbert space $\mathcal{H} = L^2(\Lambda)$, the Fock space $\mathcal{F}$ over $\mathcal{H}$ is unitarily equivalent to the product of two Fock spaces 
\begin{equation} \label{eqn-Fock-space-decomposition}
\mathcal{F}(\mathcal{H}) \cong \mathcal{F}(\chi_{r,\xi} \mathcal{H}) \otimes \mathcal{F}((\chi_{r,\xi}\mathcal{H})^\perp).
\end{equation}
Any state on $\mathcal{F}$ can be restricted to the Fock space over $\chi_{r,\xi} \mathcal{H}$ by taking the partial trace over the second tensor factor in \eqref{eqn-Fock-space-decomposition}. The restriction of the state $\Gamma$ will be denoted by $\Gamma_{\chi_{r,\xi}}$. 

If $d(\xi,\zeta) > 2 r$ the multiplication operator $\chi_{r,\xi}+\chi_{r,\zeta}$ defines a projection and using the fact that $\omega_b(x,y) = 0$ as long as $d(x,y) > b$ we easily check that
\begin{equation}\label{eq:productstructure}
	\Omega_{b,\chi_{r,\xi}+\chi_{r,\zeta}} \cong \Omega_{b,\chi_{r,\xi}} \otimes \Omega_{b,\chi_{r,\zeta}}
\end{equation}
holds if $d(\xi,\zeta) > 2 r + b$. More precisely, we use that the one-particle density matrix of $\Omega_{b,\chi_{r,\xi}+\chi_{r,\zeta}}$ is given by $( \chi_{\xi,r} + \chi_{\zeta,r} ) \omega_b ( \chi_{\xi,r} + \chi_{\zeta,r} ) = \chi_{\xi,r} \omega_b \chi_{\xi,r} + \chi_{\zeta,r} \omega_b \chi_{\zeta,r}$. The right-hand side is nothing but the one-particle density matrix of $\Omega_{b,\chi_{r,\xi}}$ plus the one of $\Omega_{b,\chi_{r,\zeta}}$, which proves the claim. The above identity also holds for $\Omega_b^z$ because $U(z)$ has the same product structure. 

Concerning spatial localization, the relative entropy is superadditive in the following sense.
\begin{lem}
	\label{lem:superadditivityrelentropy}
	Let $X_i$, $1 \leq i \leq k$, denote $k$ mutually orthogonal projections on $\mathcal{H}$. Let $\Omega$ be a state on $\mathcal{F}$ which factorizes under restrictions as $\Omega_{\sum_{i} X_i} = \otimes_i \Omega_{X_i}$. Then, for any state $\Gamma$, we have
	\begin{equation}
		S\left(\Gamma,\Omega_{\sum_{i} X_i}  \right) \geq \sum_i S\left( \Gamma_{X_i}, \Omega_{X_i} \right).
	\end{equation}
\end{lem}
The proof Lemma~\ref{lem:superadditivityrelentropy} can be found in \cite[Section~2.8]{Seiringer2008}, see also \cite[Section~5.1]{Seiringer06B}. We emphasize that the factorization property of $\Omega$ is crucial, the relative entropy need not be superadditive, in general. This is  the reason for introducing the cut-off $b$. Without it, the state $\Omega_b^z$ would not factorize as in \eqref{eq:productstructure}.

We apply Lemma~\ref{lem:superadditivityrelentropy} with  $\Omega = \Omega_b^z$ and $X_i$ multiplication operators by characteristic functions of balls with radius $r$ that are separated by the distance $2 b$. When we average over the position of the balls (see \cite[Section~5.1]{Seiringer06B} for details), we obtain for $r \leq 2b$ and $L/(2b) \in \mathbb{N}$ the inequality
\begin{equation}
S(\Gamma,\Omega_b^z) \geq \frac{1}{(2b)^2} \int_{\Lambda} S\left( \Gamma_{\chi_{r,\xi}}, \Omega_{b,\chi_{r,\xi}}^z \right) \de \xi.
\label{eq:locrelentropy3}
\end{equation}
That is, the integral over local relative entropies of $\Gamma$ with respect to $\Omega_b^z$ can be estimated from above by their global relative entropy.  The restriction $L/(2b) \in \mathbb{N}$ is of no further importance since we take the thermodynamic limit. From \eqref{eq:locrelentropy3} for $\Gamma = \Upsilon_{\pi}^z$, we infer
\begin{align*} 	\label{eq:locrelentropy4}
\int_{\Lambda} \left\| \Upsilon^z_{\pi,\chi_{r,\xi}} - \Omega_{b,\chi_{r,\xi}}^z \right\|_1 \de \xi &\leq |\Lambda|^{1/2} \left( \int_{\Lambda} \left\| \Upsilon^z_{\pi,\chi_{r,\xi}} - \Omega_{b,\chi_{r,\xi}}^z \right\|_1^2 \de \xi \right)^{1/2}\\
&\leq \sqrt{2} |\Lambda|^{1/2} \left( \int_{\Lambda} S(\Upsilon^z_{\pi,\chi_{r,\xi}}, \Omega_{b,\chi_{r,\xi}}^z) \de \xi \right)^{1/2}\\
&\leq 2^{3/2} b \vert \Lambda \vert^{1/2} S(\Upsilon_{\pi}^z, \Omega_b^z)^{1/2} \asterisknum
\end{align*}
for any $b \geq 2 r$. This estimate follows from using the Cauchy-Schwarz inequality for the integral over $\xi$ and the fact that the relative entropy of two states $\Gamma$ and $\Gamma'$ is bounded from below by the square of the trace norm distance, by Pinsker's inequality (see \cite[Theorem~1.15]{OP04})
\begin{equation}
S(\Gamma, \Gamma') \geq \frac 1 2 \| \Gamma - \Gamma' \|_1^2.
\end{equation}
In Subsection~\ref{sec:effectofcutoff}, we will estimate the effect of the cutoff $b$ and obtain a bound on \eqref{eq:locrelentropy4} in terms of the a priori bound \eqref{eq:relativeentropyandaprioribounds4} on the relative entropy. We remark that Pinsker's inequality could not be used with benefit for the global relative entropy. This is because the relative entropy is an extensive quantity while the trace norm difference of two states is always bounded by $2$.
\subsection{Interaction energy, part I}
\label{sec:interactionenergyA}

In the following three subsections we shall derive a lower bound for $\Tr_{\mathcal{F}}[ \mathbb{W} \Upsilon_\pi^z]$.  The estimate \eqref{eq:locrelentropy4} will play an important role in this analysis. We start by giving a bound on the first term  in \eqref{eq:fillingholes13} in this subsection, and postpone the analysis of the second term to  Subsection~\ref{sec:interactionenergyB}. In Subsection~\ref{sec:interactionenergyC} we combine these bounds to obtain the final bound. A main difficulty is related to the fact that the vector $z$ is rather arbitrary, and hence the density of the particles described by the coherent states can be far from homogeneous.  

Let us give a name to the positive and the negative part of the interaction energy. We write
\begin{equation}
\mathbb{W} = \mathbb{W}_1 - \mathbb{W}_2,
\label{eq:interactionpartA1}
\end{equation}
where
\begin{equation}
\mathbb{W}_1 = \bigoplus_{n=2}^{\infty} \sum_{j=1}^n \tilde{U}_R' \left( d\left( x_j, x^{J_j}_{\mathrm{NN}}(x_j) \right) \right)
\label{eq:interactionpartA2}
\end{equation}
and
\begin{equation}
\mathbb{W}_2 = \bigoplus_{n=2}^{\infty} \sum_{j=1}^{n} \sum_{i \in J_j} \frac{1}{\epsilon\ln(R/\tilde a)} w_R(x_j - x_i).
\label{eq:interactionpartA3}
\end{equation}
We start by giving a lower bound to the expectation of $\mathbb{W}_1$ in the state $\Upsilon_{\pi}^z$. First of all, recalling the definition of $j$ from \eqref{eq:fillingholes1}, we note that since $L \geq 2 R$ we can write
\begin{equation}
j(d(x,y)/R) = \frac{32}{\pi R^2} \int_{\Lambda} \theta(R/2 - d(\xi,x)) \theta(R/2 - d(\xi,y)) \de \xi
\label{eqn-j-integral-representation}
\end{equation}
for $x, y \in \Lambda$. Inserting this into \eqref{eq:fillingholes7}, we have
\begin{equation}
\tilde{U}_R'\left( d(x,y) \right) = \frac{32}{\pi \ln(R/a') R^4} \int_{\Lambda} \theta(R/2 - d(\xi,x)) \theta(R/2 - d(\xi,y)) \de \xi.
\label{eq:interactionpartA4}
\end{equation}
This gives rise to a similar decomposition of $\mathbb{W}_1$ which we write as
\begin{equation}
\mathbb{W}_1 = \frac{32}{\pi \ln(R/a') R^4} \int_{\Lambda} w(\xi) \de \xi,
\label{eq:interactionpartA5}
\end{equation}
with
\begin{equation}
w(\xi) = \bigoplus_{n=2}^{\infty} \sum_{j=1}^n \theta\left( R/2 - d\left( \xi,x_j \right) \right) \theta\left(R/2 - d\left( \xi, x^{J_j}_{\mathrm{NN}}(x_j) \right) \right).
\label{eq:interactionpartA6}
\end{equation}
For $r>0$, define $n_{r,\xi}$ as the number operator of a disk of radius $r$ centered at $\xi \in \Lambda$, which is nothing  but the second quantization of the multiplication operator $\theta(r - d(\xi, \,\cdot\,))$ on $L^2(\Lambda)$. We claim that
\begin{equation}
w(\xi) \geq n_{R/10,\xi} \theta(n_{R/10,\xi} - 2).
\label{eq:interactionpartA7}
\end{equation}
This is the second quantized version of
\begin{align*} \label{eqn-estimate-w-2nd-quantized}
&\theta(R/2 - d(\xi,x_j)) \theta\left(R/2 - d\left( \xi, x^{J_j}_{\mathrm{NN}}(x_j)\right) \right)\\
&\quad \geq \theta(R/10 - d(\xi,x_j)) \left( 1 - \prod_{i,\,i \neq j} \theta(d(\xi,x_i) - R/10) \right), \asterisknum
\end{align*}
which can be shown using the defining property of $J_j$. More precisely, \eqref{eqn-estimate-w-2nd-quantized} says that if $x_j$ and some $x_k$ with $k \neq j$ are in a disk of radius $R/10$ centered at $\xi$ (i.e., if the right-hand side is equal to one), then the nearest neighbor of $x_j$ in the set $J_j$ is in a disk of radius $R/2$ with the same center (i.e., the left-hand side equals one). Assume therefore that $x_j$ and $x_k$ are in a disk of radius $R/10$ centered at $\xi$ and $k \in J_j$. Then we have
\begin{equation}
d \left( x_j, x^{J_j}_\mathrm{NN}(x_j) \right) \leq d(x_j, x_k) \leq \frac R 5,
\end{equation}
which implies $d(\xi, x^{J_j}_\mathrm{NN}(x_j)) \leq 3R/10$. Conversely, if $k \notin J_j$, then by definition of $J_j$, there exists $l \in J_j$ such that $d(x_l, x_k) < R/5$. Therefore
\begin{equation}
d \left( x_j, x^{J_j}_\mathrm{NN}(x_j) \right) \leq d(x_j, x_l) < \frac{2 R}{5},
\end{equation}
which implies $d(\xi, x^{J_j}_\mathrm{NN}(x_j)) < R/2$ and proves \eqref{eqn-estimate-w-2nd-quantized}.

In particular, the above implies
\begin{equation}
w(\xi) \geq \overline{w}(\xi) := w(\xi) \theta\left( 2 - n_{3R/2,\xi} \right) + n_{R/10,\xi} \theta\left( n_{R/10,\xi} - 2\right) \theta\left( n_{3R/2,\xi} - 3 \right).
\label{eq:interactionpartA8}
\end{equation}
We also have
\begin{equation}
w(\xi) \theta\left( 2 - n_{3R/2,\xi} \right) = n_{R/2,\xi} \left( n_{R/2,\xi} - 1 \right) \theta\left( 2 - n_{3R/2,\xi} \right),
\label{eq:interactionpartA9}
\end{equation}
which can be seen from the following consideration. Assume two particles $x_i$ and $x_j$ are in a disk of radius $R/2$ and no other particle is in the bigger disk of radius $3R/2$ (with the same center), then these two particles must be nearest neighbors and by construction $i \in J_j$ and $j \in J_i$, which implies \eqref{eq:interactionpartA9}.

We note that the operator in \eqref{eq:interactionpartA9} is  bounded. Its operator norm  equals  $2$ and in combination with $n_{R/10,\xi} \leq n_{3R/2,\xi}$, this implies that
\begin{equation}
| \overline{w}(\xi) - n_{R/10,\xi} | \leq 2,
\label{eq:interactionpartA10}
\end{equation}
as can be seen using \eqref{eq:interactionpartA8} and an easy counting argument. Eqs.~\eqref{eq:interactionpartA5}, \eqref{eq:interactionpartA8} and \eqref{eq:interactionpartA10} imply that
\begin{align} \label{eq:interactionpartA11}
\Tr_{\mathcal{F}}\left[ \mathbb{W}_1 \Upsilon_{\pi}^z \right] &\geq \frac{32}{\pi \ln(R/a') R^4} \int_{\Lambda} \Tr_{\mathcal{F}} \left[ \overline{w}(\xi) \Upsilon_{\pi}^z \right] \de \xi  \nonumber\\
&\geq \frac{32}{\pi \ln(R/a') R^4} \int_{\Lambda} \Tr_{\mathcal{F}} \left[ \overline{w}(\xi) \Omega_b^z + n_{R/10, \xi} \left( \Upsilon_{\pi}^z - \Omega_b^z \right) \right] \de \xi \nonumber \\
&\quad - \frac{64}{\pi \ln(R/a') R^4} \int_{\Lambda} \left\| \Upsilon^z_{\pi,\chi_{3R/2,\xi}} - \Omega_{b,\chi_{3R/2,\xi}}^z \right\|_1 \de \xi.
\end{align}
The second term on the right-hand side of \eqref{eq:interactionpartA11} can be written as
\begin{equation}
\int_{\Lambda} \Tr_{\mathcal{F}} \left[ n_{R/10, \xi} \left( \Upsilon_{\pi}^z - \Omega_b^z \right) \right] \de \xi = \pi \left( \frac{R}{10} \right)^2 \Tr_{\mathcal{F}} \left[ \mathbb{N} \left( \Upsilon_{\pi}^z - \Omega_b^z \right) \right].
\label{eq:interactionpartA12}
\end{equation}
On the other hand, Eq.~\eqref{eq:locrelentropy4} implies that
\begin{equation}
\int_{\Lambda} \left\| \Upsilon_{\pi,\chi_{3R/2,\xi}}^z - \Omega_{b,\chi_{3R/2,\xi}}^z \right\|_1 \de \xi \leq 2^{3/2} b \vert \Lambda \vert^{1/2} S(\Upsilon_{\pi}^z, \Omega_b^z)^{1/2}
\label{eq:interactionpartA13}
\end{equation}
holds as long as $3R \leq b$. 

In the following, we will derive two different lower bounds to $\Tr_{\mathcal{F}}[\overline{w}(\xi) \Omega_b^z]$ in order to have a good bound for all values of $z$. To obtain the first bound, we use \eqref{eq:interactionpartA8} (where we drop the last term for a lower bound) and \eqref{eq:interactionpartA9}. This implies
\begin{align}
\Tr_{\mathcal{F}}\left[ \overline{w}(\xi) \Omega_b^z \right] &\geq \Big[ \Tr_{\mathcal{F}} \left[ n_{R/2,\xi} \left( n_{R/2,\xi} - 1 \right) \Omega_b^z \right] \label{eq:interactionpartA14} \\
&\qquad - \Tr_{\mathcal{F}} \left[ n_{3R/2,\xi} \left( n_{3R/2,\xi} - 1 \right)  \left( n_{3R/2,\xi} - 2 \right) \Omega_b^z \right] \Big]_+, \nonumber
\end{align}
where we take the positive part of this bound since the right-hand side can become negative, in which case we simply estimate the left-hand side by zero. The advantage of the right-hand side of \eqref{eq:interactionpartA14} is that all terms can be evaluated explicitly because $\Omega_b^z$ is a combination of a coherent and a quasi-free state. Let $\Phi_z$ denote the one-particle wave function $| \Phi_z \rangle = \sum_{|p| < p_{\text{c}}} z_p | p \rangle$.  With the aid of $U(z)^\dagger a_x U(z) =  a_x + \Phi_z(x)$ and  Wick's theorem we compute
\begin{align}
\Tr&_{\mathcal{F}} \left[ n_{3R/2,\xi} \left( n_{3R/2,\xi} - 1 \right) \left( n_{3R/2,\xi} - 2 \right) \Omega_b^z \right]  \nonumber \\
&= \left( \Tr_{\mathcal{F}} \left[ n_{3R/2,\xi} \Omega_b^z \right] \right)^3 + 2 \tr \left( \chi_{3R/2,\xi} \omega_b \right)^3 + 6 \langle \Phi_z \big| \left( \chi_{3R/2,\xi} \omega_b \chi_{3R/2,\xi} \right)^2 \big| \Phi_z \rangle \nonumber \\
&\hspace{0.3cm} + 3 \Tr_{\mathcal{F}}\left[ n_{3R/2,\xi} \Omega_b^z \right] \left( 2 \langle \Phi_z \big| \chi_{3R/2,\xi} \omega_b \chi_{3R/2,\xi} \big| \Phi_z  \rangle + \tr \left( \chi_{3R/2,\xi}  \omega_b \right)^2 \right) \nonumber \\
&\leq 6 \left( \Tr_{\mathcal{F}}\left[ n_{3R/2,\xi} \Omega_b^z \right] \right)^3,\label{eq:interactionpartA17} 
\end{align}
with $\omega_b$ the one-particle density matrix of $\Omega_b$ in \eqref{eqn-def-Omega-b-z}. 
Here the symbol $\tr$ denotes the trace over the one-particle Hilbert space $L^2(\Lambda)$. The first lower bound is thus given by
\begin{equation}
\Tr_{\mathcal{F}} [ \overline{w}(\xi) \Omega^z_b] \geq \left[ \Tr_{\mathcal{F}} \left[ n_{R/2,\xi} \left( n_{R/2,\xi} - 1 \right) \Omega^z_b \right] - 6 \left( \Tr_{\mathcal{F}} \left[ n_{3R/2,\xi} \Omega^z_b \right] \right)^3 \right]_+.
\end{equation}

To obtain the second lower bound for $\Tr_{\mathcal{F}}[\overline{w}(\xi) \Omega_b^z]$ we use
\begin{equation}
\Tr_{\mathcal{F}} \left[ \overline{w}(\xi) \Omega_b^z \right] \geq \Tr_{\mathcal{F}} \left[ n_{R/10,\xi} \theta\left( n_{R/10,\xi} - 2 \right) \Omega_b^z \right],
\label{eq:interactionpartA19}
\end{equation}
which follows from \eqref{eq:interactionpartA7}. Let us denote by $\Pi_0^{\mathcal{F}}$ the vacuum state on $\mathcal{F}$. The state $\Omega_{b,\chi_{R/10,\xi}}$ is a particle number conserving quasi-free state, whose vacuum expectation is given by
\begin{align}
\Tr_{\mathcal{F}(\chi_{R/10,\xi} \mathcal{H})} [ \Omega_{b,\chi_{R/10,\xi}} \Pi^{\mathcal{F}}_{0,\chi_{R/10,\xi}}] &=\exp\left( -\tr \ln \left( 1 + \chi_{R/10,\xi} \omega_b \chi_{R/10,\xi} \right) \right) \nonumber \\
&\geq \exp\left( - \tr \chi_{R/10,\xi} \omega_b \chi_{R/10,\xi} \right) = \exp\left( -\pi (R/10)^2 \rho_{\omega} \right), \label{eq:interactionpartA20}
\end{align}
where $\rho_\omega$ was defined after \eqref{eqn-def-Omega-b-z} to be the density of $\Omega_b$.  
Hence,
\begin{equation}
\Omega_{b,\chi_{R/10,\xi}} \geq \exp\left( -\pi (R/10)^2 \rho_{\omega} \right) \Pi_{0,{\chi_{R/10,\xi}}}^{\mathcal{F}}, 
\label{eq:interactionpartA21}
\end{equation}
as well as
\begin{equation}
\Omega_{b,\chi_{R/10,\xi}}^z \geq \exp\left( -\pi (R/10)^2 \rho_{\omega} \right) \left( U(z) \Pi_0^{\mathcal{F}} U(z)^{\dagger} \right)_{\chi_{R/10,\xi}}.
\label{eq:interactionpartA22}
\end{equation}
This in particular implies
\begin{equation}
\Tr_{\mathcal{F}} \left[ \overline{w}(\xi) \Omega_b^z \right] \geq \e^{ -\pi (R/10)^2 \rho_{\omega} } \Tr_{\mathcal{F}}\left[ n_{R/10,\xi} \theta\left( n_{R/10,\xi} - 2 \right) U(z) \Pi_0^{\mathcal{F}} U(z)^{\dagger} \right].
\label{eq:interactionpartA22a}
\end{equation}
The state $U(z) \Pi_0^{\mathcal{F}} U(z)^{\dagger}$ as well as its restriction to the Fock space over $\chi_{R/10,\xi} \mathcal{H}$ are coherent states. In the Fock space sector with $n$ particles, the latter is given by the projection onto the $n$-fold tensor product of the wave function $\chi_{R/10,\xi} \Phi_z$ times a normalization factor. We therefore have
\begin{align} 
&\Tr_{\mathcal{F}} \left[ n_{R/10,\xi} \theta\left( n_{R/10,\xi} - 2 \right) U(z) \Pi_0^\mathcal{F} U(z)^{\dagger} \right] = \e^{- \langle \Phi_z | \chi_{R/10,\xi} | \Phi_z \rangle } \sum_{n \geq 2} n \frac{\langle \Phi_z | \chi_{R/10,\xi} | \Phi_z \rangle^n}{n!}  \nonumber \\
&\hspace{3cm} = \langle \Phi_z | \chi_{R/10,\xi} | \Phi_z \rangle \left( 1 - \e^{- \langle \Phi_z | \chi_{R/10,\xi} | \Phi_z \rangle } \right) \geq \frac{\langle \Phi_z | \chi_{R/10,\xi} | \Phi_z \rangle^2}{1+\langle \Phi_z | \chi_{R/10,\xi} | \Phi_z \rangle}. \label{eq:interactionpartA23}
\end{align}
To arrive at the last line, we used the estimate $x(1-e^{-x}) \geq x^2/(1+x)$ for $x \geq 0$.

We combine the estimates from Eqs.~\eqref{eq:interactionpartA11}, \eqref{eq:interactionpartA13}, \eqref{eq:interactionpartA14}, \eqref{eq:interactionpartA17}, \eqref{eq:interactionpartA22a} and \eqref{eq:interactionpartA23} to see that for any $0 \leq \lambda \leq 1$ we have
\begin{align} 
\Tr_{\mathcal{F}} \left[ \mathbb{W}_1 \Upsilon_{\pi}^z \right] \geq& \frac{8}{25 \ln(R/a') R^2} \Tr_{\mathcal{F}}\left[ \mathbb{N} \left( \Upsilon_{\pi}^z - \Omega_b^z \right) \right] - \frac{128 \sqrt{2} b \vert \Lambda \vert^{1/2}}{\pi \ln(R/a') R^4} S(\Upsilon_{\pi}^z, \Omega_b^z)^{1/2} \nonumber \\
&+\frac{32 \lambda}{\pi \ln(R/a') R^4} \int_{\Lambda}\left[  \Tr_{\mathcal{F}} \left[ n_{R/2,\xi} \left( n_{R/2,\xi} - 1 \right) \Omega_b^z \right] - 6 \left( \Tr_{\mathcal{F}} \left[ n_{3R/2,\xi} \Omega_b^z \right] \right)^3 \right]_+ \de \xi \nonumber \\
&+ \frac{32 (1-\lambda) \e^{ -\pi (R/10)^2 \rho_{\omega} }}{\pi \ln(R/a') R^4} \int_{\Lambda} \frac{\langle \Phi_z | \chi_{R/10,\xi} | \Phi_z \rangle^2}{1+\langle \Phi_z | \chi_{R/10,\xi} | \Phi_z \rangle} \de \xi. \label{eq:interactionpartA24}
\end{align}
The choice of $\lambda$ will depend on the function $| \Phi_z |$. If it is approximately a constant, in a sense to be defined in Subsection~\ref{sec:interactionenergyC} below, we will choose $\lambda = 1$, otherwise we choose $\lambda = 0$.
\subsection{Interaction energy, part II}
\label{sec:interactionenergyB}
In this Section we give an upper bound for the expectation value of $\mathbb{W}_2$ in \eqref{eq:interactionpartA3}. The two-dimensional version of \cite[Lemma~5]{Seiringer2008} is the following statement\footnote{In \cite[Lemma~5]{Seiringer2008} the corresponding bound in three dimensions is incorrectly claimed with $C_n = 1$.}.
\begin{lem}
	\label{lem:Fourierdecay}
	Let $o:\mathbb{R}^2 \to \mathbb{C}$ be a smooth function, supported in a square of side length $4$, and for $s>0$, let $u(x) = L^{-2} \sum_{p\in \frac{2\pi}L \mathbb{Z}^2} o(sp) \e^{-ipx}$. Then for any nonnegative integer $n$, there exists a constant $C_n>0$ such that
	\begin{equation}
	\vert u(x) \vert \leq \left( \frac{s}{d(x,0)} \right)^{2n} C_n \max_{|\alpha| = 2n} \| \partial^\alpha o \|_{\infty} \left( \frac{2}{\pi s} + \frac{2n+1}{L} \right)^2.
	\label{eq:interactionpartB1}
	\end{equation} 
	Here  $\partial^\alpha o$ denotes the partial derivative of $o$ with respect to the multiindex $\alpha$.
	\begin{proof}
		For $x \in \Rr^2$ we write $x = (x_1,x_2)$. We have
		\begin{equation} \label{eqn-lem-fourierdecay-1}
		u(x) \left( 2 L^2 \left( 2 - \cos \left( \frac{2 \pi x_1}{L} \right) - \cos \left( \frac{2 \pi x_2}{L} \right) \right) \right)^n = \frac{1}{|\Lambda|} \sum_p \e^{- i p x} (-\Delta_\text{d})^n[o(sp)],
		\end{equation}
		where $(- \Delta_\text{d})f(p) = L^2(4 f(p) - \sum_{|e|=1} f(p+2 \pi e/L))$ denotes the discrete Laplacian in momentum space. It is easy to check that the discrete Laplacian can be estimated by maximizing over the second partial derivatives as
		\begin{equation} \label{eqn-discrete-laplacian-estimate}
		|(-\Delta_\text{d})^n f(p)| \leq C_n \max_{|\alpha| = 2n} \| \partial^\alpha f \|_{\infty}
		\end{equation}
		for an $n$-dependent constant $C_n$ independent of $f$. Note also that if $f$ is supported in a square of side length $\ell$, then after $n$-fold application of $-\Delta_\text{d}$ the support is contained in a square of side length $\ell + 4 \pi n/L$. An easy counting argument then allows us to estimate
		\begin{align*}
		|\eqref{eqn-lem-fourierdecay-1}| &\leq \frac{C_n}{|\Lambda|} \max_{|\alpha| = 2n} \| \partial^\alpha o(s \,\cdot\,) \|_{\infty} \sum_p \mathds{1}_{\supp (-\Delta_\text{d})^n o(sp)} \leq \frac{C_n s^{2n}}{|\Lambda|} \max_{|\alpha| = 2n} \| \partial^\alpha o \|_{\infty} \left( 1 + \frac{2L}{\pi s} + 2 n \right)^2\\
		&= C_n s^{2n} \max_{|\alpha| = 2n} \| \partial^\alpha o\|_{\infty} \left( \frac{2}{\pi s} + \frac{2n + 1}{L} \right)^2. \asterisknum
		\end{align*}
		We also estimate
		\begin{equation}
		1 - \cos \left( \frac{2 \pi x_i}{L} \right) \geq \frac{8}{L^2} \min_{k \in \Zz} | x_i - k L|^2
		\end{equation}
		and obtain
		\begin{equation}
		2L^2 \left( 2 - \cos \left( \frac{2 \pi x_1}{L} \right) - \cos \left( \frac{2 \pi x_2}{L} \right) \right) \geq 16 d(x,0)^2.
		\end{equation}
		Absorbing the factor $16$ into the constant $C_n$, we arrive at \eqref{eq:interactionpartB1}.
	\end{proof}
\end{lem}
By the definition of $f_R$ in~\eqref{eqn-Dyson-lemma-def-f-w}, we have
\begin{equation}
f_R(x) \leq R \sup_{d(x,y) \leq R} |\nabla h(y)| \leq R \sup_{d(x,y) \leq s} \vert \nabla h(y) \vert,
\label{eq:interactionpartB2}
\end{equation}
where we used $R \leq s$. By applying Lemma~\ref{lem:Fourierdecay} to $\nabla h$ we conclude that for $L$ large enough there exists a smooth function $g$ of rapid decay (i.e., $g$ decays like an arbitrary power) that is independent of $L$ such that the function $w_R$ defined in \eqref{eqn-Dyson-lemma-def-f-w} satisfies
\begin{equation}
w_R(x-y) \leq \frac{R^2}{s^4} g(d(x,y)/s).
\label{eq:interactionpartB3}
\end{equation}
For $\mathbb{W}_2$ this implies
\begin{equation}
\mathbb{W}_2 \leq \bigoplus_{n=2}^{\infty} \sum_{j=1}^n \sum_{i \in J_j} \frac{1}{\epsilon \ln(R/\tilde{a})} \frac{R^2}{s^4} g\left( \frac{d(x_j,x_i)}{s} \right).
\label{eq:interactionpartB4}
\end{equation}
Next we decompose the function $g$ into an integral over characteristic functions of disks. For this purpose, we use \cite[Theorem~1]{HS02} which allows us to write
\begin{equation}
g(t) = \int_0^{\infty} m(r) j(t/r) \de r
\label{eq:interactionpartB7}
\end{equation}
with
\begin{equation}
m(r) = -\frac{r}{16} \int_r^{\infty} g'''(s) s \left(s^2-r^2 \right)^{-1/2} \de s
\label{eq:interactionpartB8}
\end{equation}
and $j$ defined in \eqref{eq:fillingholes1}. 
Since the third derivative of $g$, denoted here by $g'''$, is of rapid decay, the same is true for $m$. As $j$ is a decreasing function, we have
\begin{equation}
g(t) \leq j(t) \int_0^1 \vert m(r) \vert \de r + \int_1^{\infty} \vert m(r) \vert j(t/r) \de r,
\label{eq:interactionpartB9}
\end{equation}
which implies
\begin{equation}\label{eq:interactionpartB10}
g\left( \frac{d(x_i,x_j)}{s} \right) \leq \int_s^{\infty}\left(   \delta(r-s)  \int_0^1  \vert m(t) \vert \de t + s^{-1} \vert m(r/s) \vert\right) j\left( \frac{d(x_i,x_j)}{r} \right) \de r.
\end{equation}
The integral over the $\delta$ function is understood as evaluation at $r = s$, i.e., the right-hand side of \eqref{eq:interactionpartB10} is nothing but the right-hand side of \eqref{eq:interactionpartB9} with  $t=d(x_i,x_j)/s$. As noted before in \eqref{eqn-j-integral-representation}, we can write
\begin{equation}
j(d(x_i,x_j)/r) = \frac{32}{\pi r^2} \int_{\Lambda} \chi_{r/2,\xi}(x_i) \chi_{r/2,\xi}(x_j) \de \xi
\label{eq:interactionpartB11}
\end{equation}
as long as $L \geq 2 r$. Eqs.~\eqref{eq:interactionpartB4} and \eqref{eq:interactionpartB10} together with Eq.~\eqref{eq:interactionpartB11} show that
\begin{align}
\mathbb{W}_2 \leq& \frac{32}{\pi \epsilon \ln(R/\tilde{a})} \frac{R^2}{s^6} \int_s^{b} \de r \left\{ \delta(r-s) \int_0^1 \vert m(t) \vert \de t + s^{-1} \vert m(r/s) \vert \right\} \nonumber \\
&\hspace{5cm} \times \int_{\Lambda} \de \xi \bigoplus_{n=0}^{\infty} \sum_{j=1}^{n} \sum_{i \in J_j} \chi_{r/2,\xi}(x_j) \chi_{r/2,\xi}(x_i) \nonumber \\
&\qquad + \frac{1}{\epsilon \ln(R/\tilde{a})} \frac{R^2}{s^4} \int_{b}^{\infty} s^{-1} \vert m(r/s) \vert \bigoplus_{n=0}^{\infty} \sum_{j=1}^{n} \sum_{i \in J_j} j\left( \frac{d(x_i,x_j)}{r} \right) \de r \label{eq:interactionpartB12}
\end{align}
holds. Here, we have split the integral over $r$ into two parts, one with $s \leq r \leq b$ and one with $b \leq r$. In the second part we do not have the same representation of $j$ as in \eqref{eq:interactionpartB11} as eventually $2r \geq L$. The cutoff parameter $b$ is chosen the same as in the definition of $\Omega^z_b$ in \eqref{eqn-def-Omega-b-z}.

Let $v_r(\xi)$ denote the integrand of the integral over $\xi$ in \eqref{eq:interactionpartB12}. Because $d(x_i,x_k) \geq R/5$ for $i,k \in J_j$, the number of $x_i$ inside a disk of radius $r/2$ is bounded from above by $(1+5r/R)^2$. Hence,
\begin{equation}
v_r(\xi) \leq n_{r/2,\xi} \left( 1+\frac{5r}{R} \right)^2.
\label{eq:interactionpartB13}
\end{equation}
On the other hand, we trivially have
\begin{equation}
v_r(\xi) \leq n_{r/2,\xi} \left( n_{r/2,\xi} - 1 \right).
\label{eq:interactionpartB13b}
\end{equation}
Combining these two bounds gives
\begin{equation}
v_r(\xi) \leq f\left( n_{r/2,\xi} \right) \quad \text{ where } \quad f(n) = n \min\left\{ (n-1), \left( 1+\frac{5r}{R} \right)^2 \right\}. 
\label{eq:interactionpartB13c}
\end{equation}
We use the above bounds and $| f(n) - n  ( 1+\frac{5r}{R} )^2 | \leq ( 1 + ( 1+\frac{5r}{R} )^2 )^2/4$ to estimate
\begin{align}
\Tr_{\mathcal{F}}\left[ v_r(\xi) \Upsilon^z_{\pi} \right] &\leq \Tr_{\mathcal{F}} \left[ f\left( n_{r/2,\xi} \right) \Upsilon^z_{\pi} \right] \nonumber \\
&\leq  \Tr_{\mathcal{F}} \left[ f\left( n_{r/2,\xi} \right) \Omega_b^z \right] + \left( 1+\frac{5r}{R} \right)^2 \Tr_{\mathcal{F}} \left[ n_{r/2,\xi} \left( \Upsilon^z_{\pi} - \Omega_b^z \right) \right] \nonumber \\
&\hspace{0.5cm} + \frac{1}{4} \left( 1 + \left( 1+\frac{5r}{R} \right)^2 \right)^2 \left\| \Upsilon^z_{\pi, \chi_{r/2,\xi}} - \Omega^z_{b,\chi_{r/2,\xi}} \right\|_1. \label{eq:interactionpartB14}
\end{align}
When integrated over $\xi$, the second and the third term on the right-hand side of \eqref{eq:interactionpartB14} can be estimated as in \eqref{eq:interactionpartA12} and \eqref{eq:interactionpartA13}, respectively. Using Wick's rule and a similar estimate as in \eqref{eq:interactionpartA17}, we bound the first term from above by
\begin{align} \label{eq:interactionpartB15}
\Tr_{\mathcal{F}} \left[ f\left( n_{r/2,\xi} \right) \Omega_b^z \right] &\leq \min \left\{ \Tr_{\mathcal{F}} \left[ n_{r/2,\xi} \left( n_{r/2,\xi} - 1 \right) \Omega_b^z \right], \left( 1+\frac{5r}{R} \right)^2 \Tr_{\mathcal{F}} \left[ n_{r/2,\xi} \Omega_b^z \right] \right\} \nonumber \\
&\leq \min \left\{ 2 \left( \Tr_{\mathcal{F}} \left[ n_{r/2,\xi} \Omega_b^z \right] \right)^2, \left( 1+\frac{5r}{R} \right)^2 \Tr_{\mathcal{F}} \left[ n_{r/2,\xi} \Omega_b^z \right] \right\} \nonumber\\
&\leq \frac{4 \left( \Tr_{\mathcal{F}} \left[ n_{r/2,\xi} \Omega_b^z \right] \right)^2}{1+ 2\Tr_{\mathcal{F}} \left[ n_{r/2,\xi} \Omega_b^z \right]/\left( 1+5r/R \right)^2}.
\end{align}
Moreover,
\begin{equation}
\Tr_{\mathcal{F}} \left[ n_{r/2,\xi} \Omega_b^z \right] = \frac{\pi r^2}{4} \rho_{\omega} + \langle \Phi_z | \chi_{r/2,\xi} | \Phi_z \rangle. 
\label{eq:interactionpartB15a}
\end{equation}
Using convexity of the function $x \mapsto x^2/(1+x)$, we obtain
\begin{equation}
\Tr_{\mathcal{F}} \left[ f\left( n_{r/2,\xi} \right) \Omega_b^z \right] \leq \frac{1}{2} \left( \pi r^2 \rho_{\omega} \right)^2 + \frac{8 \langle \Phi_z | \chi_{r/2,\xi} | \Phi_z \rangle^2}{1 + 4 \langle \Phi_z | \chi_{r/2,\xi} | \Phi_z \rangle/ \left( 1+5r/R \right)^2}.
\label{eq:interactionpartB15b}
\end{equation}
Putting these considerations together we find (for $R \leq s \leq r \leq b$) 
\begin{align} 
\int_{\Lambda} \Tr_{\mathcal{F}} \left[ v_r(\xi) \Upsilon^z_\pi \right] \de \xi \leq&  \frac{| \Lambda |}{2} \left( \pi r^2 \rho_{\omega} \right)^2 + \int_{\Lambda} \frac{8 \langle \Phi_z | \chi_{r/2,\xi} | \Phi_z \rangle^2}{1 + 4 \langle \Phi_z | \chi_{r/2,\xi} | \Phi_z \rangle/ \left( 1+5r/R \right)^2} \de \xi \nonumber \\
&+ \frac{9\pi r^4}{R^2} \Tr_{\mathcal{F}} \left[ \mathbb{N} \left( \Upsilon^z_{\pi} - \Omega_b^z \right) \right] + \frac{b | \Lambda |^{1/2}}{\sqrt{2}} 37^2  \left( \frac{r}{R} \right)^4 S\left( \Upsilon^z_{\pi}, \Omega_b^z \right)^{1/2}. \label{eq:interactionpartB16}
\end{align}

In order to be able to compare the second term on the right-hand side of the above inequality to the last term in \eqref{eq:interactionpartA24}, we use the pointwise bound
\begin{equation}
\chi_{r/2,\xi}(x) \leq \frac{\left( 1 + 5r/R \right)^2 }{\pi (r/2 + R/10)^2} \int_{|a| \leq r/2 + R/10} \chi_{R/10,\xi+a}(x) \de a.
\label{eq:interactionpartB16c}
\end{equation}
We first use the monotonicity of the map $x \mapsto x^2/(1+x)$ to replace $\chi_{r/2,\xi}(x)$ by the right-hand side of the above equation in the second term on the right-hand side of \eqref{eq:interactionpartB16}. Afterwards we use the convexity of the same map and Jensen's inequality to see that
\begin{align}\label{eq:interactionpartB16d}
&\frac{8 \langle \Phi_z | \chi_{r/2,\xi} | \Phi_z \rangle^2}{1 + 4 \langle \Phi_z | \chi_{r/2,\xi} | \Phi_z \rangle/ \left( 1+5r/R \right)^2} \nonumber\\
&\quad \leq \frac{\left( 1+5r/R \right)^4}{\pi (r/2 + R/10)^2} \int_{|a| \leq r/2 + R/10} \frac{8 \langle \Phi_z | \chi_{R/10,\xi + a} | \Phi_z \rangle^2}{1 + 4 \langle \Phi_z | \chi_{R/10,\xi+a} | \Phi_z \rangle} \de a
\end{align}
holds. Now we integrate in $\xi$ over $\Lambda$ and obtain
\begin{align} 
&\frac{(1+5r/R)^4}{\pi(r/2 + R/10)^2} \int_\Lambda \int_{|a| \leq r/2 + R/10} \frac{8 \braket{\Phi_z | \chi_{R/10,\xi + a} | \Phi_z}^2}{1 + 4 \braket{\Phi_z | \chi_{R/10,\xi + a} | \Phi_z}} \de a \de \xi 	\label{eq:interactionpartB16e}  \\
&\hspace{2cm} = \left( 1+5r/R \right)^4 \int_{\Lambda} \frac{8 \langle \Phi_z | \chi_{R/10,\xi} | \Phi_z \rangle^2}{1 + 4 \langle \Phi_z | \chi_{R/10,\xi} | \Phi_z \rangle} \de \xi \leq \left( 6r/R \right)^4 \int_{\Lambda} \frac{8 \langle \Phi_z | \chi_{R/10,\xi} | \Phi_z \rangle^2}{1 +  \langle \Phi_z | \chi_{R/10,\xi} | \Phi_z \rangle} \de \xi. \nonumber
\end{align}
The integral in the first term on the right-hand side of \eqref{eq:interactionpartB12} is therefore bounded from above by
\begin{align}
&\int_s^{b}  \left\{ \delta(r-s) \int_0^1 \vert m(t) \vert \de t + s^{-1} \vert m(r/s) \vert \right\} \int_{\Lambda} \Tr_{\mathcal{F}}\left[ v_r(\xi) \Upsilon^z_\pi \right]  \de \xi \de r \nonumber \\
&\hspace{4cm}\leq c \Bigg[ \frac{\pi}{4} s^2 \left( \frac{6s}{R} \right)^2 \Tr_{\mathcal{F}}\left[ \mathbb{N} \left( \Upsilon_{\pi}^z - \Omega_b^z \right) \right] + \frac{b | \Lambda |^{1/2}}{\sqrt{2}} 37^2 \left( \frac{s}{R} \right)^4 S\left( \Upsilon_{\pi}^z, \Omega_b^z \right)^{1/2}  \nonumber \\
&\hspace{4.6cm}+ \left( \frac{6s}{R} \right)^4 \int_{\Lambda} \frac{8 \langle \Phi_z | \chi_{R/10,\xi} | \Phi_z \rangle^2}{1 + \langle \Phi_z | \chi_{R/10,\xi} | \Phi_z \rangle} \de \xi  + \frac{ | \Lambda |}{2}  \left( \pi s^2 \rho_{\omega} \right)^2   \Bigg], \label{eq:interactionpartB19a}
\end{align}
where
\begin{equation}
c = \int_0^1 \vert m(t) \vert \de t + \int_1^{\infty} \vert m(t) \vert t^4 \de t.
\label{eq:interactionpartB19}
\end{equation}

It remains to bound the second term on the right-hand side of \eqref{eq:interactionpartB12} where $r \geq b$. We use \eqref{eq:fillingholes2} and the same argument that led to \eqref{eq:interactionpartB13} to see that
\begin{equation}
\sum_{i \in J_j} j\left( \frac{d(x_i,x_j)}{r} \right) \leq 8 \left( 1 + \frac{5 r}{R} \right)^2.
\label{eq:interactionpartB16a} 
\end{equation}
This implies
\begin{equation}
\int_{b}^{\infty} s^{-1} | m(r/s) | \bigoplus_{n=2}^{\infty} \sum_{j=1}^{n} \sum_{i \in J_j} j\left( \frac{d(x_i,x_j)}{r} \right) \de r \leq \mathbb{N} \left( \frac{6 s}{R} \right)^2 8 \int_{b/s}^{\infty} | m(r) | r^2 \de r. 
\label{eq:interactionpartB16b}
\end{equation}
In the following we denote
\begin{equation}
J(b/s) = \int_{b/s}^{\infty} | m(r) | r^2 \de r.
\label{eq:interactionpartB17}
\end{equation}
Since $|m|$ decays like  an arbitrary power, the same holds true for $J$. The contribution to $\Tr_{\mathcal{F}} \left[ \mathbb{W}_2 \Upsilon_{\pi}^z \right]$ from this part (except for the prefactor) is therefore bounded from above by
\begin{align}
\left( \frac{6 s}{R} \right)^2 8 J(b/s)  \Tr_{\mathcal{F}} \left[ \mathbb{N} \Upsilon_{\pi}^z  \right] \,.
\label{eq:interactionpartB18}
\end{align}
In combination, \eqref{eq:interactionpartB12}, \eqref{eq:interactionpartB19a} and \eqref{eq:interactionpartB18} show that
\begin{align}
\Tr_{\mathcal{F}} \left[ \mathbb{W}_2 \Upsilon^z_\pi \right] &\leq \frac{32 R^2}{\epsilon \pi s^2\ln(R/\tilde{a})} \Bigg( \frac{ 9 \pi \Tr_{\mathcal{F}}\left[ \mathbb{N} \left( \Upsilon_{\pi}^z - \Omega_b^z \right) \right] }{R^2} \left( c + J(b/s) \right) \nonumber \\
&\quad + \frac{9 \pi \Tr_{\mathcal{F}}\left[  \mathbb{N} \Omega_b^z \right] }{R^2} J(b/s)  + \frac{37^2 c b}{\sqrt{2}R^4} |\Lambda|^{1/2} S(\Upsilon_{\pi}^z,\Omega_b^z)^{1/2} + \frac{| \Lambda | c \pi^2 \rho_{\omega}^2}{2}  \nonumber \\
&\quad + \left( \frac{6}{R} \right)^4 8 c \int_{\Lambda} \frac{ \langle \Phi_z | \chi_{R/10,\xi} | \Phi_z \rangle^2}{1 + \langle \Phi_z | \chi_{R/10,\xi} | \Phi_z \rangle} \de \xi \Bigg)\label{eq:interactionpartB20}
\end{align}
holds. This is the equivalent\footnote{We note that in  \cite[Eq.~(2.10.27)]{Seiringer2008} the first factor $J(b/s)$ on the right side is missing. This is of no consequence, however, as $J(b/s)$ is small for $s\ll b$.} of \cite[Eq.~(2.10.27)]{Seiringer2008}.
\subsection{Interaction energy, part III}
\label{sec:interactionenergyC}
In this subsection we will put the bounds of the previous two subsections together in order to obtain the final lower bound for $\Tr_{\mathcal{F}}\left[ \mathbb{W} \Upsilon_{\pi}^z \right]$. To do so we will distinguish two cases depending on the value of a certain function of $\Phi_z$. 

Assume first that
\begin{equation}
\int_{\Lambda} \frac{ \langle \Phi_z | \chi_{R/10,\xi} | \Phi_z \rangle^2}{1 + \langle \Phi_z | \chi_{R/10,\xi} | \Phi_z \rangle} \de \xi \geq \frac{\pi^2}{8} | \Lambda | \left( R^2 \rho \right)^2 
\label{eq:interactionpartC1a}
\end{equation} 
holds. Essentially, this condition means that $\Phi_z$ is far from being a constant. In this case, we choose $\lambda = 0$ in \eqref{eq:interactionpartA24}. Using the condition \eqref{eq:interactionpartC1a}, we check that the difference of the last term in \eqref{eq:interactionpartA24} and the last term in \eqref{eq:interactionpartB20} is bounded from below by
\begin{equation}
\frac{4 \pi | \Lambda | \rho^2}{\ln ( R/a' )} \left( 1 - \pi \left( \frac{R}{10} \right)^2 \rho_{\omega} - 8 c \frac{6^4 R^2 \ln(R/a')}{\epsilon s^2 \ln(R/\tilde a)} \right).
\label{eq:interactionpartC1b}
\end{equation}
Here we  used that for our choice of parameters the term in parentheses will be positive (in fact, close to $1$). 

Next we consider the case when \eqref{eq:interactionpartC1a} does not hold, in which case we choose $\lambda=1$ in \eqref{eq:interactionpartA24}. We start by proving some bounds that will turn out to be helpful below. Using \eqref{eq:interactionpartB16c} with the choice $r = 3R$ and the monotonicity as well as the convexity of the map $x \mapsto x^2/(1+x)$, we see that
\begin{equation}
\int_{\Lambda} \frac{ \langle \Phi_z | \chi_{3R/2,\xi} | \Phi_z \rangle^2}{1 + 16^{-2} \langle \Phi_z | \chi_{3 R/2,\xi} | \Phi_z \rangle} \de \xi \leq 16^4 \frac{\pi^2}{8} | \Lambda | \left( R^2 \rho \right)^2
\label{eq:interactionpartC1c}
\end{equation}
holds in this case. Pick some $D>0$ and let $\mathcal{B} \subset \Lambda$ be the set
\begin{equation}
\mathcal{B} = \left\{ \xi \in \Lambda \ \big| \ \langle \Phi_z | \chi_{3 R/2,\xi} | \Phi_z \rangle \geq 16^2 D R^2 \rho \right\} .
\label{eq:interactionpartC1d}
\end{equation}
Using \eqref{eq:interactionpartC1c} as well as monotonicity of the map $x \mapsto x/(1+x)$, we obtain
\begin{equation}
\int_{\mathcal{B}} \langle \Phi_z | \chi_{3R/2,\xi} | \Phi_z \rangle \de \xi \leq 32 \pi^2 |\Lambda| R^2 \rho \left( \frac 1 D + R^2 \rho \right).
\label{eq:interactionpartC1e}
\end{equation}
We proceed similarly to find an estimate for the volume of $\mathcal{B}$:
\begin{equation}
| \mathcal{B} | \leq \frac{\pi^2 | \Lambda | }{8 D^2} \left( 1 + D R^2 \rho \right).
\label{eq:interactionpartC1f}
\end{equation}
We choose $\lambda = 1$ in \eqref{eq:interactionpartA24} and estimate the relevant term from below by
\begin{align}
&\int_{\Lambda}\left[ \left( \Tr_{\mathcal{F}} \left[ n_{R/2,\xi} \left( n_{R/2,\xi} - 1 \right) \Omega_b^z \right] \right)  - 6 \left( \Tr_{\mathcal{F}} \left[ n_{3R/2,\xi} \Omega_b^z \right] \right)^3 \right]_+ \de \xi \nonumber \\
&\quad\geq \int_{\Lambda \backslash \mathcal{B}} \left( \left( \Tr_{\mathcal{F}} \left[ n_{R/2,\xi} \left( n_{R/2,\xi} - 1 \right) \Omega_b^z \right] \right) - 6 \left( \Tr_{\mathcal{F}} \left[ n_{3R/2,\xi} \Omega_b^z \right] \right)^3 \right) \de \xi. \label{eq:interactionpartC1g}
\end{align}
Recall that we defined $\Omega_b^z = U(z) \Omega_b U(z)^\dagger$, where $U(z)$ is the Weyl operator from \eqref{eq:coherentstates8} and $\Omega_b$ is the quasi-free state with one-particle density matrix $\omega_b$ defined in \eqref{eq:locrelentropy2}. In order to derive a bound on the second term on the right-hand side, we note that $\Tr_{\mathcal{F}} \left[ n_{3R/2,\xi} \Omega_b^z \right] = \pi (3R/2)^2 \rho_{\omega} + \langle \Phi_z | \chi_{3R/2,\xi} | \Phi_z  \rangle$. Together with the convexity of the map $x \mapsto x^3$ and \eqref{eq:interactionpartC1d} we conclude that
\begin{align} \label{eq:interactionpartC1h}
\int_{\Lambda \backslash \mathcal{B}} \left( \Tr_{\mathcal{F}} \left[ n_{3R/2,\xi} \Omega_b^z \right] \right)^3 \de \xi &\leq 4 | \Lambda | \left( \pi (3R/2)^2 \rho_{\omega} \right)^3 + 4 \int_{\Lambda \backslash \mathcal{B}}  \langle \Phi_z | \chi_{3R/2,\xi} | \Phi_z  \rangle^3 \de \xi \nonumber \\
&\leq 4 | \Lambda | \left( \pi (3R/2)^2 \rho_{\omega} \right)^3 + \left( 16^2 D R^2 \rho \right)^2 9 \pi R^2 | z | ^2
\end{align} 
holds. 

Now we investigate the first term on the right-hand side of \eqref{eq:interactionpartC1g}. Similarly to \eqref{eq:interactionpartA17} above, we have
\begin{align}
&\Tr_{\mathcal{F}}\left[ n_{R/2,\xi} \left( n_{R/2,\xi} - 1 \right) \Omega_b^z \right] = \Tr_{\mathcal{F}}\left[ n_{R/2,\xi} \left( n_{R/2,\xi} - 1 \right) \Omega_b \right] \nonumber \\
&\qquad+ 2 \langle \Phi_z | \chi_{R/2,\xi} \omega_b \chi_{R/2,\xi} | \Phi_z \rangle + \frac{\pi}{2} R^2 \rho_{\omega} \langle \Phi_z | \chi_{R/2,\xi} | \Phi_z \rangle + \langle \Phi_z | \chi_{R/2,\xi} | \Phi_z \rangle^2. \label{eq:interactionpartC1i}
\end{align}
Note that we have used the translation invariance of the state $\Omega_b$. Since $\Omega_b$ is quasi-free the first term on the right-hand side can be expressed in terms of the one-particle density matrix $\omega_b$ and its density $\rho_\omega$. It reads  
\begin{equation}\label{eq:interactionpartC2}
\Tr_{\mathcal{F}} \left[ n_{R/2,\xi} \left( n_{R/2,\xi} - 1 \right) \Omega_b \right] =  ( \pi R^2 \rho_{\omega}/4 )^2 + \tr \left[  \chi_{R/2,\xi} \omega_b  \chi_{R/2,\xi} \omega_b \right].
\end{equation}
In order to quantify how much the integral of the first term on the right-hand side of \eqref{eq:interactionpartC1g} differs from the one with $\Lambda \backslash \mathcal{B}$ replaced by $\Lambda$, we estimate
\begin{equation}
\int_{\mathcal{B}} \Tr_{\mathcal{F}} \left[ n_{R/2,\xi} \left( n_{R/2,\xi} - 1 \right) \Omega_b \right] \de \xi \leq 2 | \mathcal{B} | ( \pi R^2 \rho_{\omega}/4 )^2.
\label{eq:interactionpartC2a}
\end{equation}
To arrive at the right-hand side, we used that the second term in the second line of \eqref{eq:interactionpartC2} is bounded from above by the first one. Since $\langle \Phi_z | \chi_{R/2,\xi} \omega_b \chi_{R/2,\xi} | \Phi_z \rangle  \leq \tr \chi_{R/2,\xi} \omega_b \langle \Phi_z | \chi_{R/2,\xi} | \Phi_z \rangle$, we also have
\begin{align}
&\int_{\mathcal{B}} \left( 2  \langle \Phi_z | \chi_{R/2,\xi} \omega_b \chi_{R/2,\xi} | \Phi_z \rangle + \frac{\pi}{2} R^2 \rho_{\omega} \langle \Phi_z | \chi_{R/2,\xi} | \Phi_z \rangle \right) \de \xi \leq \pi R^2 \rho_{\omega} \int_{\mathcal{B}}  \langle \Phi_z | \chi_{R/2,\xi} | \Phi_z \rangle \de \xi \nonumber \\
&\quad \leq \pi R^2 \rho_{\omega} 32 \pi^2 | \Lambda | R^2 \rho \left( \frac 1 D + R^2 \rho \right). \label{eq:interactionpartC2b}
\end{align}
For the last inequality, we used \eqref{eq:interactionpartC1e} and the fact that $\int_{\mathcal{B}}  \langle \Phi_z | \chi_{R/2,\xi} | \Phi_z \rangle \de \xi$ is bounded from above by $\int_{\mathcal{B}}  \langle \Phi_z | \chi_{3R/2,\xi} | \Phi_z \rangle \de \xi$. For the last term in \eqref{eq:interactionpartC1i} we use Schwarz's inequality and \eqref{eq:interactionpartC1e} to estimate
\begin{align}
\int_{\Lambda \backslash \mathcal{B}}  \langle \Phi_z | \chi_{R/2,\xi} | \Phi_z \rangle^2 \de \xi &\geq \frac{1}{| \Lambda |} \left( \int_{\Lambda \backslash \mathcal{B}}  \langle \Phi_z | \chi_{R/2,\xi} | \Phi_z \rangle \de \xi \right)^2 \label{eq:interactionpartC2c} \nonumber\\
&\geq | \Lambda | \frac{\pi^2}{16} R^4 \left[ \rho_z^2 - \pi \rho_z \rho \frac{16^2}{D} \left( 1 + D R^2 \rho \right) \right].
\end{align}
Here we have again used the notation $\rho_z = | z |^2 / | \Lambda |$. Putting all these estimates together, we have the lower bound
\begin{align} \label{eq:interactionpartC2d}
&\int_{\Lambda \backslash \mathcal{B}} \Tr_{\mathcal{F}} \left[ n_{R/2,\xi} \left( n_{R/2,\xi} - 1 \right) \Omega_b^z \right] \de \xi \geq \frac{| \Lambda | \pi^2 R^4 \rho_{\omega}^2}{16} \left( 1 - \frac{\pi^2}{4 D^2} \left( 1 + D R^2 \rho \right) \right) \nonumber\\
&\quad + \int_{\Lambda} \tr \left[ \chi_{R/2,\xi} \omega_b \chi_{R/2,\xi} \omega_b \right] \de \xi + 2 \int_{\Lambda} \langle \Phi_z | \chi_{R/2,\xi} \omega_b \chi_{R/2,\xi} | \Phi_z \rangle \de \xi \nonumber \\
&\quad  + | \Lambda | \frac{\pi^2}{16} R^4 \left[ 2 \rho_z \rho_{\omega} + \rho_z^2 - \pi \rho_z \rho \frac{16^2}{D} \left( 1 + D R^2 \rho \right) \right] - 32 | \Lambda | \pi^3 R^4 \rho_{\omega} \rho \left( \frac 1 D + R^2 \rho \right).
\end{align}

We denote $\omega_b(x) = \omega_b(x,0) = \omega_{\pi}(x,0) \eta_{b}\left(d(x,0)\right)$. The first term in the second line of \eqref{eq:interactionpartC2d} can be written as
\begin{align}
&\int_{\Lambda^3} \chi_{R/2,\xi}(x) \chi_{R/2,\xi}(y) \vert \omega_b(x,y) \vert^2 \text{d}(x,y,\xi) = \int_{\Lambda^3} \chi_{R/2,\xi}(x+y) \chi_{R/2,\xi}(y) \vert \omega_b(x) \vert^2 \de (x,y,\xi) \nonumber  \\
&\quad = \frac{ \vert \Lambda \vert \pi R^2}{32} \int_{\Lambda} j(d(x,0)/R) \vert \omega_b(x) \vert^2 \de x. \label{eq:interactionpartC3}
\end{align} 
An application of the Cauchy-Schwarz inequality implies 
\begin{equation}
\frac{ \vert \Lambda \vert \pi R^2}{32} \int_{\Lambda} j(d(x,0)/R) \vert \omega_b(x) \vert^2 \de x \geq \frac{\vert \Lambda \vert \pi^2 R^4}{16} \gamma_b^2,
\label{eq:interactionpartC5}
\end{equation}
where we defined
\begin{equation}
\gamma_b = \frac{1}{2 \pi R^2} \int_{\Lambda} \omega_b(x) j(d(x,0)/R) \de x. 
\label{eq:interactionpartC4}
\end{equation}
We note that $\gamma_b \sim \rho_{\omega}$ for $b \gg R$ and $\beta^{1/2} \gg R$. Below we will give more precise estimates (see \eqref{eqn-gamma_b-estimate}). It remains to give a lower bound on the second term in the second line of \eqref{eq:interactionpartC2d}. We claim that
\begin{equation}
\int_{\Lambda} \langle \Phi_z | \chi_{R/2,\xi} \omega_b \chi_{R/2,\xi} | \Phi_z \rangle \de \xi \geq |z|^2 \frac{\pi^2 R^4}{16} \left( \gamma_b - \rho_{\omega} p_{\text{c}} R \right).
\label{eq:interactionpartC5a}
\end{equation}
To see this, we write
\begin{align}
&\frac{32}{\pi R^2} \int_{\Lambda} \langle \Phi_z | \chi_{R/2,\xi} \omega_b \chi_{R/2,\xi} | \Phi_z \rangle \de \xi - |z|^2 \int_{\Lambda} \omega_b(x) j\left( \frac{d(x,0)}{R} \right) \de x \label{eq:interactionpartC5b} \nonumber\\
&\quad = \int_{\Lambda \times \Lambda} \left( \Phi_z^\dagger(x+y) - \Phi_z^\dagger(y) \right) \Phi_z(y) \omega_b(x) j\left( \frac{d(x,0)}{R} \right) \de (x,y) \nonumber \\
&\quad \geq - \left\| \Phi_z \right\| \int_{\Lambda} \left\| \Phi_z(x + \cdot) - \Phi_z(\cdot) \right\| | \omega_b(x) | j\left( \frac{d(x,0)}{R} \right) \de x. 
\end{align}
We estimate $| \omega_b(x) | \leq \omega_b(0) = \rho_{\omega}$. Moreover, writing the relevant norm in momentum space one easily checks that $\left\| \Phi_z(x + \cdot) - \Phi_z(\cdot) \right\| \leq \left\| \Phi_z \right\| p_{\text{c}} d(x,0)$. Since the support of $j(\cdot /R)$ is the interval $[0,R]$, the integral over $\Lambda$ can be estimated as 
\begin{equation}
\int_{\Lambda} j(d(x,0)/R) d(x,0) \de x \leq 2 \pi R^3.
\label{eq:interactionpartC5c}
\end{equation}
This proves \eqref{eq:interactionpartC5a}. Combining these estimates with \eqref{eq:interactionpartC1h} and \eqref{eq:interactionpartC2d} we see that
\begin{align} \label{eq:interactionpartC5d} 
&\frac{32}{\pi \ln(R/a') R^4} \int_{\Lambda}\left[ \left( \Tr_{\mathcal{F}} \left[ n_{R/2,\xi} \left( n_{R/2,\xi} - 1 \right) \Omega_b^z \right] \right)  - 6 \left( \Tr_{\mathcal{F}} \left[ n_{3R/2,\xi} \Omega_b^z \right] \right)^3 \right]_+ \de \xi \nonumber\\
&\quad \geq \frac{2 \pi | \Lambda | \rho_{\omega}^2}{\ln(R/a')} \left( 1 - \frac{\pi^2}{4 D^2} \left( 1 + D R^2 \rho \right) \right) + \frac{2 \pi | \Lambda | \gamma_b^2}{\ln(R/a')} + \frac{4 \pi | \Lambda | \rho_z}{\ln(R/a')} \left( \gamma_b - \rho_{\omega} p_{\text{c}} R \right) \nonumber \\
&\qquad + \frac{2 \pi | \Lambda |}{\ln(R/a')} \left[ 2 \rho_z \rho_{\omega} + \rho_z^2 - \pi \rho_z \rho \frac{16^2}{D}\left( 1 + D R^2 \rho \right) \right] - \frac{12 \cdot 3^6 \pi^2 | \Lambda | \rho_{\omega}^3 R^2}{\ln(R/a') }  \nonumber \\
&\qquad - \frac{32^2 \pi^2 | \Lambda | \rho_{\omega} \rho}{\ln(R/a')} \left( \frac 1 D + R^2 \rho \right) - \frac{1728 \cdot 16^4 | \Lambda | (D R^2 \rho)^2 \rho_z}{\ln(R/a') R^2}.
\end{align}
Now we put the results of this subsection and the two previous ones together. More precisely, we combine the estimates from Eqs.~\eqref{eq:interactionpartA24}, \eqref{eq:interactionpartB20}, \eqref{eq:interactionpartC1b} and \eqref{eq:interactionpartC5d} to obtain
\begin{align} \label{eq:interactionpartC7}
\Tr_{\mathcal{F}}\left[ \mathbb{W} \Upsilon_{\pi}^z \right] &\geq \Tr_{\mathcal{F}} \left[ \mathbb{N}\left( \Upsilon_{\pi}^z - \Omega_b^z \right) \right] \left( \frac{8}{25 \ln(R/a') R^2} - \frac{288}{\epsilon \ln(R/\tilde{a})s^2} \left( c + J(b/s) \right) \right) \nonumber \\
&\quad -\frac{\sqrt{2}}{\pi \ln(R/\tilde{a}) R^4} \left( b^2 | \Lambda | S(\Upsilon_{\pi}^z,\Omega_b^z) \right)^{1/2} \left( 128 + \frac{16 \cdot 37^2 c R^2}{\epsilon s^2 } \right) \\
&\quad - \frac{2 \pi | \Lambda |}{\ln(R/\tilde{a})} \left( \frac{144  \left( \rho_{\omega} + \rho_z \right) }{\pi \epsilon s^2}  J(b/s) +\frac{ 8 c \rho_{\omega}^2 R^2}{\epsilon s^2} \right) + \frac{2 \pi | \Lambda |}{\ln(R/a')} \min\{ \mathcal{A}_1, \mathcal{A}_2 \}. \nonumber
\end{align}
To arrive at this result we used that $a' \leq \tilde{a}$, and we defined
\begin{equation}
\mathcal{A}_1 = 2 \rho^2 \left( 1 - \pi \left( \frac{R}{10} \right)^2 \rho_{\omega} - 8 c \frac{6^4 R^2 \ln(R/a')}{\epsilon s^2 \ln(R/\tilde a)} \right)
\label{eq:interactionpartC8}
\end{equation}
and
\begin{align} \label{eq:interactionpartC9}
\mathcal{A}_2 &=  \rho_{\omega}^2 + \gamma_b^2 + 2 \rho_z ( \gamma_b +  \rho_{\omega}) + \rho_z^2 \nonumber\\
&\quad  - \rho_{\omega}^2 \left[ \frac{\pi^2}{4 D^2} \left( 1+D R^2 \rho \right) + 6 \cdot 3^6 \pi \rho_{\omega} R^2 \right] -2 \rho_z \rho_{\omega} p_{\text{c}} R - 2\rho_{\omega} \rho \frac{16^2 \pi}{D} \left( 1+ D R^2 \rho \right) \nonumber\\
&\quad - \rho \rho_z \left[ \frac{864}{\pi} \cdot 16^4 D^2 R^2 \rho + \pi \frac{16^2}{D} \left( 1 + D R^2 \rho \right) \right] -16 c \frac{6^4 R^2 \rho^2 \ln(R/a')}{\epsilon s^2 \ln(R/\tilde a)}.
\end{align}
Later we will choose the parameters such that $\ln(R/\tilde{a})$ and $\ln(R/a')$ are equal to leading order in the dilute limit. We will also choose $\epsilon s^2/R^2$ large enough such that the factor multiplying $\Tr_{\mathcal{F}} \left[ \mathbb{N}\left( \Upsilon_{\pi}^z - \Omega_b^z \right) \right]$ in \eqref{eq:interactionpartC7} is positive. Hence, it will be sufficient to give a lower bound for the difference of the expected particle numbers of $\Upsilon_{\pi}^z$ and $\Omega_b^z$, which will be done in the next subsection. 

To simplify the expressions, we make a choice of the parameters $\epsilon$ and $D$ and restrict the range of $R$.  We claim that all the terms with a negative sign appearing in $\mathcal{A}_1$ and $\mathcal{A}_2$ (together with the prefactor)  can be bounded from below by
\begin{equation}
- \text{const. } \frac{|\Lambda| \rho^2}{|\ln a^2 \rho|} \left( (R^2 \rho)^{1/3} + \frac{R}{s} + p_{\text{c}} R \right).
\end{equation}
To see this we, we employ the bound on $\rho_z$ derived in \eqref{eq:relativeentropyandaprioribounds7} as well as the following bound on $\rho_{\omega}$. Recall that $\ell(p)$ was defined in \eqref{eqn-def-l(p)} and satisfies $\ell(p) \geq \beta(p^2 - \mu_0)$ for all $p$. This implies
\begin{equation}\label{bound:rhoo}
\rho_{\omega} = \frac{1}{|\Lambda|} \sum_p \frac{1}{\e^{\ell(p)} - 1} \leq \frac{1}{|\Lambda|} \sum_p \frac{1}{\e^{\beta(p^2 - \mu_0)} - 1} = \rho + o(1)
\end{equation}
in the thermodynamic limit. In order to minimize the error terms in $\mathcal{A}_2$, we choose $D = (R^2 \rho)^{-1/3}$. On the other hand, note that in the definition of $1/\ln(R/a')$ in \eqref{eq:fillingholes6} there is a factor $1-\epsilon$ which means there is competition between $\epsilon$ and $R^2/(\epsilon s^2)$ to leading order and thus the optimal choice is $\epsilon = R/s$. We also use that $a' \leq \tilde a \leq a$ and make the assumption
\begin{equation} \label{eqn-lnR/a-ub}
\frac{1}{\ln(R/a)} \lesssim \frac{1}{|\ln a^2 \rho|}.
\end{equation}
In combination, these considerations prove the claim. 

Now we give upper and lower bounds to $\gamma_b$ in terms of $\rho_{\omega}$ as promised above. We claim that
\begin{equation} \label{eqn-gamma_b-estimate}
\rho_{\omega} \geq \gamma_b \geq \rho_{\omega} \left( 1 - \frac{\text{const. } R^2}{b^2} \right) - \frac{\text{const. } R^2}{\beta^2} - o(1),
\end{equation}
where the $o(1)$ contribution vanishes in the thermodynamic limit. The upper bound can be obtained by noting that $|\omega_b(x)| \leq \omega_b(0) = \rho_{\omega}$. For the lower bound, recall that $\omega_b(x) = \omega_{\pi}(x,0)\eta_b(d(x,0))$. We use $\cos(x) \geq 1 - \frac 1 2 x^2$ to estimate
\begin{equation}
\omega_\pi(x) = \frac{1}{|\Lambda|} \sum_p \frac{\cos(px)}{\e^{\ell(p)}- 1} \geq \rho_{\omega} - \frac{d(x,0)^2}{2 |\Lambda|} \sum_p \frac{p^2}{\e^{\ell(p)} - 1}.
\end{equation}
We further use that $|\eta| \leq 1$ and $\eta(t) \geq 1 - \text{const.} \, t^2$. With the support of $j$ being contained in a disk of radius one, we can estimate $d(x,0) \leq R$ inside the integral in \eqref{eq:interactionpartC4}. Additionally, we use $\ell(p) \geq \beta p^2$. In combination, the above facts allow us to bound
\begin{align*}
\gamma_b &\geq \frac{\rho_{\omega}}{2 \pi R^2} \int_\Lambda \eta ( d(x,0)/b ) j ( d(x,0)/R ) \de x\\
&\qquad - \frac{1}{4 \pi |\Lambda| R^2} \sum_p \frac{p^2}{\e^{\ell(p)} - 1} \int_\Lambda d(x,0)^2 \eta(d(x,0)/b) j(d(x,0)/R) \de x\\
&\geq \frac{\rho_{\omega}}{2 \pi R^2} \left( \int_\Lambda j(d(x,0)/R) \de x - \text{const. } \int_\Lambda \frac{d(x,0)^2}{b^2} j(d(x,0)/R) \de x \right)\\
&\qquad - \frac{1}{8 \pi^2 \beta^2} \int_{\Rr^2} \frac{p^2}{\e^{p^2} - 1} \de p \int_\Lambda j(d(x,0)/R) \de x - o(1)\\
&= \rho_{\omega} \left( 1 - \text{const. } \frac{R^2}{b^2} \right) - \text{const. } \frac{R^2}{\beta^2} - o(1). \asterisknum
\end{align*}
This proves \eqref{eqn-gamma_b-estimate}.

To estimate the terms in $\mathcal{A}_1$ and $\mathcal{A}_2$ with a positive sign, we apply the lower bound from \eqref{eqn-gamma_b-estimate} to $\gamma_b$ and find
\begin{equation}
\rho_{\omega}^2 + \gamma_b^2 + 2 \rho_z (\gamma_b + \rho_{\omega}) + \rho_z^2 \geq 2 \rho_{\omega}^2 + 4 \rho_z \rho_{\omega} + \rho_z^2 - \text{const.} \left( \rho^2 \frac{R^2}{b^2} + \rho \frac{R^2}{\beta^2} \right) - o(1). \label{eq:6}
\end{equation}
In combination, our considerations imply 
\begin{align*}
\frac{2\pi |\Lambda|}{\ln(R/a')} \min \{ \mathcal{A}_1, \mathcal{A}_2 \} &\geq \frac{2 \pi |\Lambda|}{\ln(R/a')} \min \{ 2 \rho^2, \rho_z^2 + 4 \rho_z \rho_{\omega} + 2 \rho_{\omega}^2 \} \asterisknum\\
&\quad - \text{const. } \frac{|\Lambda| \rho^2}{|\ln a^2 \rho |} \left( (R^2 \rho)^{1/3} + \frac{R}{s} + p_{\text{c}} R + \frac{R^2}{b^2} + \frac{R^2}{\beta^2 \rho} \right) - o(|\Lambda|). 
\end{align*}
Here, we can drop the terms $R^2/b^2$ and $R^2/(\beta^2 \rho)$ as they are dominated by $R/s$ and $(R^2 \rho)^{1/3}$, respectively. This follows from the assumptions $b > s > R$, $\beta \rho \gtrsim 1$ and $R^2 \rho \ll 1$. Using Lemma~\ref{lem-scattering-length} with the choice $\delta = \sqrt{\ln(R/a)/\varphi}$ as well as the definition of $a'$ in \eqref{eq:fillingholes6}, we estimate
\begin{equation}
\frac{1}{\ln(R/a')} \geq \frac{1}{\ln(R/a)} - \text{const. } \frac{1}{\ln(R/a)} \left( \frac{R}{s} + \kappa + \frac{1}{\sqrt{\varphi \ln(R/a)}} - \frac{R_0^2}{R^2} \ln(R/a) \right).
\end{equation}
We will choose $R^2 \rho \ll 1$ and, in particular,  $R^2 \rho \leq 1$, i.e.,
\begin{equation} \label{eqn-lnR/a-lb}
\frac{1}{\ln(R/a)} \geq \frac{2}{| \ln a^2 \rho |}.
\end{equation}
We thus finally arrive at
\begin{align*} \label{eqn-lb-A1A2}
&\frac{2 \pi |\Lambda|}{\ln(R/a')} \min \{ \mathcal{A}_1, \mathcal{A}_2 \} \geq \frac{4 \pi |\Lambda|}{|\ln a^2 \rho |} \min \{ 2 \rho^2, \rho_z^2 + 4 \rho_z \rho_{\omega} + 2 \rho_{\omega}^2 \} \\
&\quad - \text{const. } \frac{|\Lambda| \rho^2}{|\ln a^2 \rho |} \left( (R^2 \rho)^{1/3} + \frac{R}{s} + p_{\text{c}} R + \kappa + \frac{1}{\sqrt{\varphi |\ln a^2 \rho |}} + \frac{R_0^2}{R^2} |\ln a^2 \rho | \right). \asterisknum
\end{align*}
\subsection{A bound on the number of particles}
\label{sec:bound-particle-number}
In this section we give a lower bound on the terms involving the number operator and its square. More precisely, we consider the sum of the first term from \eqref{eq:interactionpartC7} and the term $\frac 1 2 \Tr_{\mathcal{F}}[ \mathbb{K} \Upsilon^z]$ from \eqref{eq:fillingholes17}. Recalling that we already chose $\epsilon = R/s$ and that $\mathbb{K}$ was defined in \eqref{eq:fockspace2a}, we seek a lower bound on the expression
\begin{align} \label{eqn-bd-n-1}
\mathcal{N} &= \left( \frac{8}{25 \ln(R/a') R^2} - \frac{288}{\ln(R/\tilde{a}) R s} (c + J(b/s)) \right)\Tr_{\mathcal{F}} [ \Nn ( \Upsilon_\pi^z - \Omega_b^z)] \nonumber \\
&\quad + \frac{2 \pi C}{|\Lambda| |\ln a^2 \rho |} \Tr_{\mathcal{F}} [(\Nn - N)^2 \Upsilon^z]. 
\end{align}
The fact that we need to give a bound for the first term on the right-hand side is one of the reasons for introducing the operator $\mathbb{K}$ in Subsection~\ref{sec:fockspace}.

Using the definition of $\Omega_b$ and $\Omega_{\pi}$ in \eqref{def:Omegapi}--\eqref{eqn-def-Omega-b-z} and the fact that they have the same density, we conclude
\begin{equation}
\Tr_{\mathcal{F}}[ \Nn ( \Upsilon^z_\pi - \Omega_b^z)] = \Tr_{\mathcal{F}_>} [\Nn^> (\Gamma^z - \Gamma_0)],
\end{equation}
where
\begin{equation}\label{def:Nup}
\Nn^> = \sum_{|p| \geq p_\mathrm{c}} a_p^\dagger a_p.
\end{equation}
For the quadratic term, we use the inequality
\begin{equation}
(\Nn - N)^2 \geq ( |z|^2 + \Tr_{\mathcal{F}_>}[\Nn^> \Gamma_0] - N)^2 + 2 (|z|^2 + \Tr_{\mathcal{F}_>}[\Nn^> \Gamma_0] - N)(\Nn - |z|^2 - \Tr_{\mathcal{F}_>}[\Nn^> \Gamma_0]).
\end{equation}
This implies
\begin{align*}
\Tr_{\mathcal{F}}[(\Nn - N)^2 \Upsilon^z] &\geq (|z|^2 + \Tr_{\mathcal{F}_>}[\Nn^> \Gamma_0] - N)^2\\
&\quad + 2(|z|^2 + \Tr_{\mathcal{F}_>}[\Nn^> \Gamma_0] - N) \Tr_{\mathcal{F}_>}[\Nn^>(\Gamma^z - \Gamma_0)]. \asterisknum
\end{align*}
Hence, we obtain the following expression as a lower bound
\begin{align*} \label{eqn-bd-n-2}
\mathcal{N} &\geq \frac{2 \pi C}{|\Lambda| |\ln a^2 \rho|} (|z|^2 + \Tr_{\mathcal{F}_>} [\Nn^> \Gamma_0] - N)^2\\
&\qquad + \Tr_{\mathcal{F}_>} [ \Nn^>(\Gamma^z - \Gamma_0)] \left[ \left( \frac{8}{25 \ln(R/a') R^2} - \frac{288}{\ln(R/\tilde{a}) Rs} (c + J(b/s)) \right) \right.\\
&\qquad \qquad \left.+ \frac{4 \pi C}{|\Lambda| |\ln a^2 \rho |} ( |z|^2 + \Tr_{\mathcal{F}_>}[ \Nn^> \Gamma_0] - N) \right]. \asterisknum
\end{align*}
We will choose the parameters $R$, $s$ and $C$ satisfying the conditions $C \ll 1/(R^2 \rho)$ and $R \ll s$ such that the term in square brackets on the right-hand side of \eqref{eqn-bd-n-2} is always positive (for any value of $|z|$) and therefore we need a lower bound on the expression $\Tr_{\mathcal{F}_>}[\Nn^>(\Gamma^z - \Gamma_0)]$. 

Let
\begin{equation}
\tilde f(\mu) = \frac 1 \beta \sum_{|p| \geq p_{\text{c}}} \ln \left( 1 - \e^{- \beta(p^2 - \mu_0 - \mu)} \right).
\end{equation}
Using the definition of the relative entropy in \eqref{eq:11} and the Gibbs variational principle for the ideal gas, we see that for any $\mu \leq 0$
\begin{equation}
S(\Gamma^z,\Gamma_0) - \beta \mu \Tr_{\mathcal{F}_>} [\Nn^> \Gamma^z] \geq \beta( \tilde f(\mu) - \tilde f(0)).
\end{equation}
From the absolute monotonicity
of $\tilde f$ (i.e., all derivatives being negative), 
we obtain
\begin{equation}
\tilde f(\mu) \geq \tilde f(0) + \mu \tilde f'(0) + \frac 1 2 \mu^2 \tilde f''(0).
\end{equation}
This implies
\begin{equation}
\Tr_{\mathcal{F}_>}[\Nn^> (\Gamma^z - \Gamma_0)] \geq - \frac{1}{\beta |\mu|} S(\Gamma^z,\Gamma_0) - \frac{\beta |\mu|}{4} \sum_{|p| \geq p_{\text{c}}} \frac{1}{\cosh(\beta(p^2 - \mu_0))-1}
\label{eq:3}
\end{equation}
as well as 
\begin{equation}
\Tr_{\mathcal{F}_>}[\Nn^>(\Gamma^z - \Gamma_0)] \geq - \left( S(\Gamma^z,\Gamma_0) \sum_{|p| \geq p_{\text{c}}} \frac{1}{\cosh(\beta(p^2 - \mu_0)) - 1} \right)^{1/2}
\end{equation}
when we optimize the right-hand side of \eqref{eq:3} over $\mu$.

We can use the a priori bound from \eqref{eq:relativeentropyandaprioribounds4} to bound the relative entropy, while for the sum over $p$ we use the bound $\cosh x - 1 \geq x^2/2$. Thus,
\begin{equation}
\sum_{|p| \geq p_{\text{c}}} \frac{1}{\cosh(\beta(p^2 - \mu_0))} \leq \frac{2}{\beta^2} \sum_{|p| \geq p_{\text{c}}} \frac{1}{(p^2 - \mu_0)^2} = \frac{|\Lambda|}{2 \beta^2 \pi^2} \int_{|p| \geq p_{\text{c}}} \frac{\de p}{(p^2 - \mu_0)^2} + o(|\Lambda|).
\end{equation}
The integral equals 
\begin{equation}
\int_{|p| \geq p_\text{c}} \frac{\de p}{(p^2 - \mu_0)^2} = \frac{\pi}{p_c^2 - \mu_0}.
\end{equation}
In conclusion, we have shown that
\begin{equation}
\Tr_{\mathcal{F}_>}[\Nn^> (\Gamma^z - \Gamma_0)] \geq - \left( \frac{4 |\Lambda|^2 \rho^2}{|\ln a^2 \rho | (\beta p_c^2 - \beta \mu_0)} \right)^{1/2} - o(|\Lambda|)
\end{equation}
holds. We now insert this into \eqref{eqn-bd-n-2} and obtain
\begin{equation} \label{eqn-bd-n-3}
\mathcal{N} \geq \frac{2 \pi C}{|\Lambda| |\ln a^2 \rho|} ( |z|^2 + \Tr_{\mathcal{F}_>} [\Nn^> \Gamma_0] - N)^2 - Z^{(3)} - o(|\Lambda|),
\end{equation}
where
\begin{equation} \label{eqn-def-Z3}
Z^{(3)} := \text{ const. } \frac{|\Lambda| \rho^2}{|\ln a^2 \rho|^{3/2} (\beta p_{\text{c}}^2 - \beta \mu_0)^{1/2}} \left[ |\ln a^2 \rho| \left( \frac{8}{25 \ln(R/a') R^2 \rho}  \right)
+  C \left( \frac{2}{\sqrt{C}} + \frac{\rho_{\omega}}{\rho}\right) \right]. 
\end{equation}
Note that we used \eqref{eq:relativeentropyandaprioribounds7} to bound $\rho_z$ as well as $|\Lambda|^{-1} \Tr_{\mathcal{F}_>}[\Nn^> \Gamma_0] \leq \rho_{\omega}$. Using also \eqref{bound:rhoo}, the assumption \eqref{eqn-lnR/a-ub} on $R$ and choosing  $C \ll 1/(R^2 \rho)$, this simplifies to
\begin{equation}
Z^{(3)} \lesssim \frac{|\Lambda| \rho^2}{|\ln a^2 \rho|} \frac{1}{(|\ln a^2 \rho | (\beta p_{\text{c}}^2 - \beta \mu_0))^{1/2} R^2 \rho}.
\end{equation}

\subsection{Relative entropy, effect of cutoff}
\label{sec:effectofcutoff}
In this section we quantify the effect of the cutoff parameter $b$ on the relative entropy $S(\Upsilon^z_\pi, \Omega^z_b)$ appearing in \eqref{eq:interactionpartC7}. The goal is to estimate $S(\Upsilon^z_\pi,\Omega_b^z)$ in terms of $S(\Pi \otimes \Gamma^z,\Omega_{\pi}) = S(\Gamma^z, \Gamma_0)$. For the latter expression we have the a priori bound \eqref{eq:relativeentropyandaprioribounds4}. To obtain such an estimate it will  be important that the vacuum state $\Pi_0$ has been replaced by the more general quasi-free state $\Pi$ in Subsection~\ref{sec:replcaingvacuum}. 

For any quasi-free state $\Omega_\omega$ with one-particle density matrix $\omega$ and any state $\Gamma$ it is easy to check that the relative entropy $S(\Gamma,\Omega_\omega)$ is convex in $\omega$. The one-particle density matrix of $\Omega_b$ is given by the following convex combination
\begin{equation}
\omega_b = \frac{1}{\vert \Lambda \vert} \sum_q \hat{\eta}_b(q) \frac{1}{2} \sum_p \left( \omega_{\pi}(p+q) + \omega_{\pi}(p-q) \right) \vert p \rangle\langle p \vert.
\label{eq:effectofcutoff1}
\end{equation}
Convexity of the map $\omega \mapsto S(\Gamma,\Omega_\omega)$ therefore implies
\begin{equation}
S\left( \Pi \otimes \Gamma^z, \Omega_b \right) \leq \frac{1}{| \Lambda |} \sum_q \hat{\eta}_b(q) S\left( \Pi \otimes \Gamma^z, \Omega_q \right),
\label{eq:effectofcutoff2}
\end{equation}
where $\Omega_q$ is the quasi-free state corresponding to the $1$-particle density matrix with eigenvalues $\frac{1}{2} \left( \omega_{\pi}(p+q) + \omega_{\pi}(p-q) \right)$. Further arguments based on convexity (see \cite[Eqs.~(5.15) and (5.16)]{Seiringer06B}) yield
\begin{align}
S(\Pi \otimes \Gamma^z, \Omega_q) &\leq \left( 1+t^{-1} \right) S(\Gamma^z, \Gamma_0) \nonumber \\
&\quad + \sum_p \left( h_q(p) - h_0(p) \right) \left( \frac{1}{\e^{h_0(p) + t(h_0(p) - h_q(p))} - 1} - \frac{1}{\e^{h_q(p)} - 1} \right) \label{eq:effectofcutoff3}
\end{align}
for any $t>0$. Here we defined
\begin{equation}
h_q(p) = \ln\left( \frac{2 + \omega_{\pi}(p+q) + \omega_{\pi}(p-q)}{\omega_{\pi}(p+q) + \omega_{\pi}(p-q)} \right).
\label{eq:effectofcutoff4}
\end{equation}
To estimate \eqref{eq:effectofcutoff3} from above, we require the following lemma. Since the proof of the analogous \cite[Lemma~6]{Seiringer2008} does not explicitly depend on the dimension of the configuration space it translates to the two-dimensional case without changes. We therefore omit the proof of Lemma~\ref{lem-ell}. 
\begin{lem} \label{lem-ell}
	Let $\ell : \mathbb{R}^2 \rightarrow \mathbb{R}_+$, and let $L_{\pm} = \pm \sup_p \sup_{| q | = 1} \pm (q\cdot  \nabla)^2 \ell(p)$ denote the supremum (infimum) of the largest (smallest) eigenvalue of the Hessian of $\ell$. Let $\omega_{\pi}(p) = [\e^{\ell(p)}-1]^{-1}$, and let $h_q(p)$ be given as in \eqref{eq:effectofcutoff4}. Then
	\begin{equation}
	h_q(p) - h_0(p) \leq L_{+} q^2,
	\label{eq:effectofcutoff5}
	\end{equation}  
	and 
	\begin{align} \label{eq:effectofcutoff6}
	&h_q(p) - h_0(p) \nonumber\\
	&\geq q^2 L_{-} + q^2 \min\{ L_{-}, 0 \} - 4 q^2 \sup_p [ | \nabla \ell(p) |^2 \omega_{\pi}(p) ] -2q^2 (| q | + | p | )^2 \sup_{p} [ |  \nabla \ell(p) |^2/p^2 ]. 
	\end{align}
\end{lem}
Recall that the $\ell(p)$ in question was defined in \eqref{eqn-def-l(p)}. Now we choose the parameters $\pi_p$ which determine $\ell(p)$ for $|p| < p_{\text{c}}$. For that purpose let $g : \Rr^2 \to [0,1]$ be a smooth radial function that is supported in a disk of radius one and assume that $g(p) \geq \frac 1 2$ for $|p| \leq \frac 1 2$. Then we set
\begin{equation} \label{eqn-choice-of-l}
\ell(p) = \beta(p^2 - \mu_0) + \beta p_{\text{c}}^2 g(p/p_{\text{c}}).
\end{equation}
This corresponds to the choice
\begin{equation} \label{eqn-choice-pi-p}
\pi_p = \frac{1}{\e^{\beta(p^2 - \mu_0) + \beta p_{\text{c}}^2 g(p/p_{\text{c}})} - 1}.
\end{equation}
Note that this choice indeed satisfies our earlier assumption on $\ell(p)$, which was $\ell(p) \geq \beta(p^2 - \mu_0)$. Furthermore, we can estimate $\pi_p \lesssim 1/(\beta (p_{\text{c}}^2 - \mu_0))$. This can be seen by considering $|p| \geq p_{\text{c}}/2$ and $|p| < p_{\text{c}}/2$ separately and using $\ell(p) \geq \beta(p^2 - \mu_0)$ in the first case and $g(p/p_{\text{c}}) \geq 1/2$ in the second case. Using this and $M \lesssim p_{\text{c}}^2 |\Lambda|$, we can bound $P$ from Subsection~\ref{sec:replcaingvacuum} as
\begin{equation} \label{eqn-P-estimate}
P = \sum_{|p| \leq p_{\text{c}}} \pi_p \lesssim \frac{M}{\beta (p_{\text{c}}^2 - \mu_0)} \lesssim \frac{|\Lambda| p_\text{c}^2}{\beta (p_\text{c}^2 - \mu_0)}.
\end{equation}
The bound on $P$ is needed for estimating  $Z^{(2)}$ in \eqref{eq:replacingvacuum4}.

For our choice of $\ell$ it is easy to see that both $L_+/\beta$ and $L_-/\beta$ are bounded independently of all parameters. We further have the bounds $|\nabla \ell(p)| \lesssim \beta |p|$ and $\omega_{\pi}(p) \leq \ell(p)^{-1} \leq (\beta p^2)^{-1}$, and together with Lemma~\ref{lem-ell}, this implies
\begin{equation}
- B \beta q^2 \left( 1 + \beta \left( |p| + |q| \right)^2 \right) \leq h_q(p) - h_0(p) \leq B \beta q^2
\label{eq:effectofcutoff7}
\end{equation}
for some $B>0$. Using $\sinh(x)/x \leq \cosh(x)$ for $x \in \Rr$, we estimate
\begin{align}
&\left( h_q(p) - h_0(p) \right) \left( \frac{1}{\e^{h_0(p) + t(h_0(p) - h_q(p))} - 1} - \frac{1}{\e^{h_q(p)} - 1} \right) \label{eq:effectofcutoff8} \nonumber\\
& \leq \frac{1}{2} (1+t) \left( h_q(p) - h_0(p) \right)^2 \frac{\e^{-h_q(p)} + \e^{-h_0(p) + t( h_q(p) - h_0(p) )}}{\left( 1-\e^{-h_0(p) + t( h_q(p) - h_0(p) )} \right)\left( 1 - \e^{-h_q(p)} \right)}.
\end{align}
We use
\begin{equation}
\left( h_q(p) - h_0(p) \right)^2 \leq B^2 \left( \beta q^2 \right)^2 \left( 1 + \beta \left( |p| + |q| \right)^2  \right)^2 \label{eq:effectofcutoff9}
\end{equation}
as well as the fact that the last fraction on the right-hand side of \eqref{eq:effectofcutoff8} is bounded from above by
\begin{equation}
\frac{\e^{-h_q(p)} + \e^{-h_0(p) + t \beta B q^2 }}{\left( 1-\e^{-h_0(p) + t \beta B q^2 } \right)\left( 1 - \e^{-h_q(p)} \right)} = \omega^t(p) + \frac{1}{2} \left( \omega_{\pi}(p+q) + \omega_{\pi}(p-q) \right) \left( 1+ 2 \omega^t(p) \right), \label{eq:effectofcutoff10} 
\end{equation} 
where $ \omega^t(p) = [ \e^{h_0(p) - B \beta t q^2} -1]^{-1} $. To obtain this result, we assumed that $t$ is small enough such that $h_0(p) - B \beta t q^2 > 0$ for all $p$. Since sums converge to integrals in the thermodynamic limit we need to bound
\begin{equation}
\int_{\mathbb{R}^2} \left( 1 + \beta \left( |p| + |q| \right)^2  \right)^2 \left( \omega^t(p) + \frac{1}{2} \left( \omega_{\pi}(p+q) + \omega_{\pi}(p-q) \right) \left( 1+ 2 \omega^t(p) \right) \right) \de p.
\label{eq:effectofcutoff11}
\end{equation}
We replace $\omega_{\pi}(p-q)$ by $\omega_{\pi}(p+q)$ without changing the value of the integral. Then we use $\omega_{\pi}(p) \leq \omega^t(p)$, change variables $p \rightarrow p-q$ and use Schwarz's inequality to see that \eqref{eq:effectofcutoff11} is bounded from above by
\begin{align*} \label{eq:effectofcutoff12}
\eqref{eq:effectofcutoff11} &\leq \int_{\Rr^2} \left(1+ \beta (|p| + |q|)^2 \right)^2 \left( \omega^t(p) + \omega^t(p+q) ( 1 + 2 \omega^t(p)) \right) \de p\\
&\leq 2 \int_{\Rr^2} \left( 1 + \beta (|p| + 2|q|)^2 \right)^2 \omega^t(p) \de p  + \left( \int_{\Rr^2} \left( 1 + \beta (|p| + |q|)^2 \right)^2 (\omega^t(p+q))^2 \de p \right)^{1/2}\\
&\qquad \quad \times \left( 4 \int_{\Rr^2} \left( 1 + \beta (|p| + |q|)^2 \right)^2 (\omega^t(p))^2 \de p\right)^{1/2}\\
&\leq 2 \int_{\mathbb{R}^2} \left( 1 + \beta \left( |p| + 2|q| \right)^2  \right)^2 \omega^t(p) \left( 1 + \omega^t(p) \right)  \de p. \asterisknum
\end{align*}
We choose $t = \min\{ 1, (b^2 q^2)^{-1} \}$, which implies $tq^2 \leq b^{-2}$. We also have
\begin{equation}
\ell(p) - B \beta t q^2 \geq \beta \left[ \frac{p^2}{2} - \mu_0 + p_{\text{c}}^2 \left( \frac 1 8  - \frac{B}{b^2 p_{\text{c}}^2} \right) \right] \geq \beta \left[ \frac{p^2}{2} - \mu_0 + \frac{p_c^2}{16} \right],
\end{equation}
which can be seen by considering, similarly to before when estimating $P$ in \eqref{eqn-P-estimate}, $|p| \geq p_{\text{c}}/2$ and $|p| < p_{\text{c}}/2$ separately. For the last inequality, we already assumed that $b$ and $p_\text{c}$ will be chosen in such a way that $b^2 p_\text{c}^2 \gg 1$ and, in particular, $B/(b^2 p_\text{c}^2) \leq 1/16$ holds. Denoting
\begin{equation} \label{eqn-def-tau}
\tau = - \beta \mu_0 + \frac{\beta p_{\text{c}}^2}{16},
\end{equation}
we thus have the bound 
\begin{equation}
\omega^t \leq \left( \e^{\tau + \beta p^2/2} - 1 \right)^{-1} \leq \e^{-\tau - \beta p^2/2} \left[ 1 + \frac{1}{\tau + \beta p^2/2} \right].
\label{eq:effectofcutoff13}
\end{equation}
Inserting \eqref{eq:effectofcutoff13} into \eqref{eq:effectofcutoff12}, we find
\begin{align*} \label{eq:effectofcutoff14}
\eqref{eq:effectofcutoff12} & \leq 2 \int_{\Rr^2} \left( 1 + \beta (|p| + 2 |q|)^2 \right)^2 \e^{- \tau - \beta p^2/2} \left[ 1 + \frac{1}{\tau + \beta p^2/2} \right]\\
&\qquad \times \left( 1+ \e^{-\tau - \beta p^2/2} \left[ 1 + \frac{1}{\tau + \beta p^2/2} \right] \right) \de p\\
&\lesssim \frac{\e^{- \tau}}{\beta} \left( 1 + \beta^2 q^4  \right) \int_{\mathbb{R}^2} \left( 1 + p^4 \right) \e^{- p^2/2} \left[ 1 + \frac{1}{(\tau + p^2/2)^2} \right] \de p\\
&\lesssim \frac{\e^{- \tau}}{\beta} \left( 1 + \beta^2 q^4  \right) \left( 1 + \tau^{-1} \right) . \asterisknum
\end{align*}
We combine the above equations and use $t^{-1} \leq 1 + b^2 q^2$ to see that
\begin{equation}
S(\Pi \otimes \Gamma^z, \Omega_q) \lesssim \left( 2 + b^2 q^2 \right) S(\Gamma^z, \Gamma_0) +  \frac{| \Lambda | }{\tau} \beta q^4 \left( 1 + \beta^2 q^4 \right) + o(| \Lambda |) \label{eq:effectofcutoff15}
\end{equation}
holds. Using \eqref{eq:effectofcutoff2} and $\eta_b(0) = 1$, we therefore have
\begin{equation}
S\left( \Pi \otimes \Gamma^z, \Omega_b \right) \lesssim S(\Gamma^z, \Gamma_0) +  \frac{\beta}{\tau} \sum_q \hat{\eta}_b(q) q^4  \left( 1 + \beta^2 q^4 \right) + o(| \Lambda |).
\label{eq:effectofcutoff16}
\end{equation}
We will choose $b$ such that $b^2 \gg \beta$ and this implies, in particular, that $\beta b^{-2} \lesssim 1$. We therefore have
\begin{equation}
S(\Pi \otimes \Gamma^z,\Omega_b) \lesssim S(\Gamma^z,\Gamma_0) +  \frac{\beta |\Lambda|}{\tau b^4} + o(|\Lambda|).
\label{eq:effectofcutoff17}
\end{equation}

The above inequality quantifies the effect of the cutoff. From \eqref{eq:interactionpartC7}, we know that we still have to multiply the relative entropy term by $b^2$. Using also the a priori bound from \eqref{eq:relativeentropyandaprioribounds4}, we obtain
\begin{align*}
b^2 S(\Upsilon_{\pi}^z, \Omega_b^z) &\lesssim b^2 \left( S(\Gamma^z, \Gamma_0) + \frac{\beta |\Lambda|}{\tau b^4} + o(|\Lambda|) \right) \lesssim \beta |\Lambda| \left( \frac{b^2 \rho^2}{|\ln a^2 \rho |} + \frac{1}{\tau b^2} + o(1) \right). \asterisknum
\end{align*}
From this expression it is easy to read off the optimal choice of $b$ which is given by 
\begin{equation}
b = \left( \frac{|\ln a^2 \rho |}{\tau \rho^2} \right)^{1/4}.
\end{equation}
The result of this subsection is therefore the following bound on the relative entropy
\begin{equation} \label{eqn-rel-entropy-cutoff}
b^2 S(\Upsilon_\pi^z, \Omega_b^z) \lesssim |\Lambda| \left( \frac{\beta \rho}{(\tau |\ln a^2 \rho |)^{1/2}} + o(1) \right).
\end{equation}

\subsection{Final lower bound}

In this section we collect the above estimates to give a lower bound on $F_z(\beta)$, which in turn will give  a lower bound on the free energy. 
Recall from Subsections~\ref{sec:fockspace} and~\ref{sec:coherentstates} that
\begin{equation} \label{eqn-final-lb-1}
-\beta^{-1} \ln \Tr_{\mathcal{H}_N} \e^{-\beta H_N}  \geq  \mu_0 N - \frac 1 \beta \ln \int_{\Cc^M} \e^{- \beta F_z(\beta)} \de z - Z^{(1)} 
\end{equation}
with $Z^{(1)}$  defined in \eqref{eq:coherentstates5}. We combine the estimates from \eqref{eq:fillingholes17}, \eqref{eq:interactionpartC7}, \eqref{eqn-lb-A1A2} as well as \eqref{eqn-bd-n-3} and \eqref{eqn-rel-entropy-cutoff} to obtain the final lower bound to $F_z(\beta)$, which reads
\begin{align*} \label{eqn-final-lb-2}
F_z(\beta) &\geq - \frac 1 \beta \ln \Tr_{\mathcal{F}_>} \left[ \e^{- \beta \mathbb{T}^c_s(z)} \right] - Z^{(2)} - Z^{(3)} - Z^{(4)} - o(|\Lambda|)\\
&\quad + \frac{2 \pi C}{ |\Lambda| | \ln a^2 \rho |} \left( |z|^2 + \Tr_{\mathcal{F}_>} [\Nn^> \Gamma_0] - N \right)^2 + \frac{4 \pi |\Lambda|}{|\ln a^2 \rho|} \min \left\{ \rho_z^2 + 4 \rho_z \rho_{\omega} + 2 \rho_{\omega}^2, 2 \rho^2 \right\}. \asterisknum
\end{align*}
Here, the error terms $Z^{(2)}$ and $Z^{(3)}$ are defined in \eqref{eq:replacingvacuum4} and \eqref{eqn-def-Z3}, respectively. The error term $Z^{(4)}$ contains the remaining errors and is defined by
\begin{align*}
Z^{(4)} &:= \text{const. } \frac{|\Lambda| \rho^2}{|\ln a^2 \rho |} \left( \frac{1}{R^4 \rho^2} \frac{(\beta \rho)^{1/2}}{\tau^{1/4} |\ln a^2 \rho |^{1/4}} + \frac{1}{R s \rho} J\left( \frac{|\ln a^2 \rho |^{1/4}}{\tau^{1/4} \rho^{1/2} s} \right) + \frac{R}{s} \right.\\
&\qquad + \left.(R^2 \rho)^{1/3} + p_{\text{c}} R + \kappa + \frac{1}{\sqrt{\varphi |\ln a^2 \rho|}} + \frac{R_0^2}{R^2} |\ln a^2 \rho | \right) + \text{const. } \frac{|\Lambda| p_{\text{c}}^2}{\beta} \frac{R_0^2}{R^2}. \asterisknum
\end{align*}
To obtain this form of the error term, we also used \eqref{eqn-lnR/a-ub} to replace the logarithmic factors $\ln(R/a)$ by the desired factor $|\ln a^2 \rho|$ and inserted the choices $\epsilon = R/s$ and $b = (|\ln a^2 \rho|/(\tau \rho^2))^{1/4}$  made earlier. The last term in $Z^{(4)}$ originates from the term $(\kappa-\kappa') \sum_p p^2 \pi_p$ in \eqref{eq:fillingholes17} using \eqref{eq:fillingholes8} and \eqref{eqn-P-estimate}. 

Let us have a closer look at the two terms in the second line of \eqref{eqn-final-lb-2}. We define
\begin{equation}
\rho^0 = \frac{1}{|\Lambda|} \Tr_{\mathcal{F}_>}[ \Nn^> \Gamma_0] = \rho_{\omega} - \frac{P}{|\Lambda|},
\end{equation}
where $P = \tr \pi = \sum_{|p| < p_\mathrm{c}} \pi_p$ was defined in Subsection~\ref{sec:replcaingvacuum}. Using $\rho^0 \leq \rho_{\omega}$, we replace $\rho_{\omega}$ in the second term in the second line of \eqref{eqn-final-lb-2} by $\rho^0$ for a lower bound. When we minimize over $\rho_z$ we find
\begin{align*} \label{eqn-final-lower-bound-min-z}
\frac{C}{2} \left( \rho_z - (\rho - \rho^0) \right)^2 + \rho_z^2 + 4 \rho_z \rho^0 + 2 (\rho^0)^2 \geq \frac{1}{1 + 2/C} \left( 2 \rho^2 - (\rho - \rho^0)^2 - \frac 4 C (\rho^0)^2 \right). \asterisknum 
\end{align*}
Note that the right-hand side of \eqref{eqn-final-lower-bound-min-z} is bounded from above by $2 \rho^2$. This implies in particular that the minimum in \eqref{eqn-final-lb-2} will be attained by the first term when we minimize over $\rho_z$. Therefore, we have the lower bound
\begin{align*}
F_z(\beta) &\geq - \frac 1 \beta \ln \Tr_{\mathcal{F}_>} \e^{- \beta \mathbb{T}^c_s(z)} + \frac{4 \pi |\Lambda|}{|\ln a^2 \rho|} \left( 2 \rho^2 - (\rho - \rho^0)^2 - \frac{4}{C} \rho^2 \right) - \sum_{i=2}^4 Z^{(i)} - o(|\Lambda|), \asterisknum
\end{align*}
where we used  
\begin{equation} \label{eqn-rho0-estimate}
\rho^0 =  \frac{1}{4\pi^2} \int_{|p| > p_c} \frac{\de p}{\e^{\beta(p^2 - \mu_0)} - 1} + o(1) \leq \rho(1 + o(1))
\end{equation}
 in the $1/C$ correction term. 
The only remaining $z$ dependence is then in the first term
\begin{equation}
- \frac 1 \beta \ln \Tr_{\mathcal{F}_>} \e^{- \beta \mathbb{T}^{\text{c}}_s(z)} = \sum_{|p| < p_{\text{c}}} \epsilon(p) |z_p|^2 + \frac 1 \beta \sum_{|p| \geq p_{\text{c}}} \ln \left( 1 - \e^{-\beta \epsilon(p)} \right)\,,
\end{equation}
where $\epsilon(p)$ was defined in \eqref{eq:fillingholes12} as $\epsilon(p) = \kappa' p^2 + (1 - \kappa) p^2( 1 - \chi(p)^2) - \mu_0$, with $\chi$  a cutoff function at the scale $s \geq R$. We  evaluate the integral over $\Cc^M$ in \eqref{eqn-final-lb-1} to give
\begin{equation}
\int_{\Cc^M} \e^{- \beta \sum_{|p| < p_{\text{c}}} \epsilon(p) |z_p|^2} \de z 
= \prod_{|p| < p_{\text{c}}} \frac{1}{\beta \epsilon(p)}.
\end{equation}
Now we estimate the term that contributes to the free part of the free energy. Using the fact that $x \geq 1 - \e^{-x}$ for $x \geq 0$, we find
\begin{align*}
&\frac{1}{\beta |\Lambda|} \sum_{|p| < p_{\text{c}}} \ln(\beta \epsilon(p)) + \frac{1}{\beta |\Lambda|} \sum_{|p| \geq p_{\text{c}}} \ln \left(1 - \e^{-\beta \epsilon(p)} \right)\\
&\quad \geq \frac{1}{\beta |\Lambda|} \sum_{p} \ln \left(1 - \e^{-\beta \epsilon(p)} \right) \geq \frac{1}{4 \beta \pi^2} \int_{\Rr^2} \ln \left(1 - \e^{-\beta \epsilon(p)}\right) \de p - o(1). \asterisknum
\end{align*}
We split the integral into the two parts $|p| \leq s^{-1}$ and $|p| \geq s^{-1}$. In the first part we have $\epsilon(p) = (1 - \kappa + \kappa') p^2 - \mu_0$, while in the second part we have the bound $\epsilon(p) \geq \kappa' p^2$. Hence,
\begin{align*} \label{eqn-final-lb-exp-small}
&\int_{\Rr^2} \ln \left( 1 - \e^{-\beta \epsilon(p)} \right) \de p\\
&\quad \geq \frac{1}{1 - \kappa + \kappa'} \int_{\Rr^2} \ln \left( 1 - \e^{- \beta(p^2 - \mu_0)} \right) \de p + \frac{1}{\kappa' \beta} \int_{|p|^2 \geq \kappa' \beta / s^2} \ln \left( 1 - \e^{- p^2} \right) \de p. \asterisknum
\end{align*}
The parameter $s$ will be chosen such that $s^2 \ll \kappa' \beta$; the second integral is then exponentially small in the parameter $s^2 / (\kappa' \beta)$.

Define
\begin{equation}
\rho_\text{s} := \rho \left[ 1 - \frac{\ln |\ln a^2 \rho |}{4 \pi \beta \rho} \right]_+.
\end{equation}
Our goal is to bound $\rho - \rho^0$ by $\rho_\text{s}$ plus an error term. This will be achieved by introducing a new parameter $\tilde p_{\text{c}}$ that satisfies
\begin{equation}
\frac{1}{4\pi^2} \int_{|p| \leq \tilde p_{\text{c}}} \frac{\de p}{\e^{\beta(p^2 - \mu_0)} - 1} = \rho_\text{s}.
\end{equation}
By an explicit computation, we find
\begin{equation} \label{eqn-def-tilde-pc}
\beta \tilde p_{\text{c}}^2 = \frac{1}{\e^{4 \pi \beta \rho} - 1} \left[ \frac{\e^{4\pi \beta \rho}}{|\ln a^2 \rho |} - 1 \right]_+.
\end{equation}
We remark that $p_\text{c}$ will be chosen such that $p_\text{c} \geq \tilde p_\text{c}$ holds, and we use  \eqref{eqn-rho0-estimate} to write
\begin{equation}
\rho - \rho^0 = \rho_\text{s} + \frac{1}{4 \pi^2} \int_{\tilde p_\text{c} \leq |p| \leq p_\text{c}} \frac{\de p}{\e^{\beta(p^2 - \mu_0)} - 1} + o(1).
\end{equation}
The remaining correction term can be estimated as
\begin{equation} \label{eqn-pc-pc-tilde-corr}
\frac{1}{4 \pi^2} \int_{\tilde p_\text{c} \leq |p| \leq p_\text{c}} \frac{\de p}{\e^{\beta(p^2 - \mu_0)} - 1} 
\leq
\frac{1}{4 \pi^2 \beta} \int_{\tilde p_\text{c} \leq |p| \leq p_\text{c}} \frac{\de p}{ p^2 -\mu_0 }
  = \frac{1} {4\pi \beta} \ln \left( \frac{ p_\text{c}^2 -  \mu_0}{ \tilde p_\text{c}^2 -  \mu_0} \right).
\end{equation}
In combination, the above estimates show that
\begin{align*} \label{eqn-final-lb-result}
-\frac 1{\beta |\Lambda|}  \ln \Tr_{\mathcal{H}_N} \e^{-\beta H_N}
&\geq \mu_0 \rho + \frac{1}{4 \beta \pi^2} \int_{\Rr^2} \ln \left( 1 - \e^{-\beta(p^2 - \mu_0)} \right) \de p - \frac{1}{|\Lambda|} \sum_{i=1}^5 Z^{(i)} - o(1)\\
&\qquad + \frac{4 \pi}{|\ln a^2 \rho |} \left( 2 \rho^2 - \rho_\text{s}^2 \right), \asterisknum
\end{align*}
where
\begin{align*} \label{eqn-def-Z5}
Z^{(5)} &:= \, \text{const.} \, (\kappa - \kappa') \frac{|\Lambda|}{\beta} \int_{\Rr^2} \ln \left( 1 - \e^{- \beta(p^2 - \mu_0)} \right) \de p - \frac{|\Lambda|}{\kappa' \beta^2} \int_{|p|^2 \geq \kappa' \beta/s^2} \ln \left( 1 - \e^{-p^2} \right) \de p\\
&\qquad + \frac{\text{const.} \, |\Lambda| \rho^2}{|\ln a^2 \rho |} \left[ \frac{1}{C} + \frac{1}{\beta \rho} \ln \left( \frac{ p_{\text{c}}^2 -  \mu_0}{ \tilde p_{\text{c}}^2 -  \mu_0} \right) + \frac{1}{(\beta \rho)^2} \ln^2 \left( \frac{ p_{\text{c}}^2 -  \mu_0}{ \tilde p_{\text{c}}^2 -  \mu_0} \right) \right]. \asterisknum
\end{align*}
Note that the right-hand side of \eqref{eqn-final-lb-result} has the desired form: The sum of the first two terms on the right-hand side equals $f_0(\beta,\rho)$, the free energy of non-interacting bosons, since $\mu_0$ is given by \eqref{eqn-chemical-potential-ideal-gas}. The last term in \eqref{eqn-final-lb-result} is the desired interaction energy.   It remains to choose the parameters in the error terms and show that they are of lower order than this interaction energy.

\subsection{Minimizing the error terms}

In this section we show how to choose the parameters in order to optimize the error terms of the lower bound.

To simplify the notation, we replace the factor $1/16$ in the definition of $\tau$ from \eqref{eqn-def-tau} by one, i.e., we redefine
\begin{equation}\label{eq:8}
\tau = - \beta \mu_0 + \beta p_{\text{c}}^2 \quad \text{ and denote } \quad \tilde \tau = - \beta \mu_0 + \beta \tilde p_{\text{c}}^2. 
\end{equation}
For brevity, let us also introduce the notation
\begin{equation}
\sigma := |\ln a^2 \rho|.
\end{equation}
Similarly as in the three-dimensional case the following terms are relevant for the minimization: $p_\mathrm{c}^4$ from $Z^{(1)}$, $ \rho^2 \sigma^{-1} ( \kappa + R/s )$ and $ \rho^2 \sigma^{-1} (\beta \rho)^{1/2}(R^2 \rho)^{-2} (\tau \sigma)^{-1/4} $ from $Z^{(4)}$ as well as
\begin{equation}
	- \frac{1}{\kappa' \beta^2} \int_{|p|^2 \geq \kappa' \beta/s^2} \ln \left( 1 - \e^{- p^2 } \right) \de p \label{eq:4}
\end{equation}
from $Z^{(5)}$. In turns out, however, that in the two-dimensional case the additional error terms $ \rho^2 \sigma^{-1} (R^2 \rho)^{1/3}$ from $Z^{(4)} $ and $ \rho^2 \sigma^{-1} \ln(\tau/\tilde \tau)/(\beta \rho) $ from $Z^{(5)}$ are also relevant for choosing the parameters. The constraints on the parameters, that is, $p_{\text{c}} \leq 1/s$, $s \gg R$, $s^2 \ll \kappa \beta$, $R_0^2/R^2 \ll \kappa$, $b \gg 1/p_{\text{c}}$, $b \gg R$ and $b \gg \beta^{1/2}$ will be automatically satisfied with the choice of the parameters below. The same is true for \eqref{eqn-lnR/a-ub} and \eqref{eqn-lnR/a-lb}, which have to be obeyed by the parameter $R$. Since $R$ appears in these expression only in the argument of a logarithm, we still have quite some freedom in its choice.

In order for \eqref{eq:4} to be small, we require that $s^2 \ll \kappa' \beta$, with $\kappa'$ defined in \eqref{eq:fillingholes8}. This is equivalent to $s^2 \ll \kappa \beta$, since we will choose $R_0^2/R^2 \ll \kappa$. It we take  $\kappa' = (1+\delta)s^2 \beta^{-1}\ln \sigma$ for some   $\delta > 0$, \eqref{eq:4} is bounded by  
 $(s^2\beta)^{-1} \sigma^{-1-\delta}$, which will be negligible compared to the other terms. We can now optimize the term $|\Lambda| \rho^2 \sigma^{-1} ( \kappa + R/s )$ over $s$ resulting in the choice 
\begin{equation}
s = \left( \frac{\beta R}{\ln \sigma} \right)^{1/3}.
\end{equation}
With this choice of $s$ the error term becomes
\begin{equation} \label{eqn-minimization-s}
\frac{|\Lambda| \rho^2}{\sigma} \left( (1+\delta)\frac{s^2 \ln \sigma}{\beta} + \frac{R}{s} \right) \sim \frac{|\Lambda| \rho^2}{\sigma} \left( \frac{R^2 \ln \sigma}{\beta} \right)^{1/3}.
\end{equation}
Among the main terms there are now only three terms left that depend on $R$, namely  \eqref{eqn-minimization-s}, $|\Lambda| \rho^2 \sigma^{-1} (R^2 \rho)^{1/3}$ and $|\Lambda| \rho^2 \sigma^{-1} (\beta \rho)^{1/2}(R^2 \rho)^{-2} (\tau \sigma)^{-1/4}$. Denoting
\begin{equation}\label{def:d}
d = 1 + \left( \frac{\ln \sigma}{\beta \rho} \right)^{1/3} \sim 1 + \left( \frac{\beta_{\mathrm{c}}}{\beta} \right)^{1/3},
\end{equation}
we write the sum of the first two terms as $|\Lambda| \rho^2 \sigma^{-1} (R^2 \rho)^{1/3} d$. Hence, the optimal choice of $R$ is
\begin{equation}
(R^2 \rho)^{1/3} = \frac{(\beta \rho)^{1/14}}{d^{1/7} (\tau \sigma)^{1/28}}
\end{equation}
and the resulting error term reads
\begin{equation}
\frac{ \rho^2}{\sigma} (R^2 \rho)^{1/3} d = \frac{ \rho^2}{\sigma} d^{6/7} \left( \frac{(\beta \rho)^2}{\tau \sigma} \right)^{1/28}.
\end{equation}

We are thus left with the following three error terms
\begin{align*}
A_1 &= \frac{|\Lambda| \rho^2}{\sigma} \frac{1}{\beta \rho} \ln \left( \frac{\tau}{\tilde \tau} \right) = \frac{|\Lambda| \rho^2}{\sigma} \frac{1}{\beta \rho} \ln \left( \frac{\beta p_\text{c}^2 - \ln ( 1 - \e^{- 4 \pi \beta \rho} )}{\beta \tilde p_\text{c}^2 - \ln ( 1 - \e^{- 4 \pi \beta \rho} )} \right),\\
A_2 &= |\Lambda| p_{\text{c}}^4,\\
A_3 &= \frac{|\Lambda| \rho^2}{\sigma} d^{6/7} \left( \frac{(\beta \rho)^2}{\tau \sigma} \right)^{1/28} = \frac{|\Lambda| \rho^2}{\sigma} \left( 1 + \left( \frac{\beta_{\mathrm{c}}}{\beta} \right)^{1/3} \right)^{6/7} \left( \frac{(\beta \rho)^2}{(\beta p_c^2 - \ln ( 1 - \e^{- 4 \pi \beta \rho} )) \sigma} \right)^{1/28}. \asterisknum
\end{align*}
They depend solely on $p_\text{c}$, $\beta \rho$ and $\sigma$, as $\tilde p_\text{c}$ is given explicitly in \eqref{eqn-def-tilde-pc}. By minimizing over $p_\text{c}$ we therefore obtain the final error rate $\min_{p_\text{c}} \{ A_1 + A_2 + A_3 \}$, which depends only on $\beta \rho$ and $\sigma$. Optimization  turns out  to lead to the choice
\begin{equation} \label{eqn-choice-of-pc}
\beta p_{\text{c}}^2 = \begin{cases}
0 & \text{if } 1 \lesssim 4 \pi \beta \rho \leq \ln \left( \frac{\sigma}{(\ln \sigma)^{30}} \right),\\
\frac{(\beta \rho)^{30}}{\sigma \ln^{28}((\beta \rho)^{30} /(\sigma \tilde \tau)) } & \text{if } \ln \left( \frac{\sigma}{(\ln \sigma)^{30}} \right) \leq 4 \pi \beta \rho \lesssim \sigma^{1/59},\\
\left( \frac{(\beta \rho)^2}{\sigma} \right)^{29/57} & \text{if } \sigma^{1/59} \lesssim \beta \rho \lesssim \sigma^{1/2}.
\end{cases}
\end{equation}
The upper limit $\beta \rho \lesssim \sigma^{1/2}$  is a natural restriction, since the interaction term is comparable to the non-interacting free energy if $\beta \rho \sim \sigma^{1/2}$ (compare with \eqref{eqn-freegas-asymptotics}), and hence the perturbative argument, on which the proof of the lower bound is based, cannot be expected to work anymore in this regime. 
For $\beta \rho$ of the order $\sigma^{1/2}$ or larger, an additional argument using the result at $T=0$ \cite{LY2001} as a crucial ingredient will be given in Subsection~\ref{sec:uniformityinT} to complete the proof of the lower bound.

The parameters $\varphi$ and $C$ in the remaining error terms (which we did not need to consider for the choice of $p_\text{c}$) may be chosen according to
\begin{equation}
\frac 1 \sigma \ll \varphi \ll \frac{\beta \rho}{\sigma}, \quad 1 \ll C \ll \sigma
\end{equation}
if $\beta \rho$ is such that $p_\text{c} \neq 0$. In case $\beta \rho$ is so small that $p_\text{c} = 0$, we find that the upper restrictions to $\varphi$ and $C$ do not apply anymore and their choice only needs to satisfy the lower ones. 

We now explain how to arrive at the choice \eqref{eqn-choice-of-pc} of $p_\text{c}$. We start by  discussing what can be expected. For $\beta \rho$ far below $\beta_{\mathrm{c}} \rho$, in a sense to be made precise below, we have that the (absolute value of the) chemical potential $-\beta \mu_0$ is large enough compared to $\sigma^{-1}$ to control the term $A_3$ and even allows for the choice $p_{\text{c}} = 0$, which means that $A_1$ and $A_2$ both vanish. This changes when $\beta \rho$ comes close to $\beta_{\mathrm{c}} \rho$, where we need that $\beta p_{\text{c}}^2$ is larger than $\sigma^{-1}$. Here, only $A_1$ and $A_3$ have to be considered for the optimization, while $A_2$ is subleading. For $\beta \rho$ far above $\beta_{\mathrm{c}} \rho$, the optimal error rate changes as the term $A_1$ becomes irrelevant and we optimize using the terms $A_2$ and $A_3$.

Consider first the case $p_\text{c} = 0$, which means $\tilde p_\text{c} = 0$ by the assumption $p_\mathrm{c} \geq \tilde p_\mathrm{c}$, which also means $\e^{4 \pi \beta \rho} \leq \sigma$ or $\beta\leq \beta_\mathrm{c}$. This implies $A_1 = A_2 = 0$ as well as $\tau = - \beta \mu_0 = - \ln(1 - \e^{-4 \pi \beta \rho})$. The remaining error term is given by
\begin{equation} \label{eqn-error-pc=0}
A_3 \lesssim \frac{|\Lambda| \rho^2}{\sigma}  \left( \frac{\beta_{\mathrm{c}}}{\beta} \right)^{2/7} \left( \frac{(\beta \rho)^2}{\sigma \e^{- 4\pi \beta \rho}} \right)^{1/28}\,.
\end{equation}
It can be read off that $\e^{4 \pi \beta \rho} \lesssim \sigma/(\ln \sigma)^2$ is the upper limit for this error to be smaller than the interaction scale, which is  much smaller than the value of that function at the inverse critical temperature, $\e^{4\pi \beta_{\mathrm{c}} \rho} =  \sigma$. Hence, we need to choose a non-zero $p_\text{c}$ already well above the critical temperature.

Next, we consider the case $p_c \neq 0$. This will be the case only in the regime $\beta \gtrsim \beta_c$,   hence $d$ in \eqref{def:d} satisfies $d\sim 1$.  Since we have three main error terms to consider, there are three different possibilities of how to obtain the optimal $p_\text{c}$, out of which only two will be relevant. The first way of choosing $p_\text{c}$ is obtained by optimizing $A_1$ and $A_3$.   This leads to the equation
\begin{equation}
\frac{1}{\beta \rho} \ln \left( \frac{\tau}{\tilde \tau} \right) = \left( \frac{(\beta \rho)^2}{\sigma \tau} \right)^{1/28},
\end{equation}
which, to leading order, is solved by
\begin{equation} \label{eqn-def-pc-intermediate}
\tau = \beta p_\text{c}^2 - \beta \mu_0 = \frac{(\beta \rho)^{30}}{\sigma \ln^{28} \left( \frac{(\beta \rho)^{30}}{\sigma \tilde \tau} \right)}.
\end{equation}
As mentioned before, the reason for switching to $p_\mathrm{c} \neq 0$ is that $- \beta \mu_0$ becomes too small in order to control the term $A_3$ (i.e., to ensure that $A_3$ is smaller than the interaction scale $|\Lambda| \rho^2 / \sigma$). Therefore, we can take the right-hand side of \eqref{eqn-def-pc-intermediate} as the defining equation for $\beta p_\text{c}^2$ and neglect the term $-\beta \mu_0$. The error terms with this choice of $p_\text{c}$ become
\begin{align*} \label{eqn-error-beta-rho-intermediate}
A_1 \sim A_3 &\lesssim \frac{|\Lambda| \rho^2}{\sigma} \frac{1}{\beta \rho} \ln \left( \frac{(\beta \rho)^{30}}{\sigma \tilde \tau \ln^{28}((\beta \rho)^{30} /(\sigma \tilde \tau))} \right),\\
A_2 &\lesssim \frac{|\Lambda| \rho^2}{\sigma} \frac{(\beta \rho)^{58}}{\sigma \ln^{56}((\beta \rho)^{30}/(\sigma \tilde \tau))}. \asterisknum
\end{align*}
Note that $A_3 = A_1$ to leading order by our choice of $p_\text{c}$ and that $A_2$ is indeed of lower order than $A_1$ or $A_3$ for $\beta \rho \sim \beta_{\mathrm{c}} \rho$. 

Now we can compare the term $A_1$ from \eqref{eqn-error-beta-rho-intermediate} to the term $A_3$ we obtained by choosing $p_\text{c} = 0$ (from \eqref{eqn-error-pc=0}) to determine the point at which we switch to $p_\text{c} \neq 0$ as given in \eqref{eqn-def-pc-intermediate}. This gives
\begin{equation}
\left( \frac{(\beta\rho)^2}{\sigma \e^{-4\pi \beta \rho}} \right)^{1/28} = \frac{1}{\beta \rho} \ln \left( \frac{(\beta \rho)^{30}}{\sigma \tilde \tau \ln^{28}((\beta \rho)^{30}/(\sigma \tilde \tau))} \right),
\end{equation}
which we solve to leading order by
\begin{equation}
4 \pi \beta \rho = \ln \left( \frac{\sigma}{(\ln \sigma)^{30}} \right).
\end{equation}
For this value of $\beta \rho$ we switch to $p_\text{c}$ as given in \eqref{eqn-def-pc-intermediate}. 

It is clear, however, that for larger $\beta \rho$ the term $A_2$ from \eqref{eqn-error-beta-rho-intermediate} will become larger than $A_1$ or $A_3$ as it is increasing in $\beta \rho$. The point at which this happens is given by the solution of the equation
\begin{equation}
\frac{1}{\beta \rho} \ln \left( \frac{(\beta \rho)^{30}}{\ln^{28} (\beta \rho)^{30}} \right) = \frac{(\beta \rho)^{58}}{\sigma \ln^{56} (\beta \rho)^{30}}.
\end{equation}
To leading order we solve it by $\beta \rho = \sigma^{1/59}$. From here on, we use the second way of optimizing $p_\text{c}$ by considering the terms $A_2$ and $A_3$ with the result
\begin{equation}
\beta p_\text{c}^2 = \left( \frac{(\beta \rho)^2}{\sigma} \right)^{29/57}\,.
\end{equation}
The error terms then become
\begin{align*} \label{eqn-error-beta-rho-large}
A_1 &\lesssim \frac{|\Lambda| \rho^2}{\sigma} \frac{1}{\beta \rho} \ln \left( (\beta \rho)^{58/57} \sigma^{28/57} \right),\\
A_2 &\lesssim \frac{|\Lambda| \rho^2}{\sigma} \left( \frac{(\beta \rho)^2}{\sigma} \right)^{1/57}. \asterisknum
\end{align*}
Note that from this form of $A_2$ we can also read off the natural upper limit $\beta \rho \ll \sigma^{1/2}$ for the error terms to be small.

\subsection{Uniformity in the temperature}
\label{sec:uniformityinT}

For $\beta \rho$ of the order $\sigma^{1/2}$ or larger we apply a technique that uses in an essential way the result for the ground state energy \cite{LY2001}. This will allow us to obtain the desired uniformity in $\beta \rho$, as already mentioned in the previous subsection.

Starting from the original Hamiltonian with potential $v$ (which we denoted by $H_N$), we use Lemma~\ref{lem-Dyson-lemma} to obtain
\begin{align*} \label{eqn-gse-argument-dyson}
H_N &\geq \sum_{j=1}^N \Biggl[ - \nabla_j \left(1 - (1-\kappa) \chi(p_j)^2 \right) \nabla_j + (1-\epsilon)(1-\kappa) U_R(d(x_j,x_{\text{NN}}^{J_j}(x_j)))\\
&\qquad - \frac{1}{\epsilon} \int_{\Rr_+} U_R(t) t \de t \sum_{i \in J_j} w_R(x_j- x_i) \Biggr]. \asterisknum
\end{align*}
Strictly speaking we should work with a symmetrization of the right-hand side of \eqref{eqn-gse-argument-dyson} since the potential that we obtained from Lemma~\ref{lem-Dyson-lemma} is not permutation symmetric. As already mentioned  before, this does not need to concern us since we only consider expectation values in bosonic states. The last term  in \eqref{eqn-gse-argument-dyson} can be estimated using the integral condition on $U_R$ (from \eqref{eqn-U_R-cond}), the decay property of $g$ (which was introduced in \eqref{eq:interactionpartB3}) as well as the definition of $J_j$:
\begin{align*}
&\sum_{j=1}^N \frac{1}{\epsilon} \int_{\Rr_+} U_R(t) t \de t \sum_{i \in J_j} w_R(x_j - x_i)\\
&\quad \leq \frac{1}{\epsilon \ln(R/a)} \sum_{j=1}^N \sum_{i\in J_j} \frac{R^2}{s^4} g(d(x_i,x_j)/s) \lesssim \frac{N}{\epsilon \ln(R/a) s^2}. \asterisknum
\end{align*}

To find a lower bound for the remaining terms, we use the main result from \cite{LY2001} (for the choice $\kappa = \sigma^{-1/5}, R \rho^{1/2} = \sigma^{-1/10}$) and find
\begin{equation}
\sum_{j=1}^N \left( - \frac{\kappa}{2} \Delta_j + (1 - \epsilon)(1-\kappa) U_R(d(x_j,x_{\text{NN}}^{J_j}(x_j))) \right) \geq \frac{4 \pi N \rho}{\sigma} \left( 1 - \epsilon - \frac{\text{const}}{\sigma^{1/5}} \right).
\end{equation}
Even though the result in \cite{LY2001} was for Neumann boundary conditions and the full nearest-neighbor interaction, it is straight-forward to check that it also holds in our case. The ground state of the non-interacting system for periodic boundary conditions is also a constant,  and  the difference between the nearest-neighbor interaction in that paper and our interaction can be bounded by a constant times $N^2 (R^2/L^2)^2 \| U_R \|_\infty$. A term like this is already contained in the original estimate in \cite[Eqs.~(3.18) and (3.19)]{LY2001}. In \cite{LY2001}  the potential $U_R(d(x_j,x_\text{NN}(x_j)))$ is used, where the nearest neighbor was determined among all other particles while here we only look for the nearest neighbor in the set $J_j$. The related error can be controlled with an estimate for the probability of finding a particle coordinate that is not contained in the set $J_j$. It is straightforward to check that this probability is bounded by a constant times $N^2 (R^2/L^2)^2$ times the $L^\infty$ norm of the potential~$U_R$.

The above considerations allow us to show that
\begin{equation}\label{eq:7}
H_N \geq \sum_{j=1}^N \ell \left(\sqrt{-\Delta_j} \right) + \frac{4 \pi N \rho}{\sigma} \left( 1 - \epsilon - \frac{\text{const.}}{\sigma^{1/5}} - \frac{\text{const.}}{\epsilon s^2 \rho} \right),
\end{equation}
where $\ell(p) = p^2( 1 - \sigma^{-1/5}/2 - (1-\sigma^{-1/5}) \chi(p)^2)$. We already inserted the choice $\kappa = \sigma^{-1/5}$ from above. Next, we consider the free energy related to $H_N$, introduce the chemical potential $\mu_0$ and drop the restriction on the particle number. When we also take the thermodynamic limit we find
\begin{align}
f(\beta,\rho) &\geq f_0(\beta,\rho) + \text{const. } \frac{1}{\beta \sigma^{1/5}} \int_{\Rr^2} \ln \left( 1 - \e^{-\beta(p^2 - \mu_0)} \right) \de p \label{eq:5} \\
&\quad + \frac{1}{\beta^2 \sigma^{1/5}} \int_{p^2 \geq \beta/(s^2 \sigma^{1/5})} \ln \left( 1 - \e^{-p^2/2} \right) \de p + \frac{4 \pi \rho^2}{\sigma} \left( 1 - \epsilon - \frac{\text{const. }}{\sigma^{1/5}} - \frac{\text{const. }}{\epsilon s^2 \rho} \right). \nonumber
\end{align}
As before, we require $s^2 \sigma^{1/5}/\beta \ll 1$ for the correction term to the non-interacting free energy to be small. If we choose
\begin{equation}
\frac{s^2}{\beta} = \frac{1}{2 \delta \sigma^{1/5}  \ln \sigma }
\end{equation}
for some $\delta>0$ this error term is bounded from above by a constant times $\beta^{-2} \sigma^{-1/5 - \delta}$ and will be negligible compared to other terms. Optimization over $\epsilon$ yields
\begin{equation}
\epsilon = \sqrt{\frac{1}{s^2 \rho}}.
\end{equation}
Therefore, we have
\begin{align*} \label{eqn-gse-argument-error}
f(\beta,\rho) &\geq f_0(\beta,\rho) + \frac{4 \pi \rho^2}{\sigma} \left( 1 - \text{const.} \left[ \frac{\sigma^{4/5}}{(\beta \rho)^2} + \frac{1}{\sigma^{1/5}} + \frac{\sigma^{1/10} ( \ln \sigma)^{1/2}}{(\beta \rho)^{1/2}} \right] \right). \asterisknum
\end{align*}
It remains to estimate the term depending on the critical temperature as
\begin{equation}
\frac{4 \pi \rho^2}{\sigma} \left( 1 - \left[ 1 - \frac{\beta_{\mathrm{c}}}{\beta} \right]_+^2 \right) \lesssim \frac{\rho^2}{\sigma} \frac{\beta_{\mathrm{c}}}{\beta}.
\end{equation}
Hence the total error to consider is bounded from above by a constant times
\begin{equation} \label{eqn-error-gse-regime}
\frac{\rho^2}{\sigma} \left( \frac{\sigma^{4/5}}{(\beta \rho)^2} + \frac{1}{\sigma^{1/5}} + \frac{\ln \sigma}{\beta \rho} + \frac{\sigma^{1/10} ( \ln \sigma)^{1/2}}{(\beta \rho)^{1/2}}\right).
\end{equation}
The optimal point at which we switch from the error given in \eqref{eqn-error-beta-rho-large} to this error is determined by comparing the term $A_2$ with the first term in \eqref{eqn-error-gse-regime}. This leads to the equation
\begin{equation}
\frac{\sigma^{4/5}}{(\beta \rho)^2} = \left( \frac{(\beta \rho)^2}{\sigma} \right)^{1/57},
\end{equation}
which is solved by $\beta \rho = \sigma^{233/580}$. If $\beta \rho$ is larger than or equal to this value we use the result derived in this section. 

In conclusion, by combining the results from the previous estimates in \eqref{eqn-error-pc=0}, \eqref{eqn-error-beta-rho-intermediate}, \eqref{eqn-error-beta-rho-large} and \eqref{eqn-error-gse-regime}, we have shown that the bound
\begin{equation} \label{eqn-statement-lower-bound}
f(\beta, \rho) \geq f_0(\beta,\rho) + \frac{4 \pi \rho^2}{\sigma} \left( 2 - \left[ 1 - \frac{\beta_{\mathrm{c}}}{\beta} \right]_+^2 \right) ( 1 - o(1))
\end{equation}
holds uniformly in $\beta \rho \gtrsim 1$, where
\begin{equation} \label{eqn-total-error}
o(1) \lesssim \begin{cases}
 \left( \frac{\ln \sigma}{\beta\rho} \right)^{2/7}  \left( \frac{(\beta \rho)^2}{- \sigma \ln(1 - \e^{- 4\pi \beta \rho})} \right)^{1/28} & \text{if } 1 \lesssim 4 \pi \beta \rho \leq \ln(\sigma/(\ln \sigma)^{30}),\\
\frac{1}{\beta \rho} \ln \left( \frac{(\beta \rho)^{30}}{\sigma \tilde \tau \ln^{28}((\beta \rho)^{30} /(\sigma \tilde \tau))} \right) + \frac{(\beta \rho)^{58}}{\sigma \ln^{56}((\beta \rho)^{30}/(\sigma \tilde \tau))} & \text{if } \ln(\sigma/(\ln \sigma)^{30}) \leq 4 \pi \beta \rho \lesssim \sigma^{1/59},\\
\frac{1}{\beta \rho} \ln \left( (\beta \rho)^{58/57} \sigma^{28/57} \right) + \left( \frac{(\beta \rho)^2}{\sigma} \right)^{1/57} & \text{if } \sigma^{1/59} \lesssim \beta \rho \lesssim \sigma^{233/580},\\
\frac{\sigma^{4/5}}{(\beta \rho)^2} + \frac{1}{\sigma^{1/5}} + \frac{\sigma^{1/10} (\ln \sigma)^{1/2}}{(\beta \rho)^{1/2}} & \text{if } \sigma^{233/580} \lesssim \beta \rho.
\end{cases}
\end{equation}
The largest error occurs in the second regime if $\beta \rho \sim \beta_{\mathrm{c}} \rho$, and is given by
\begin{equation} \label{eqn-largest-error-rate}
\frac{1}{\ln \sigma} \ln \left( \frac{(\ln \sigma)^{30}}{\ln^{28}((\ln \sigma)^{30})}\right) + \frac{(\ln \sigma)^{58}}{\sigma \ln^{56}((\ln \sigma)^{30})} \lesssim \frac{\ln \ln \sigma}{\ln \sigma}
\end{equation}
for $\sigma$ large. We note that $\tilde{\tau} \sim \sigma^{-1}$ in this case, which follows from \eqref{eqn-chemical-potential-ideal-gas}, \eqref{eqn-def-tilde-pc} and \eqref{eq:8}.  This concludes the proof of Theorem~\ref{thm-lb}.
\appendix
\section{Proof of Dyson Lemma in two dimensions} \label{sec:proof-dysonlemma}
\renewcommand\thefigure{\thesection.\arabic{figure}}
\numberwithin{equation}{section}

	The proof of Lemma~\ref{lem-Dyson-lemma} can be obtained by combining the ideas of the proofs of \cite[Lemma~7]{LSS05} and \cite[Lemma~2]{Seiringer2008}. Since the proof of the two-dimensional version of the relevant Lemma in \cite{LSS05} is not spelled out explicitly, we give the proof of Lemma~\ref{lem-Dyson-lemma} here. For simplicity of the notation, we shall drop the \textasciitilde{} for $v$ and $a$.
	
	Given the points $y_i$, we partition the torus $\Lambda$ into Voronoi cells
	\begin{equation}
	\mathcal{B}_i = \{ x \in \Lambda : d(x,y_i) \leq d(x,y_k) \text{ for all } k \neq i \}.
	\end{equation}
	For any periodic $\psi \in H^1(\Lambda)$ denote by $\xi$ the function with Fourier coefficients $\hat \xi (p) = \chi(p) \hat \psi(p)$. To obtain \eqref{eqn-Dyson-lemma-statement}, it is enough to show that
	\begin{align*} \label{eqn-Dyson-lemma-show}
	\int_{\mathcal{B}_i} |\nabla \xi(x)|^2 + \frac 1 2 v(d(x,y_i)) |\psi(x)|^2 \de x &\geq (1-\epsilon) \int_{\mathcal{B}_i} U_R(d(x,y_i)) |\psi(x)|^2 \de x \asterisknum\\
	&\quad - \frac{1}{\epsilon} \int_{\Rr_+} U_R(t) t \de t \int_\Lambda w_R(x-y_i) |\psi(x)|^2 \de x.
	\end{align*}
	Using the positivity of $v$ and summing over $i$, as well as realizing that for $x \in \mathcal{B}_i$ we have $y_i = y_{\mathrm{NN}}(x)$, we obtain \eqref{eqn-Dyson-lemma-statement}:
	\begin{align}
	\int_\Lambda & | \nabla \xi(x)|^2 + \frac 1 2 \sum_i v(d(x,y_i)) |\psi(x)|^2 \de x = \sum_i \int_{\mathcal{B}_i} \left( |\nabla \xi(x)|^2 + \frac 1 2 v(d(x,y_i)) |\psi(x)|^2 \right) \de x\\
	&\geq \sum_i (1-\epsilon) \int_{\mathcal{B}_i} U_R(d(x,y_i)) |\psi(x)|^2 \de x -  \frac{1}{\epsilon} \int_{\Rr_+} U_R(t) t \de t \sum_i \int_\Lambda w_R(x-y_i) |\psi(x)|^2 \de x \nonumber \\
	&= (1-\epsilon) \int_\Lambda U_R(d(x,y_{\mathrm{NN}}(x))) |\psi(x)|^2 \de x - \frac{1}{\epsilon} \int_{\Rr_+} U_R(t) t \de t \int_\Lambda \sum_i w_R(x-y_i) |\psi(x)|^2 \de x. \nonumber
	\end{align}

	\begin{figure}[htb]
		\centering
		\includegraphics{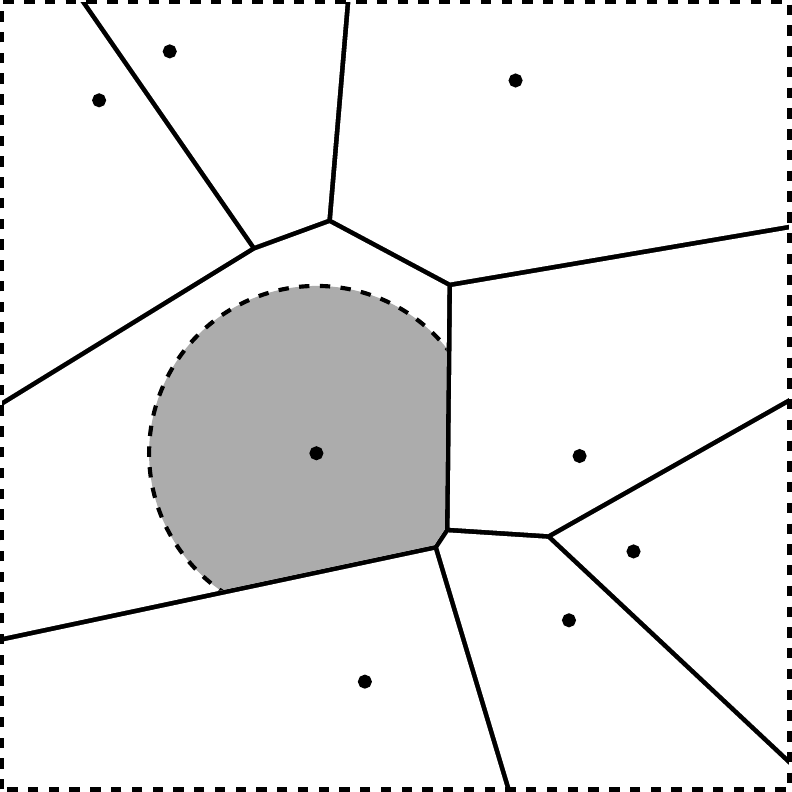}
		\caption{An example of a partition of a subset of $\Lambda$ into Voronoi cells given by the $y_i$ for $n=8$. For one of the $y_i$ the region $\mathcal{B}_R$ is shaded. Note that this image does not show the whole of $\Lambda$ but merely a cutout (that does not respect the periodic boundary conditions).}
		\label{fig-Voronoi}
	\end{figure}

We shall show that \eqref{eqn-Dyson-lemma-show} actually holds with $\mathcal{B}_i$ replaced by the smaller set $\mathcal{B}_R = \mathcal{B}_i \cap \{ x \in \Lambda : d(x,y_i) \leq R \}$ on the left-hand side of the inequality. Since the support of $U_R$ is contained in the interval $[R_0,R]$, the integral over $\mathcal{B}_i$ on the right-hand side is also over $\mathcal{B}_R$. See Figure~\ref{fig-Voronoi} for an illustration of the case $n=8$. 
We shall in fact prove that
	\begin{align}\nonumber
	&\int_{\mathcal{B}_R} |\nabla \xi(x)|^2 + \frac 1 2 v(d(x,y_i)) |\psi(x)|^2 \de x\\ \label{top}
	&\quad \geq \frac{1}{\ln(R/a)} \left[ \frac{1-\epsilon}{R} \int_{\partial \tilde{\mathcal{B}}_R} |\psi(x)|^2 \de \omega_R - \frac{1}{\epsilon} \int_\Lambda |\psi(x)|^2 w_R(x - y_i) \de x \right],
	\end{align}
where $\partial \tilde{\mathcal{B}}_R$ denotes the part of $\partial \mathcal{B}_R$ that is at a distance $R$ from $y_i$;  in Figure~\ref{fig-Voronoi} this set  corresponds to the dashed arc.
This proves the statement for the special case of $U_R$ being a radial $\delta$ function supported on the circle of radius $R$, i.e., $U_R(r) = (R \ln(R/a))^{-1} \delta(r-R)$. By replacing $R$ in the above inequality  by $r$, multiplying by $U_R(r) r \ln(r/a)$ and then finally integrating in $r$ from $R_0$ to $R$, we obtain
	\begin{align*}
	&\int_{\mathcal{B}_R} |\nabla \xi(x)|^2 + \frac 1 2 v(d(x,y_i)) |\psi(x)|^2 \de x \asterisknum\\
	&\quad \geq \int_{R_0}^R U_R(r) r \ln(r/a) \left[ \int_{\mathcal{B}_r} |\nabla \xi(x)|^2 + \frac 1 2 v(d(x,y_i)) |\psi(x)|^2 \de x \right] \de r\\
	&\quad \geq \int_{R_0}^R U_R(r) r \left[ \frac{1-\epsilon}{r} \int_{\partial \tilde{\mathcal{B}}_r} |\psi(x)|^2 \de \omega_r - \frac 1 \epsilon \int_\Lambda |\psi(x)|^2 w_r(x - y_i) \de x \right] \de r\\
	&\quad \geq (1 - \epsilon) \int_{R_0}^R U_R(r) \int_{\partial \tilde{\mathcal{B}}_r} |\psi(x)|^2 \de \omega_r \de r - \frac{1}{\epsilon} \int_{\Rr_+} U_R(t) t \de t \int_{\Lambda} |\psi(x)|^2 w_R(x - y_i) \de x\\
	&\quad = (1 - \epsilon) \int_{\mathcal{B}_R} U_R(d(x,y_i)) |\psi(x)|^2 \de x - \frac{1}{\epsilon} \int_{\Rr_+} U_R(t) t \de t \int_{\Lambda} |\psi(x)|^2 w_R(x - y_i) \de x,
	\end{align*}
	where we used \eqref{eqn-U_R-cond} in the first inequality and the fact that $w_r$ is monotone increasing in $r$ in the last inequality. This proves \eqref{eqn-Dyson-lemma-show}.

In order to prove \eqref{top}, we can without loss of generality assume that  $\partial \tilde{\mathcal{B}}_R$ is non-empty, and set $y_i=0$. We may also assume that $\psi\in H^1(\mathcal{B_R})$ and $\int_{\mathcal {B}_R} |\psi(x)|^2 v(|x|) \de x < \infty$. 
For $\omega \in \mathbb{S}^1$, let
	\begin{equation}\label{def:eta}
	\eta(\omega) = \begin{cases}
	\sqrt{R} \left( \int_{\partial \tilde{\mathcal{B}}_R} |\psi(x)|^2 \de \omega_R \right)^{-1/2}\psi (R \omega) & \text{if } R \omega \in \partial \tilde{\mathcal{B}}_R,\\
	0 & \text{otherwise,}
	\end{cases}
	\end{equation}
	which satisfies $\int_{\mathbb{S}^1} |\eta(\omega)|^2 \de \omega = 1$. In other words, we choose $\eta$ to attain the value of $\psi$ at those boundary points which are at a distance of $R$ from the origin and zero elsewhere, while maintaining an $L^2$-normalization.
By abuse of notation we shall use the same letter for the function on $\Rr^2$ taking values $\eta(x/|x|)$. Recall the notation $\phi_v$ for the minimizer of \eqref{eqn-scattering-length-1} with boundary condition $\phi_v|_{|x|=R} = 1$.
	
	Consider  the expression
	\begin{equation}
	A = \int_{\mathcal{B}_R} \eta(x) \left( \nabla \bar \xi(x) \cdot \nabla \phi_v(x-y_i) + \frac 1 2 v(|x|) \bar \psi(x) \phi_v(x )  \right) \de x.
	\end{equation} 
	An application of the Cauchy-Schwarz inequality gives
	\begin{align*}
	|A|^2 &\leq \int_{\mathcal{B}_R} \left( |\nabla \xi(x)|^2 + \frac 1 2 v(|x|)) |\psi(x)|^2 \right) \de x\\
	&\qquad \times \int_{\mathcal{B}_R} \left( |\nabla \phi_v(x)|^2 + \frac 1 2 v(|x|) |\phi_v(x)|^2 \right) |\eta(x)|^2 \de x. \asterisknum
	\end{align*}
	Since $\phi_v$ is radial, the angular integration over $\eta$ in the second integral contributes a factor of one. Using the definition of the scattering length, the remaining radial integration gives a factor $1/\ln(R/a)$. Thus,
	\begin{equation} \label{eqn-Dyson-lemma-A-ub}
	|A|^2 \ln(R/a) \leq \int_{\mathcal{B}_R} \left( |\nabla \xi(x)|^2 + \frac 1 2 v(|x|) |\psi(x)|^2 \right)\de x.
	\end{equation}
	
	For a lower bound, we note first that by integrating by parts we obtain
	\begin{align*} \label{eqn-Dyson-lemma-ibp}
	\int_{\mathcal{B}_R} \eta(x) \nabla \bar \xi(x) \cdot \nabla \phi_v(x) \de x &= - \int_{\mathcal{B}_R} \bar \xi(x) \eta(x) \Delta \phi_v(x) \de x \asterisknum\\
	&\quad + \int_{\partial \mathcal{B}_R} \bar \xi(x) \eta(x) n \cdot \nabla \phi_v(x)  \de \omega_R,
	\end{align*}
	where $\de \omega_R$ is the surface measure of the boundary of $\mathcal{B}_R$, $n$ is the outward unit normal, and we have used that all relevant derivatives are radial ones since $\phi_v$ is a radial function, and $\eta$ depends only on the angles $x/|x|$.  Note that $\xi(x) = \psi(x) - (2\pi)^{-1} h * \psi(x)$, where $h * \psi(x) = \int_{\Lambda} h(x-y) \psi(y) \de y$, as an easy calculation using the definition of $h$ shows. If we insert this as well as \eqref{eqn-Dyson-lemma-ibp} into the definition of $A$ and use the zero-energy scattering equation \eqref{eq:2} for $\phi_v$, we obtain
	\begin{align*} \label{eqn-Dyson-lemma-lb-1}
	A &= \int_{\partial \mathcal{B}_R} \left[\bar \psi(x) - (2\pi)^{-1} (\overline{h*\psi})(x)\right] \eta(x) n \cdot \nabla \phi_v(x) \de \omega_R\\
	&\quad + \frac{1}{2\pi} \int_{\mathcal{B}_R} (\overline{h*\psi})(x) \eta(x ) \Delta \phi_v(x) \de x\\
	&= \int_{\partial \mathcal{B}_R} \bar \psi(x) \eta(x ) n \cdot \nabla \phi_v(x ) \de \omega_R + \frac{1}{2\pi} \int_{\Lambda} \bar \psi(x) \int_{\mathcal{B}_R} h(y-x) \de \mu(y) \de x, \asterisknum
	\end{align*}
	where
	\begin{equation}
	\de \mu(x) = \eta(x ) \Delta \phi_v(x ) \de x - n \cdot \nabla \phi_v(x ) \eta(x ) \de \omega_R
	\end{equation}
	is a measure supported on $\mathcal{B}_R$. It satisfies
	\begin{equation}
	\int_{\mathcal{B}_R} \de \mu(x) = \int_{\mathcal{B}_R} \eta(x ) \Delta \phi_v(x ) \de x - \int_{\partial \mathcal{B}_R} n \cdot \nabla \phi_v(x ) \eta(x ) \de \omega_R = 0,
	\end{equation}
	as can be seen using again integration by parts. Moreover, since $\Delta \phi_v \geq 0$ and also $n\cdot \nabla \phi_v\geq 0$ on the boundary of $\mathcal{B}_R$, 
	\begin{equation}
	\int_{\mathcal{B}_R} \de |\mu| = 2 \int_{\mathcal{B}_R} |\eta(x )| \Delta \phi_v(x ) \de x \leq 2 \left( \int_{\mathcal{S}^1} |\eta| \right) \int_{0}^R \Delta \phi_v(r) r \de r \leq \frac{2 \sqrt{2\pi}}{\ln(R/a)},
	\end{equation}
	where we used the Cauchy-Schwarz inequality in the last step. Therefore, by invoking the definition of $f_R$ from \eqref{eqn-Dyson-lemma-def-f-w}, we obtain
	\begin{equation}
	\left| \int_{\mathcal{B}_R} h(y-x) \de \mu(y) \right| = \left| \int_{\mathcal{B}_R} (h(y - x) - h(x )) \de \mu(y) \right| \leq \frac{2 \sqrt{2\pi}}{\ln(R/a)} f_R(x ).
	\end{equation}
	This enables us to estimate the second term in \eqref{eqn-Dyson-lemma-lb-1} from below as
	\begin{align*}
	-\frac{1}{2\pi} \left| \int_{\Lambda} \bar \psi(x) \int_{\mathcal{B}_R} h(y-x) \de \mu(y) \de x \right| &\geq - \frac{1}{2\pi} \frac{2 \sqrt{2\pi}}{\ln(R/a)} \int_{\Lambda} |\psi(x)| f_R(x ) \de x\\
	&\geq - \frac{1}{\ln(R/a)} \left( \int_\Lambda |\psi(x)|^2 w_R(x ) \de x \right)^{1/2}, \asterisknum
	\end{align*}
	where we used again the Cauchy-Schwarz inequality as well as the definition of $w_R$ from \eqref{eqn-Dyson-lemma-def-f-w}. 
	Using \eqref{def:eta} as well as the explicit form of $\phi_v$ outside the support of $v$, we see that the first term in \eqref{eqn-Dyson-lemma-lb-1} equals
	\begin{equation}
	\int_{\partial \mathcal{B}_R} \bar \psi(x) \eta(x) n \cdot \nabla \phi_v(x ) \de \omega_R = \frac{1}{\sqrt{R}\ln(R/a)} \left( \int_{\partial \tilde{\mathcal{B}}_R} |\psi(x)|^2 \de \omega_R \right)^{1/2}.
	\end{equation}
	Therefore,
	\begin{equation}
	|A| \geq \frac{1}{\ln(R/a)} \left[ \frac{1}{\sqrt{R}} \left( \int_{\partial \tilde{\mathcal{B}}_R} |\psi(x)|^2 \de \omega_R \right)^{1/2} - \left( \int_\Lambda |\psi(x)|^2 w_R(x ) \de x \right)^{1/2} \right].
	\end{equation}
	Another application of the Cauchy-Schwarz inequality gives for any $\epsilon > 0$
	\begin{equation} \label{eqn-Dyson-lemma-A-lb}
	|A|^2 \ln(R/a) \geq \frac{1}{\ln(R/a)} \left[ \frac{1-\epsilon}{R} \int_{\partial \tilde{\mathcal{B}}_R} |\psi(x)|^2 \de \omega_R - \frac{1}{\epsilon} \int_\Lambda |\psi(x)|^2 w_R(x - y_i) \de x \right].
	\end{equation}
	Hence, combining \eqref{eqn-Dyson-lemma-A-ub} and \eqref{eqn-Dyson-lemma-A-lb}, we obtain \eqref{top}.
	This completes the proof.
\vspace{0.5cm}

\textit{Acknowledgments.} Financial support by the European Research Council (ERC) under the European Union's Horizon 2020 research and innovation programme (grant agreement No~694227) is gratefully acknowledged. A. D. acknowledges funding from the European Union's Horizon 2020 research and innovation programme under the Marie Sklodowska-Curie grant agreement No~836146.

\end{document}